\documentclass[floatfix,twocolumn,showpacs,preprintnumbers,amsmath,amssymb,pra,superscriptaddress,longbibliography]{revtex4-1}
\usepackage{color}
\usepackage[usenames,dvipsnames,svgnames,table]{xcolor}
\usepackage[colorlinks=true,linkcolor=blue,urlcolor=blue,citecolor=blue]{hyperref}
\usepackage{mathtools}
\usepackage{graphicx}
\usepackage{dcolumn}
\usepackage{array}
\usepackage{lipsum}
\usepackage{bm}
\usepackage{subfigure}
\usepackage{amssymb}
\usepackage{multirow}
\usepackage{tabularx}
\usepackage{amsmath}
\usepackage{braket}
\graphicspath{{plots/}}
 \usepackage{lipsum}
\usepackage{mathrsfs}

\DeclareMathOperator*{\SumInt}{%
\mathchoice%
  {\ooalign{$\displaystyle\sum$\cr\hidewidth$\displaystyle\int$\hidewidth\cr}}
  {\ooalign{\raisebox{.14\height}{\scalebox{.7}{$\textstyle\sum$}}\cr\hidewidth$\textstyle\int$\hidewidth\cr}}
  {\ooalign{\raisebox{.2\height}{\scalebox{.6}{$\scriptstyle\sum$}}\cr$\scriptstyle\int$\cr}}
  {\ooalign{\raisebox{.2\height}{\scalebox{.6}{$\scriptstyle\sum$}}\cr$\scriptstyle\int$\cr}}
}

\newcommand{\beq}{\begin{equation}}
\newcommand{\eeq}{\end{equation}}
\newcommand{\bea}{\begin{eqnarray}}
\newcommand{\eea}{\end{eqnarray}}

\begin{document}
\title{
Analyzing X-ray Thomson scattering experiments of warm dense matter in the imaginary-time domain: theoretical models and simulations
}

\author{Tobias Dornheim}
\email{t.dornheim@hzdr.de}

\affiliation{Center for Advanced Systems Understanding (CASUS), D-02826 G\"orlitz, Germany}
\affiliation{Helmholtz-Zentrum Dresden-Rossendorf (HZDR), D-01328 Dresden, Germany}

\author{Jan Vorberger}
\affiliation{Helmholtz-Zentrum Dresden-Rossendorf (HZDR), D-01328 Dresden, Germany}

\author{Zhandos A.~Moldabekov}

\affiliation{Center for Advanced Systems Understanding (CASUS), D-02826 G\"orlitz, Germany}
\affiliation{Helmholtz-Zentrum Dresden-Rossendorf (HZDR), D-01328 Dresden, Germany}


\author{Maximilian B\"ohme}

\affiliation{Center for Advanced Systems Understanding (CASUS), D-02826 G\"orlitz, Germany}
\affiliation{Helmholtz-Zentrum Dresden-Rossendorf (HZDR), D-01328 Dresden, Germany}
\affiliation{Technische  Universit\"at  Dresden,  D-01062  Dresden,  Germany}

\begin{abstract}
The rigorous diagnostics of experiments with warm dense matter (WDM) is notoriously difficult. A key method is given by X-ray Thomson scattering (XRTS), but the interpretation of XRTS measurements is usually based on theoretical models that entail various approximations. Recently, Dornheim \emph{et al.}~[arXiv:2206.12805] have introduced a new framework for temperature diagnostics of XRTS experiments that is based on imaginary-time correlation functions (ITCF). On the one hand, switching from the frequency- to the imaginary-time domain gives one direct access to a number of physical properties, which facilitates the extraction of the temperature of arbitrarily complex materials without any models or approximations. On the other hand, the bulk of theoretical works in dynamic quantum many-body theory is devoted to the frequency-domain, and, to our knowledge, the manifestation of physics properties within the ITCF remains poorly understood. In the present work, we aim to change this unsatisfactory situation by introducing a simple, semi-analytical model for the imaginary-time dependence of two-body correlations within the framework of imaginary-time path integrals. As a practical example, we compare our new model to extensive \emph{ab initio} path integral Monte Carlo results for the ITCF of a uniform electron gas, and find excellent agreement over a broad range of wave numbers, densities, and temperatures.
\end{abstract}

\maketitle

\section{Introduction\label{sec:introduction}}

The study of matter at extreme pressures ($P\sim1-10^4\,$MBar) and temperatures ($T=10^4-10^8\,$K)~\cite{fortov_review,drake2018high} constitutes a highly active frontier at the interface of a number of disciplines such as plasma physics, material science, and quantum chemistry. These conditions naturally occur in a host of astrophysical objects such as giant planet interiors~\cite{Benuzzi_Mounaix_2014,militzer1} and brown dwarfs~\cite{saumon1,becker}. In addition, they are important for technological applications such as the discovery of novel materials~\cite{Kraus2016,Kraus2017,Lazicki2021} and hot-electron chemistry~\cite{Brongersma2015,ernstorfer,ernstorfer2}. A particularly important and topical example is given by inertial confinement fusion~\cite{Betti2016,Zylstra2022} as it is realized for example at the National Ignition Facility~\cite{Moses_NIF}; here the fuel capsule is predicted to traverse the aforementioned regime on its path towards ignition~\cite{hu_ICF}.

From a theoretical perspective, this \emph{warm dense matter} (WDM) can be characterised in terms of a few dimensionless parameters that are of the order of unity~\cite{review,Ott2018}: 1) the Wigner-Seitz radius $r_s=d/a_\textnormal{B}$ is given by the ratio of the average interparticle distance to the first Bohr radius; this density parameters also serves as the quantum coupling parameter in the WDM regime~\cite{Dornheim_HEDP_2022}. 2) the degeneracy temperature $\Theta=k_\textnormal{B}T/E_\textnormal{F}$ measures the thermal energy in units of the electronic Fermi energy $E_\textnormal{F}$~\cite{quantum_theory}, with $\Theta\ll 1$ and $\Theta\gg 1$ corresponding to the fully degenerate and semi-classical regime, respectively.

Consequently, the rigorous theoretical description of WDM is notoriously difficult as there are no small parameters that can serve as the basis for a suitable expansion~\cite{wdm_book,new_POP}.
While the high current interest in the properties of WDM has sparked a surge of new developments regarding different methods~\cite{Driver_PRL_2012,Brown_PRL_2013,Sjostrom_PRB_2014,karasiev_importance,Zhang_POP_2016,Dornheim_POP_2017,manuel,ksdt,GRABOWSKI2020100905,dynamic2,Karasiev_PRL_2018,Ramakrishna_PRB_2020,Militzer_PRE_2021,Ramakrishna_PRB_2021,Dornheim_PRL_2020_ESA,Ding_PRL_2018,new_POP,pribram,Moldabekov_SciPost_2022,https://doi.org/10.48550/arxiv.2110.01034,Moldabekov_JCTC_2022,Fiedler_PRR_2022}, we are as of yet far away from having a complete understanding of the full WDM regime.

This unsatisfactory situation also constitutes a serious obstacle for the interpretation of experiments with WDM~\cite{falk_wdm}, as, due to the extreme conditions, often even basic parameters such as the temperature or the density cannot be directly measured and have to be inferred from other observations. In this regard, a very important method for the diagnostics of WDM is given by X-ray Thomson scattering (XRTS)~\cite{siegfried_review,kraus_xrts}; here an X-ray beam is produced either from backlighter sources~\cite{MacDonald_POP_2022} or using free-electron X-ray lasers (XFEL) that have become available at large research facilities such as LCLS in the USA~\cite{LCLS_2016}, SACLA in Japan~\cite{Pile2011}, or the European XFEL in Germany~\cite{Tschentscher_2017}.
More specifically, an XRTS measurement gives one access to the scattering intensity signal, which is given by the convolution of the dynamic structure factor (DSF) $S(\mathbf{q},\omega)$ with the combined source and instrument function $R(\omega)$~\cite{siegfried_review},
\begin{eqnarray}\label{eq:convolution}
I(\mathbf{q},\omega) = S(\mathbf{q},\omega) \circledast R(\omega)\ .
\end{eqnarray}
Unfortunately, the numerical deconvolution of Eq.~(\ref{eq:convolution}) is generally prevented by noise in the experimental measurement. Therefore, XRTS does not give one direct access to $S(\mathbf{q},\omega)$, which contains the sought-after physical information about the system of interest. In this situation, the most widely used approach for the interpretation of XRTS experiments is to construct an approximate model for $S(\mathbf{q},\omega)$, which is then convolved with $R(\omega)$ and subsequently compared to the experimental signal $I(\mathbf{q},\omega)$. On the one hand, one can determine a-priori unknown free parameters such as the temperature $T$ by finding the best fit between theory and experiment in this way. On the other hand, the thus inferred parameters can depend arbitrarily strongly on the employed model for $S(\mathbf{q},\omega)$, which are usually based on approximations such as the widely used Chihara decomposition~\cite{Gregori_PRE_2003,Chihara_1987,kraus_xrts}.

Very recently, Dornheim \emph{et al.}~\cite{Dornheim_T_2022} have suggested to circumvent this obstacle by switching from the usual $\omega$-representation to the imaginary-time domain. The required transformation is given by a two-sided Laplace transform
\begin{eqnarray}\label{eq:Laplace}
F(\mathbf{q},\tau) &=& \mathcal{L}\left[S(\mathbf{q},\omega)\right] \\\nonumber &=& \int_{-\infty}^\infty \textnormal{d}\omega\ e^{-\omega\tau} S(\mathbf{q},\omega)\ ,
\end{eqnarray}
which connects the DSF to the imaginary-time density--density correlation function (ITCF) $F(\mathbf{q},\tau)$. The latter naturally emerges in Feynman's imaginary-time path integral picture of statistical mechanics~\cite{Dornheim_insight_2022,Dornheim_JCP_ITCF_2021}, and corresponds to the usual intermediate scattering function evaluated at an imaginary time $t=-i\hbar\tau$
with $\tau\in[0,\beta]$ and the inverse temperature $\beta=1/k_\textnormal{B}T$.

A particular advantage of the $\tau$-domain is given by the well-known convolution theorem,
\begin{eqnarray}\label{eq:convolution_theorem}
\mathcal{L}\left[S(\mathbf{q},\omega)\right] = \frac{\mathcal{L}\left[S(\mathbf{q},\omega) \circledast R(\omega)\right]}{\mathcal{L}\left[R(\omega)\right]}\ ,
\end{eqnarray}
which makes the deconvolution trivial; in practice, it is easy to compute the Laplace transform of the XRTS intensity and to subsequently divide it by the Laplace transform of the instrument function $R(\omega)$. We note that accurate knowledge of $R(\omega)$ is usually available from source monitoring at XFEL facilities, or from the characterisation of backlighter emission spectra~\cite{MacDonald_POP_2022}.
In this way, Eq.~(\ref{eq:convolution_theorem}) gives one direct access to physical information, which allows for the accurate inference of the temperature of arbitrarily complex systems in thermodynamic equilibrium without any model, simulation, or approximation~\cite{Dornheim_T_2022}.

From a mathematical perspective, it is well-known that the two-sided Laplace transform defined in Eq.~(\ref{eq:Laplace}) constitutes a unique transformation, which means that the ITCF $F(\mathbf{q},\tau)$ contains exactly the same information as the usual DSF $S(\mathbf{q},\omega)$.
Indeed, it has subsequently been demonstrated in Ref.~\cite{Dornheim_insight_2022} that $F(\mathbf{q},\tau)$ gives one direct access to a wealth of physical information, such as the excitation energies of quasi-particles. Moreover, even complex physical processes such as the exchange--correlation induced alignment of pairs of electrons~\cite{Dornheim_Nature_2022} that lead to a \emph{roton-type} minimum in the dispersion $\omega(\mathbf{q})$ of the DSF can be observed and interpreted in the $\tau$-domain.
At the same time, we note that the bulk of dynamic quantum-many body theory has been developed in the frequency domain to describe $S(\mathbf{q},\omega)$ and related properties. 

In the present work, we aim to partly change this unsatisfacory situation. More specifically, we present a simple, semi-analytical model for the imaginary-time diffusion process, that is capable to accurately capture the dependence of the ITCF on $\tau$ over a broad range of parameters. As a practical application, we consider the uniform electron gas (UEG)~\cite{review,loos,quantum_theory}, which constitutes the archetypical model for interacting electrons and has given important insights in a number of different contexts. 
Moreover, the recent interest in the properties of the UEG at WDM conditions~\cite{review,status,Dornheim_POP_2017} allows us to compare our new model---in addition to other models such as the well-known random phase approximation (RPA)---to highly accurate \emph{ab initio} path integral Monte Carlo (PIMC) simulation results for $F(\mathbf{q},\tau)$.

We are convinced that these new insights into the imaginary-time dependence of electron--electron correlations in the WDM regime constitute an important basis for future studies of real WDM applications that include both electrons and ions. 
The paper is organised as follows:
In Sec.~\ref{sec:theory}, we introduce the relevant theoretical background, starting with an introduction to the PIMC method and its natural connection to the ITCF in Sec.~\ref{sec:PIMC}.
Sec.~\ref{sec:LRT} is devoted to a brief introduction to linear-response theory, followed by a concise overview of a few important properties of the ITCF in Sec.~\ref{sec:properties}. The theoretical background is concluded by our new imaginary-time diffusion model that is introduced in Sec.~\ref{sec:delocalization}.
In Sec.~\ref{sec:results}, we present an extensive analysis of different properties of the ITCF, starting with a discussion of its dependence on the imaginary time $\tau$ in Sec.~\ref{sec:tau}; the subsequent Sec.~\ref{sec:q}
 and Sec.~\ref{sec:temperature} are devoted to the wave number $q$ and temperature $\Theta$, respectively. 
 The paper is concluded by a discussion and outlook in Sec.~\ref{sec:summary}.

\section{Theory\label{sec:theory}}

We assume Hartree atomic units throughout. A detailed introduction to the UEG model, including the UEG Hamiltonian in different representations, can be found, e.g., in Refs.~\cite{quantum_theory,review}.

\subsection{Imaginary-time path integral Monte Carlo\label{sec:PIMC}}

Since its original inception for the description of ultacold bosonic $^4$He some six decades ago~\cite{Fosdick_PR_1966,Jordan_PR_1968}, the \emph{ab initio} PIMC method~\cite{Berne_JCP_1982,Pollock_PRB_1984,Takahashi_Imada_PIMC_1984} has emerged as one of the most successful tools for the description of nonideal quantum many-body systems in thermodynamic equilibrium. Since a detailed introduction to PIMC has been presented elsewhere~\cite{cep}, we will here restrict ourselves to outline the main idea, and how it relates to the estimation of imaginary-time correlation functions such as $F(\mathbf{q},\tau)$.

As a starting point, we express the canonical partition function of $N=N^\uparrow+N^\downarrow$ electrons in a cubic volume $\Omega=L^3$ and at an inverse temperature of $\beta=1/T$ as
\begin{widetext} 
\begin{eqnarray}\label{eq:Z}
Z_{\beta,N,\Omega} &=& \frac{1}{N^\uparrow! N^\downarrow!} \sum_{\sigma^\uparrow\in S_N^\uparrow} \sum_{\sigma^\downarrow\in S_N^\downarrow} \textnormal{sgn}(\sigma^\uparrow,\sigma^\downarrow) \int_\Omega d\mathbf{R} \bra{\mathbf{R}} e^{-\beta\hat H} \ket{\hat{\pi}_{\sigma^\uparrow}\hat{\pi}_{\sigma^\downarrow}\mathbf{R}}\ ,
\end{eqnarray}\end{widetext}
and the variable $\mathbf{R}=(\mathbf{r}_1,\dots,\mathbf{r}_N)^T$ contains the coordinates of all $N^\uparrow$ majority and $N^\downarrow$ minority electrons. We note that we restrict ourselves to the unpolarized (i.e., paramagnetic) case of $N^\uparrow=N^\downarrow=N/2$ throughout this work.
The summation over all elements $\sigma^{\uparrow,\downarrow}$ of the respective permutation group $S_{N^{\uparrow,\downarrow}}$, with $\hat{\pi}_{\sigma^{\uparrow,\downarrow}}$ being the corresponding permutation operator, takes into account the anti-symmetry of the partition function with respect to the exchange of particle coordinates of identical fermions. In particular, the sign function $\textnormal{sgn}(\sigma^\uparrow,\sigma^\downarrow)$ is positive (negative) for an even (odd) number of pair exchanges in a particular combination of $\sigma^\uparrow$ and $\sigma^\downarrow$.
For completeness, we note that the corresponding bosonic partition function can be obtained by setting the sign function to unity.

\begin{figure}\centering
\includegraphics[width=0.475\textwidth]{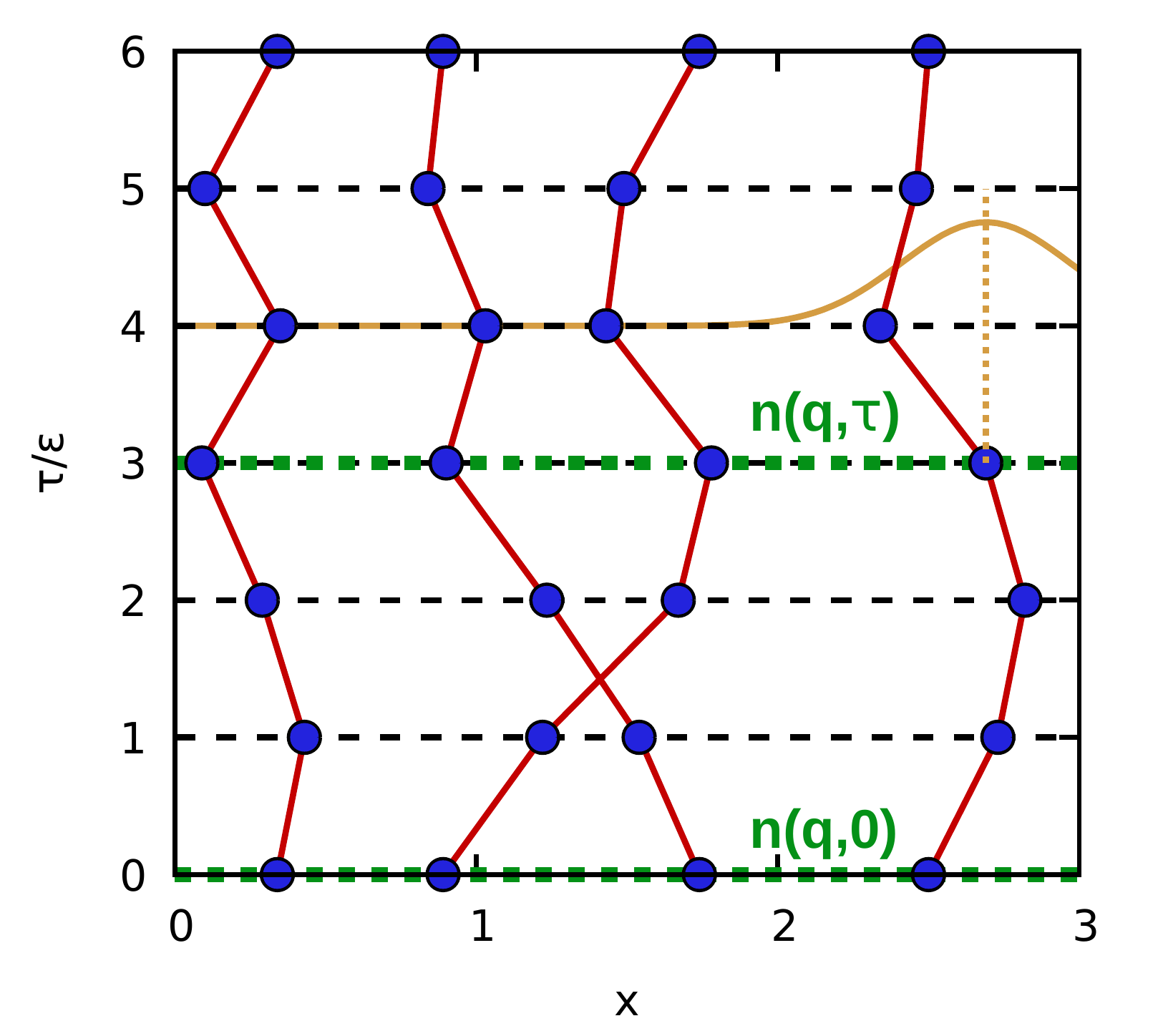}
\caption{\label{fig:illustration}
Schematic illustration of Feynman's imaginary-time path integral formalism. Shown is a configuration $\mathbf{X}$ that consists of $N=4$ electrons on $P=6$ imaginary-time slices. The yellow Gaussian illustrates the ideal kinetic density matrix [Eq.~(\ref{eq:rho_ideal})], which effectively acts as harmonic spring between beads on adjacent slices. The dashed green lines illustrate the definition of the ITCF $F(\mathbf{q},\tau)$ as the correlated evaluation between densities in reciprocal space at different imaginary times, which can naturally be estimated within the PIMC formalism. Due to the single particle exchange of the two electrons in the center, the configuration weight of the depicted configuration is actually negative, $W(\mathbf{X})<0$, thereby contributing to the fermion sign problem~\cite{dornheim_sign_problem}. 
Taken from Ref.~\cite{Dornheim_insight_2022} with the permission of the authors.
}
\end{figure}

Unfortunately, the direct evaluation of the matrix elements of the density operator $\hat\rho = e^{-\beta\hat H}$ is precluded by the noncommutativity of the kinetic ($\hat K$) and potential ($\hat V$) contributions to the total Hamiltonian $\hat H = \hat K + \hat V$.
A well-known route to overcome this obstacle is based on the utilization of the exact semi-group property of the density operator, which eventually leads to 
\begin{widetext}
\begin{eqnarray}\label{eq:Z_modified}
Z_{\beta,N,\Omega} &=& \frac{1}{N^\uparrow! N^\downarrow!} \sum_{\sigma^\uparrow\in S_{N^\uparrow}} \sum_{\sigma^\downarrow\in S_{N^\downarrow}} \textnormal{sgn}(\sigma^\uparrow,\sigma^\downarrow)\int_\Omega d\mathbf{R}_0\dots d\mathbf{R}_{P-1}
\bra{\mathbf{R}_0}e^{-\epsilon\hat H}\ket{\mathbf{R}_1} \bra{\mathbf{R}_1} \dots 
\bra{\mathbf{R}_{P-1}} e^{-\epsilon\hat H} \ket{\hat{\pi}_{\sigma^\uparrow}\hat{\pi}_{\sigma^\downarrow}\mathbf{R}_0}\ , \quad
\end{eqnarray}
\end{widetext}
In essence, we have thus replaced the evaluation of a single density matrix at the Temperature $T$ with the evaluation of $P$ density matrices at $P$ times the original temperature. Moreover, each corresponding exponential function can be viewed as a propagation in the imaginary time by an amount of $\Delta t = -i\epsilon$; this is the origin of Feynman's celebrated imaginary-time path integral formalism that is illustrated in Fig.~\ref{fig:illustration}.
Shown is a configuration of $N=4$ electrons in the $\tau$-$x$-plane, with $P=6$ high-temperature factors, corresponding to $P$ imaginary-time slices of length $\Delta\tau=\epsilon$. Evidently, each particle is represented by a closed path along the imaginary time. In practice, we have thus mapped the quantum system of interest onto a classical system of interacting ring polymers; this concept is often being referred to as \emph{classical isomorphism} in the literature~\cite{Chandler_JCP_1981}.
We note that the closed nature of the polymers is a direct consequence of the definition of the partition function as the trace of the density operator, basically leading to the same start and end points $\mathbf{R}$. The situation becomes somewhat more interesting for the cases of indistinguishable quantum particles (i.e., fermions or bosons), where the application of the permutation operator $\hat{\pi}_{\sigma^{\uparrow,\downarrow}}$ in Eq.~(\ref{eq:Z_modified}) leads to polymers with more than a single particle in it. Indeed, 
such an example is depicted in Fig.~\ref{fig:illustration}, where the two particles in the center form a single permutation cycle~\cite{Dornheim_permutation_cycles}. For completeness, we note that the presence of this single pair exchange results in a negative sign function $\textnormal{sgn}(\sigma^\uparrow,\sigma^\downarrow)$. This is the root cause of the notorious fermion sign problem~\cite{Loh_PRB_1990,troyer,dornheim_sign_problem,Dornheim_JPA_2021}, the main computational bottleneck of PIMC simulations of WDM, which is discussed in more detail below.

Let us postpone the discussion of the other elements in Fig.~\ref{fig:illustration}, and return to the partition function Eq.~(\ref{eq:Z_modified}). For sufficiently large $P$, the matrix elements of $e^{-\epsilon\hat H}$ can be evaluated using a suitable high-temperature approximation. For the parameters that are of interest in the present work, it is most convenient to make use of the primitive factorization 
$e^{-\epsilon\hat H}\approx e^{-\epsilon\hat K}e^{-\epsilon \hat V}$, which becomes exact in the limit of large $P$ as $\mathcal{O}\left(P^{-2}\right)$~\cite{kleinert2009path}. We note that the incorporation of higher-order factorizations into PIMC simulations has been discussed extensively in the literature~\cite{Takahashi_Imada_higher_order,brualla_JCP_2004,sakkos_JCP_2009,Zillich_JCP_2010}, and becomes advisable for lower temperatures.

In the end, one can express the partition function as 
\begin{eqnarray}\label{eq:Z_final}
Z = \SumInt_\Omega\textnormal{d}\mathbf{X}\ W(\mathbf{X})\ ,
\end{eqnarray}
where the configuration variable $\mathbf{X}=(\mathbf{R}_0,\dots,\mathbf{R}_{P-1})^T$ contains the coordinates of all $N$ particles on each of the $P$ imaginary-time slices. In addition, it should be noted that the symbolic notation $\SumInt_\Omega\textnormal{d}\mathbf{X}$ includes both the integration over all coordinates, as well as the summation over all possible permutation configurations.

From a practical perspective, the partition function has thus been re-cast into a sum over all possible paths (including all permutation topologies) $\mathbf{X}$, where each path contributes according to its configuration weight $W(\mathbf{X})$, which is a function that we can straightforwardly evaluate. More specifically, we can factorize the weight contribution of each imaginary-time slice $\alpha$ into a kinetic and a potential contribution, $W^K_\alpha(\mathbf{X})$ and $W^V_\alpha(\mathbf{X})$. The latter is simply determined from the the total potential energy $V(\mathbf{R}_\alpha)$ on slice $\alpha$, and we find 
\begin{eqnarray}
W^V_\alpha(\mathbf{X}) = e^{-\epsilon V(\mathbf{R}_\alpha)}\ . 
\end{eqnarray}
The kinetic part follows from the ideal single-particle density matrix for the reduced inverse temperature $\epsilon$,
\begin{eqnarray}\label{eq:rho_ideal}
\rho_0(\mathbf{r}_\alpha,\mathbf{r}_{\alpha+1};\epsilon) = \lambda_\epsilon^{-3} e^{-\frac{\pi}{\lambda_\epsilon^2}(\mathbf{r}_\alpha-\mathbf{r}_{\alpha+1})^2}\ ,
\end{eqnarray}
and is given by 
\begin{eqnarray}\label{eq:W_kinetic}
W^K_\alpha(\mathbf{X}) = \prod_{l=1}^N \rho_0(\mathbf{r}_{l,\alpha},\mathbf{r}_{l,\alpha+1};\epsilon)\ .
\end{eqnarray}
Returning once more to the illustration of the imaginary-time path integral picture in Fig.~\ref{fig:illustration}, we can say that the beads of the polymers interact with each other via the usual pair potential on a given time slice; the pair potential does not act between sets of coordinates for different $\tau$. 
In addition, beads of the same polymer on adjacent time slices are effectively linked via a harmonic spring potential by the ideal single-particle density matrix, Eq.~(\ref{eq:rho_ideal}). This is indicated by the yellow Gaussian curve at the RHS of Fig.~\ref{fig:illustration}.
In particular, the width of the Gaussian is proportional to $\lambda_\epsilon=\sqrt{2\pi\epsilon}$, with $\epsilon=\beta/P$. Therefore, the Gaussian becomes infinitely narrow in the classical limit of $\epsilon\to0$, where the paths are given as straight lines, which corresponds to point particles in coordinate space. Conversely, the Gaussian becomes increasingly broad with decreasing temperature, which leads to more extended paths. As we shall see, this imaginary-time diffusion process decisively shapes the $\tau$-dependence of the ITCF $F(\mathbf{q},\tau)$, and can thus be used to explain and interpret observations from XRTS experiments.

The basic idea of the PIMC method~\cite{cep} is to randomly generate all paths $\mathbf{X}$ with a probability that is proportional to their respective configuration weight $W(\mathbf{X})$ via the Metropolis algorithm~\cite{metropolis,Chib}. This is relatively straightforward for bosons (and also hypothetical distinguishable particles that are sometimes being referred to as boltzmannons in the literature~\cite{Dornheim_CPP_2016}. Indeed, modern sampling techniques such as the worm algorithm by Boninsegni \emph{et al.}~\cite{boninsegni1,boninsegni2} allow for the efficient exploration of different permutation topologies, and exact PIMC simulations of up to $N\sim10^4$ quantum particles are feasible.
This situation dramatically changes for the case of fermions, such as the electrons in the warm dense UEG in which we are interested in the present work. In particular, the aforementioned antisymmetry of the fermionic thermal density matrix with respect to the exchange of particle coordinates implies that the configuration weight can be both positive and negative; indeed, it holds $W(\mathbf{X})<0$ for the configuration depicted in Fig.~\ref{fig:illustration}. Therefore, $W(\mathbf{X})$ cannot correspond to a proper probability distribution, which must be strictly non-negative. While this problem can be formally circumvented~\cite{dornheim_sign_problem} by sampling the configurations proportional to the modulus weight $|W(\mathbf{X})|$ and keeping track of the respective sign of $W(\mathbf{X})$, it also means that the fermionic partition function $Z$ is given as the sum over a large number of positive and negative contributions, which might cancel to a large degree. This cancellation is the origin of the fermion sign problem~\cite{Loh_PRB_1990,troyer,dornheim_sign_problem}, and leads to an exponential increase in the required compute time with increasing $\beta$ (i.e., decreasing the temperature $T$) and with increasing the system size $N$. Indeed, the sign problem constitutes the main computational bottleneck in our simulations and limits the regime of feasible simulation parameters, in particular  with respect to $N$ and $\Theta$. We note that a detailed review to the fermion sign problem in direct PIMC simulations has been presented elsewhere~\cite{dornheim_sign_problem,Dornheim_JPA_2021} and is beyond the scope of this paper.

Due to its fundamental nature, a number of different approaches to deal with the sign problem in the simulation of quantum Fermi systems at finite temperature have been suggested in the literature~\cite{Brown_PRL_2013,Blunt_PRB_2014,Schoof_PRL_2015,Dornheim_JCP_2015,Malone_JCP_2015,Malone_PRL_2016,Joonho_JCP_2021,Yilmaz_JCP_2020,dornheim_pre,groth_jcp,Filinov_PRE_2015,Claes_PRB_2017,Filinov_CPP_2021,Hirshberg_JCP_2020,Dornheim_Bogoliubov_2020,Yilmaz_JCP_2020,Yunou_JCP_2022,Rubenstein_JCTC_2020}. A particularly important approach is given by the \emph{restricted} PIMC (RPIMC) method that has been suggested by Ceperley~\cite{Ceperley1991}. In particular, it can be shown that the fermionic partition function can be rewritten into a sum that only contains positive terms by enforcing restrictions on the nodal structure of the thermal density matrix. On the one hand, RPIMC is, therefore, completely free of the sign problem. This has allowed Militzer and co-workers to carry out a host of RPIMC studies of real materials, e.g., Refs.~\cite{Militzer_2008,Driver_PRL_2012,Driver_PRB_2016,Driver_PRE_2018}.
On the other hand, the actual nodal structure of an interacting quantum Fermi system is generally not known, and one has to rely on approximations; often, the nodal structure of an ideal Fermi gas is used. Therefore, RPIMC is afflicted with an uncontrolled approximation. Recently, Schoof \emph{et al.}~\cite{Schoof_PRL_2015} have shown that the fixed-node approximation leads to errors in the exchange--correlation energy of the warm dense UEG of $\Delta E_\textnormal{xc}\sim10\%$ for low temperatures and high densities; these results have subsequently been confirmed in independent studies~\cite{Malone_PRL_2016,Joonho_JCP_2021}. In addition, we mention the recent comparison between RPIMC and exact direct PIMC results for the momentum distribution of the UEG in Refs.~\cite{Dornheim_PRB_nk_2021,Dornheim_PRE_2021}, where it was found that the fixed-node approximation has a small impact for moderately quantum degenerate systems, $\Theta\gtrsim 1$.

In the context of the present work, the most important limitation of RPIMC is that it breaks the translation invariance of the thermal density matrix with respect to the imaginary time $\tau$. This precludes the direct estimation of the ITCF (and also of other imaginary-time correlation functions such as the Matsubara Green function~\cite{boninsegni2} and the velocity autocorrelation function~\cite{Reichman_Rabani_JCP_2002}), which is straightforward in the direct PIMC approach that we use in this work.

Let us re-call the definition of the ITCF as the intermediate scattering function~\cite{siegfried_review} evaluated at imaginary times $t=-i\tau$,
\begin{eqnarray}\label{eq:ISF}
F(\mathbf{q},\tau) = \braket{\hat{n}(\mathbf{q},0)\hat{n}(-\mathbf{q},\tau)}\ ,
\end{eqnarray}
with $\tau\in[0,\beta]$.
It is easy to see that we can evaluate Eq.~(\ref{eq:ISF}) in the path integral formalism for integer multiples of the imaginary time step $\epsilon$ by inserting the respective density operators  at a distance of $\eta\in\{0,\dots,P-1\}$ into Eq.~(\ref{eq:Z_modified}),
\begin{widetext}
\begin{eqnarray}\label{eq:F_estimator}
F(\mathbf{q},\tau_j) &=& \frac{1}{Z_{\beta,N,\Omega}} \frac{1}{N^\uparrow! N^\downarrow!} \sum_{\sigma^\uparrow\in S_{N^\uparrow}} \sum_{\sigma^\downarrow\in S_{N^\downarrow}} \textnormal{sgn}(\sigma^\uparrow,\sigma^\downarrow) \int d\mathbf{R}_0\dots d\mathbf{R}_{P-1}
\bra{\mathbf{R}_0}{\hat n}(\mathbf{q})e^{-\epsilon\hat H}\ket{\mathbf{R}_1} \bra{\mathbf{R}_1}e^{-\epsilon\hat H}\ket{\mathbf{R}_2}\dots\\\nonumber & & \dots \bra{\mathbf{R}_{j}}{\hat n}(-\mathbf{q})e^{-\epsilon\hat H}\ket{\mathbf{R}_{j+1}} \dots 
\bra{\mathbf{R}_{P-1}} e^{-\epsilon\hat H} \ket{\hat{\pi}_{\sigma^\uparrow}\hat{\pi}_{\sigma^\downarrow}\mathbf{R}_0}\ .
\end{eqnarray}
\end{widetext}
This is illustrated in Fig.~\ref{fig:illustration} by the horizontal green dashed lines. In practice, $F(\mathbf{q},\tau)$ is thus obtained by computing the correlation between density operators at different imaginary times, which, in thermodynamic equilibrium, only depends on the difference of imaginary-time arguments $\tau$.

Let us conclude this section with a note regarding the PIMC estimation of $F(\mathbf{q},\tau)$ for fermions. Specifically, it is well known that the utilization of antisymmetric imaginary-time propagators alleviates the sign problem for a small number of time slices $P$~\cite{Chin_PRE_2015,Dornheim_NJP_2015,Dornheim_CPP_2019,Filinov_CPP_2021}. Therefore, it has been shown that the combination of this idea with a higher-order factorization of the thermal density matrix allows one to obtain accurate results for parameters that are beyond the scope of direct PIMC due to the fermion sign problem~\cite{Dornheim_POP_2017,review}. Yet, the small number of time slices means that $F(\mathbf{q},\tau)$ will only be available on a sparse $\tau$-grid. Such data can still be useful for the evaluation of $\tau$-decay measures [see Eq.~(\ref{eq:decay_measure}) below], which constitute the natural analogue of the dispersion relation $\omega(\mathbf{q})$ of the dynamic structure factor in the imaginary time domain. Yet, they will likely not suffice for the full estimation of other physical properties such as the static linear density response function $\chi(\mathbf{q})$ [see Eq.~(\ref{eq:chi_static})], or for the numerical inversion of Eq.~(\ref{eq:Laplace}) to reconstruct $S(\mathbf{q},\omega)$ via an analytic continuation~\cite{JARRELL1996133,dornheim_dynamic,dynamic_folgepaper}. The in-depth investigation of the application of advanced fermionic PIMC methods for the investigation of imaginary-time properties thus constitutes an important topic for future research.

\subsection{Linear response theory\label{sec:LRT}}

Let us next consider a modified Hamiltonian that includes an external harmonic perturbation of wave vector $\mathbf{q}$ and frequency $\omega$,
\begin{eqnarray}
\hat{H} = \hat{H}_\textnormal{UEG} + A \phi_\textnormal{ext}(\mathbf{q},\omega)\ ,
\end{eqnarray}
with $\hat{H}_\textnormal{UEG}$ being the usual UEG Hamiltonian, see Refs.~\cite{review,Fraser_PRB_1996} for more details on the latter. In the limit of small perturbation amplitudes $A$, the response of the UEG is fully described by the linear density response function, which can be expressed as~\cite{kugler1,quantum_theory}
\begin{eqnarray}\label{eq:LFC}
\chi(\mathbf{q},\omega) = \frac{\chi_0(\mathbf{q},\omega)}{1-\frac{4\pi}{q^2}\left[1-G(\mathbf{q},\omega)\right]\chi_0(\mathbf{q},\omega)}\ .
\end{eqnarray}
Here $\chi_0(\mathbf{q},\omega)$ describes the density response of an ideal Fermi gas, which is sometimes known as (temperature-dependent) Lindhard function in the literature; it can readily be computed by numerically evaluating a simple one-dimensional integral~\cite{quantum_theory}. The dynamic local field correction $G(\mathbf{q},\omega)$ contains the full wave-vector- and frequency-resolved information about electronic exchange--correlation effects; it is formally equivalent to the dynamic exchange--correlation kernel $K_\textnormal{xc}(\mathbf{q},\omega)$ that is used in the context of linear-response time-dependent density functional theory simulations~\cite{marques2012fundamentals}. Indeed, setting $G(\mathbf{q},\omega)\equiv0$ in Eq.~(\ref{eq:LFC}) corresponds to the \emph{random phase approximation} (RPA), which describes the dynamic density response on the mean-field level.
Consequently, $G(\mathbf{q},\omega)$ constitutes the key input for a gamut of applications such as the construction of effective potentials~\cite{ceperley_potential,zhandos1,zhandos2}, quantum fluid models \cite{Moldabekov_SciPost_2022, zhandos_pop18, zmcpp2017}, or the construction of advanced, nonlocal exchange--correlation functionals for density functional theory~\cite{pribram,Patrick_JCP_2015}. Indeed, the development of approximate local field corrections constitutes an active field of research~\cite{stls_original,vs_original,IIT,schweng,farid,stolzmann,Dabrowski_PRB_1986,Holas_PRB_1987,arora,Tanaka_CPP_2017,tanaka_hnc,Tolias_JCP_2021,castello2021classical}. For the UEG, highly accurate results for $G(\mathbf{q},\omega)$ have been presented by Hamann \emph{et al.}~\cite{Hamann_PRB_2020} based on the framework for the analytic continuation of imaginary-time PIMC data developed in Refs.~\cite{dornheim_dynamic,dynamic_folgepaper,Dornheim_PRE_2020}.
For completeness, we note that first results for the static local field factor of real materials are currently being developed on the basis of either PIMC simulations~\cite{Bohme_PRL_2022,https://doi.org/10.48550/arxiv.2207.14716} or density functional theory~\cite{https://doi.org/10.48550/arxiv.2209.00928}.

In the context of the present work, the main utility of $\chi(\mathbf{q},\omega)$ is given by the fluctuation--dissipation theorem, which provides a straightforward connection between the density response and the dynamic structure factor~\cite{quantum_theory},
\begin{eqnarray}\label{eq:FDT}
S(\mathbf{q},\omega) = - \frac{\textnormal{Im}\chi(\mathbf{q},\omega)}{\pi n (1-e^{-\beta\omega})}\ .
\end{eqnarray}
In combination with Eq.~(\ref{eq:Laplace}), any dielectric theory for the local field correction can thus be used to compute the ITCF $F(\mathbf{q},\tau)$.

A final interesting relation from linear response theory is given by the imaginary-time version of the fluctuation--dissipation theorem~\cite{bowen2,Dornheim_insight_2022}, which states that
\begin{eqnarray}\label{eq:chi_static}
\chi(\mathbf{q},0) = - n \int_0^\beta \textnormal{d}\tau\ F(\mathbf{q},\tau)\ .
\end{eqnarray}
In practice, Eq.~(\ref{eq:chi_static}) can thus be used to compute the full wave-number dependence of the static linear density response function from a single simulation of the unperturbed system from PIMC data for $F(\mathbf{q},\tau)$ by solving a simple one-dimensional integral for each value of $q$; this approach provides the basis for a number of studies of the UEG across different regimes~\cite{dynamic_folgepaper,dornheim_ML,dornheim_electron_liquid,dornheim_HEDP,Dornheim_HEDP_2022}.
Finally, we note that Eq.~(\ref{eq:chi_static}) also constitutes a feasible route to obtain $\chi(\mathbf{q},0)$ from XRTS experiments. In particular, the direct evaluation of the inverse frequency sum-rule
\begin{eqnarray}
\braket{\omega^{-1}} &=& \int_{-\infty}^\infty \textnormal{d}\omega\ S(\mathbf{q},\omega)\ \omega^{-1} \\\nonumber &=& - \frac{1}{2n}\chi(\mathbf{q},0)\ ,
\end{eqnarray}
is usually prevented by the convolution of the DSF with the instrument function in the measured XRTS intensity. The convolution theorem of the two-sided Laplace transform [Eq.~(\ref{eq:convolution_theorem}) above], on the other hand, allows us to exactly remove the influence of $R(\omega)$, and, in this way, opens up the way to obtain the static density response of a given system.

\subsection{Properties of the imaginary-time density--density correlation function\label{sec:properties}}

Let us next revisit a number of important known properties of the ITCF. In thermodynamic equilibrium, the DSF obeys the detailed balance relation between positive and negative frequencies~\cite{quantum_theory},
\begin{eqnarray}\label{eq:detailed_balance}
S(\mathbf{q},-\omega) = S(\mathbf{q},\omega) e^{-\beta\omega}\ .
\end{eqnarray}
Inserting Eq.~(\ref{eq:detailed_balance}) into the definition of the two-sided Laplace transform [Eq.~(\ref{eq:Laplace})] gives the important symmetry relation~\cite{Dornheim_T_2022},
\begin{eqnarray}\label{eq:symmetry}
F(\mathbf{q},\tau) &=& \int_0^\infty \textnormal{d}\omega\ S(\mathbf{q},\omega)\left\{ e^{-\omega\tau} + e^{-\omega(\beta-\tau)} \right\}\\\nonumber
 &=& F(\mathbf{q},\beta-\tau)\ .
\end{eqnarray}
In particular, Dornheim \emph{et al.}~\cite{Dornheim_T_2022} have recently shown that Eq.~(\ref{eq:symmetry}) allows for an exact, simulation-free diagnostic of the temperature of any given system from an XRTS measurement by locating the minimum of $F(\mathbf{q},\tau)$ at $\tau=\beta/2=1/2T$; the further development of this idea constitutes a topic of active research~\cite{Dornheim_insight_2022}.

A further important set of properties related to the description of the dynamics of quantum many-body systems is given by the frequency moments of the DSF, which we define as
\begin{eqnarray}\label{eq:moments}
M_\alpha = \braket{\omega^\alpha} = \int_{-\infty}^\infty \textnormal{d}\omega\ S(\mathbf{q},\omega)\ \omega^\alpha\ .
\end{eqnarray}
It is straightforward to show that all positive frequency moments can be obtained from $\tau$-derivatives of $F(\mathbf{q},\tau)$ around the origin~\cite{Dornheim_insight_2022},
\begin{eqnarray}\label{eq:moments_derivative}
M_\alpha = \left( -1 \right)^\alpha \left. \frac{\partial^\alpha}{\partial\tau^\alpha} F(\mathbf{q},\tau) \right|_{\tau=0} \ .
\end{eqnarray}
We note that the odd frequency moments can alternatively be expressed in terms of nested commutator expressions~\cite{Mihara_Puff_PR_1968}, which are known as sum rules in the literature~\cite{tkachenko_book}. 
In the case of $\alpha=1$, we have the well-known f-sum rule, 
\begin{eqnarray}\label{eq:f_sum_rule}
M_1 = \braket{\omega^1} = \frac{\mathbf{q}^2}{2}\ .
\end{eqnarray}

\begin{figure}\centering
\includegraphics[width=0.475\textwidth]{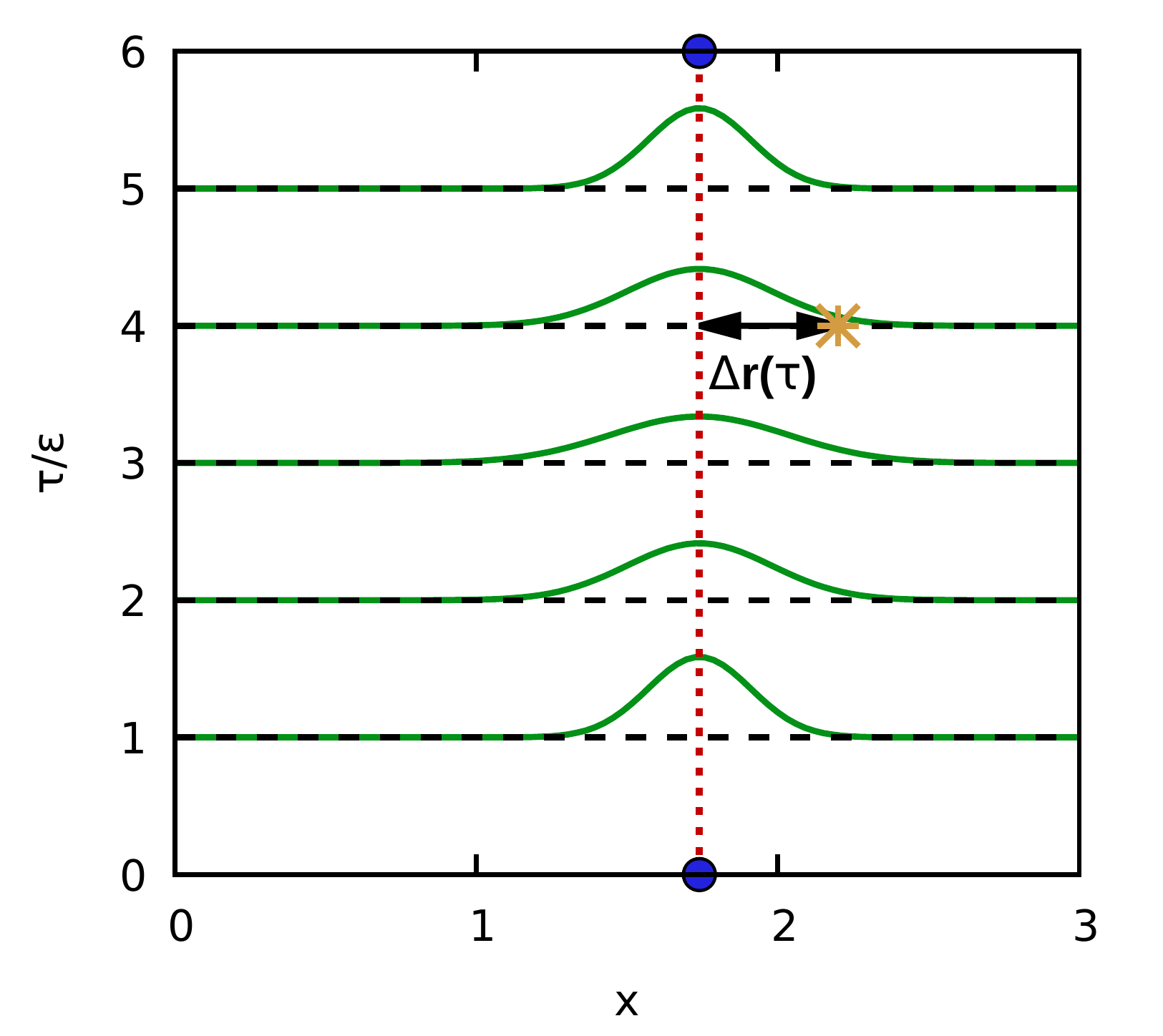}
\caption{\label{fig:ITCF_diffusion}
Schematic illustration of the single-particle diffusion through the imaginary time, and its estimation via Eq.~(\ref{eq:F_SP}). The green curves show the Gaussian probability distribution Eq.~(\ref{eq:Gaussian_modified}) of a diffused particle coordinate at a difference $\tau$ from its origin (blue beads), and the yellow star a particular realization.
}
\end{figure}

An important insight into the physical origin of the observed $\tau$-dependence of $F(\mathbf{q},\tau)$ can be obtained from the spectral representation of the DSF, which is given by~\cite{quantum_theory},
\begin{eqnarray}\label{eq:spectral}
S(\mathbf{q},\omega) = \sum_{m,l} P_m \left\|{n}_{ml}(\mathbf{q}) \right\|^2 \delta(\omega - \omega_{lm})\ .
\end{eqnarray}
From a physical perspective, Eq.~(\ref{eq:spectral}) states that the DSF is given by the sum over all possible transitions between the eigenstates $l$ and $m$ (with $\omega_{lm}$ being the energy difference) of the full $N$-body Hamiltonian, where $P_m$ is the occupation probability of the initial state, and $\left\|{n}_{ml}(\mathbf{q}) \right\|^2$ denotes the corresponding transition matrix element. Inserting Eq.~(\ref{eq:spectral}) into Eq.~(\ref{eq:Laplace}) then gives~\cite{Dornheim_insight_2022}
\begin{eqnarray}\label{eq:spectral_F}
F(\mathbf{q},\tau) &=& \sum_{m,l} P_m \left\|{n}_{ml}(\mathbf{q}) \right\|^2 e^{-\tau\omega_{lm}}\ .
\end{eqnarray}
In other words, the $\tau$-decay is shaped by the energy difference between the eigenstates for transitions that are important at a particular wave vector $\mathbf{q}$. More specifically, large energy differences as they occur in the non-collective single-particle regime where the dispersion relation of the DSF is given by $\omega(q)\sim q^2$ lead to a pronounced $\tau$-decay, with $F(\mathbf{q},\tau)$ almost vanishing around the minimum at $\tau=\beta/2$. Conversely, energetically low-lying transitions such as the roton feature of the UEG~\cite{Dornheim_Nature_2022,Dornheim_JCP_2022} manifest as a reduced $\tau$-decay; see Ref.~\cite{Dornheim_insight_2022} for an extensive discussion of this effect. In other words, the existence of a low-energy excitation is equivalent to the stability of density correlations throughout the imaginary time.
Dornheim \emph{et al.}~\cite{Dornheim_insight_2022} have recently suggested to quantify this mechanism via a relative decay measure of the form
\begin{eqnarray}\label{eq:decay_measure}
\Delta F_\tau(\mathbf{q}) = \frac{F(\mathbf{q},0)-F(\mathbf{q},\tau)}{F(\mathbf{q},0)}\ ,
\end{eqnarray}
which we will also employ in the present work.

\begin{figure*}\centering
\includegraphics[width=0.475\textwidth]{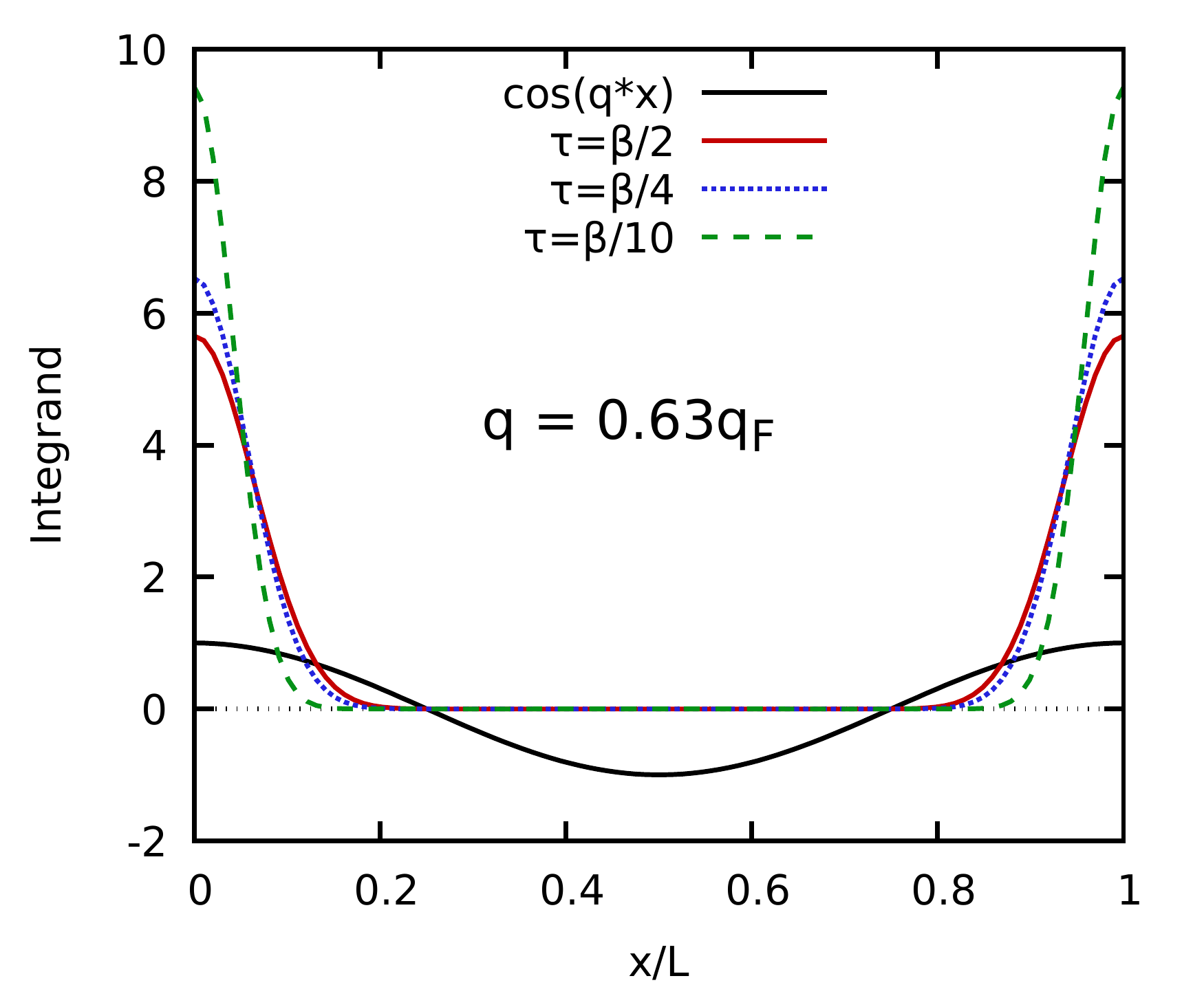}\includegraphics[width=0.475\textwidth]{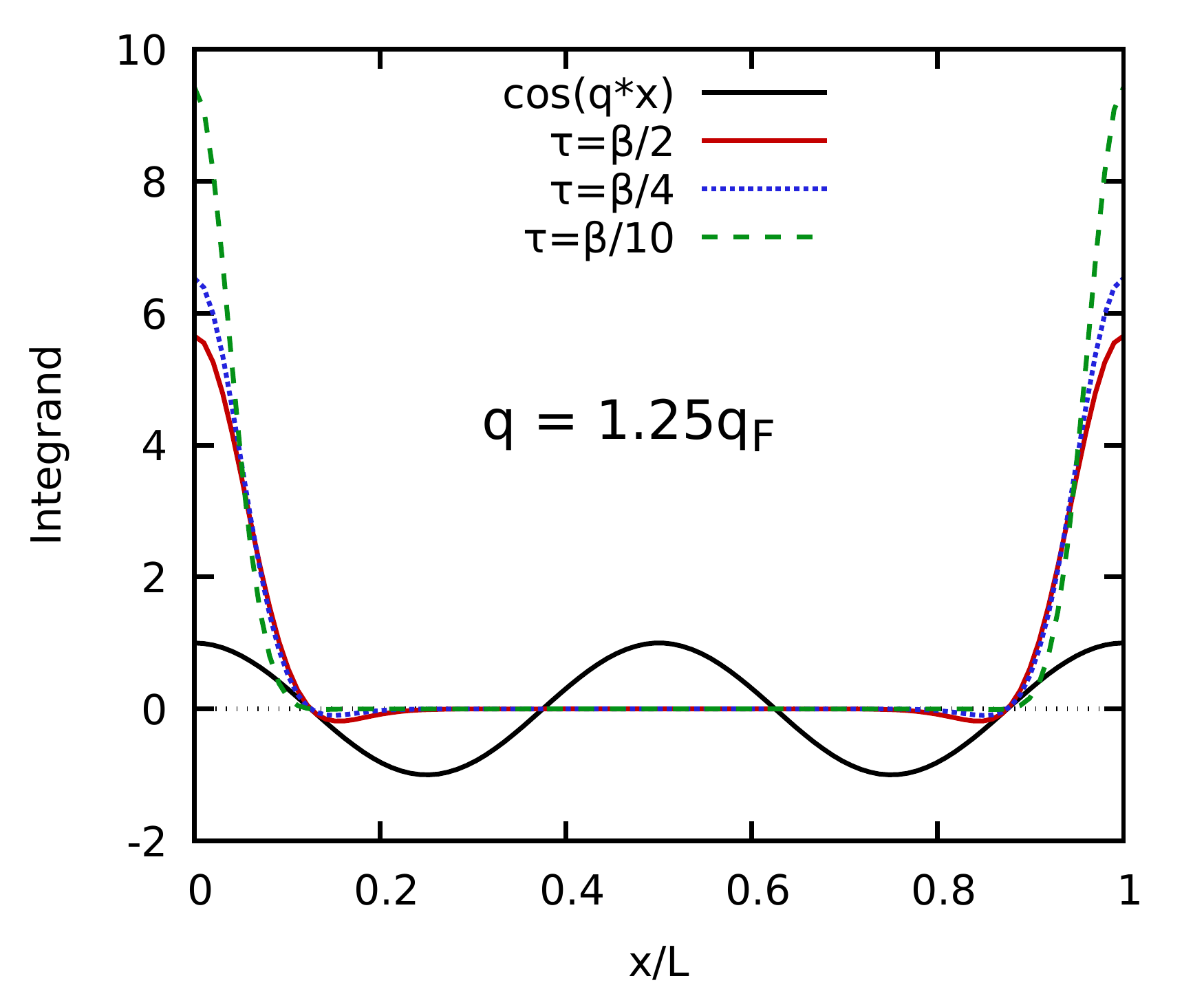}\\\vspace*{-0.5cm}\includegraphics[width=0.475\textwidth]{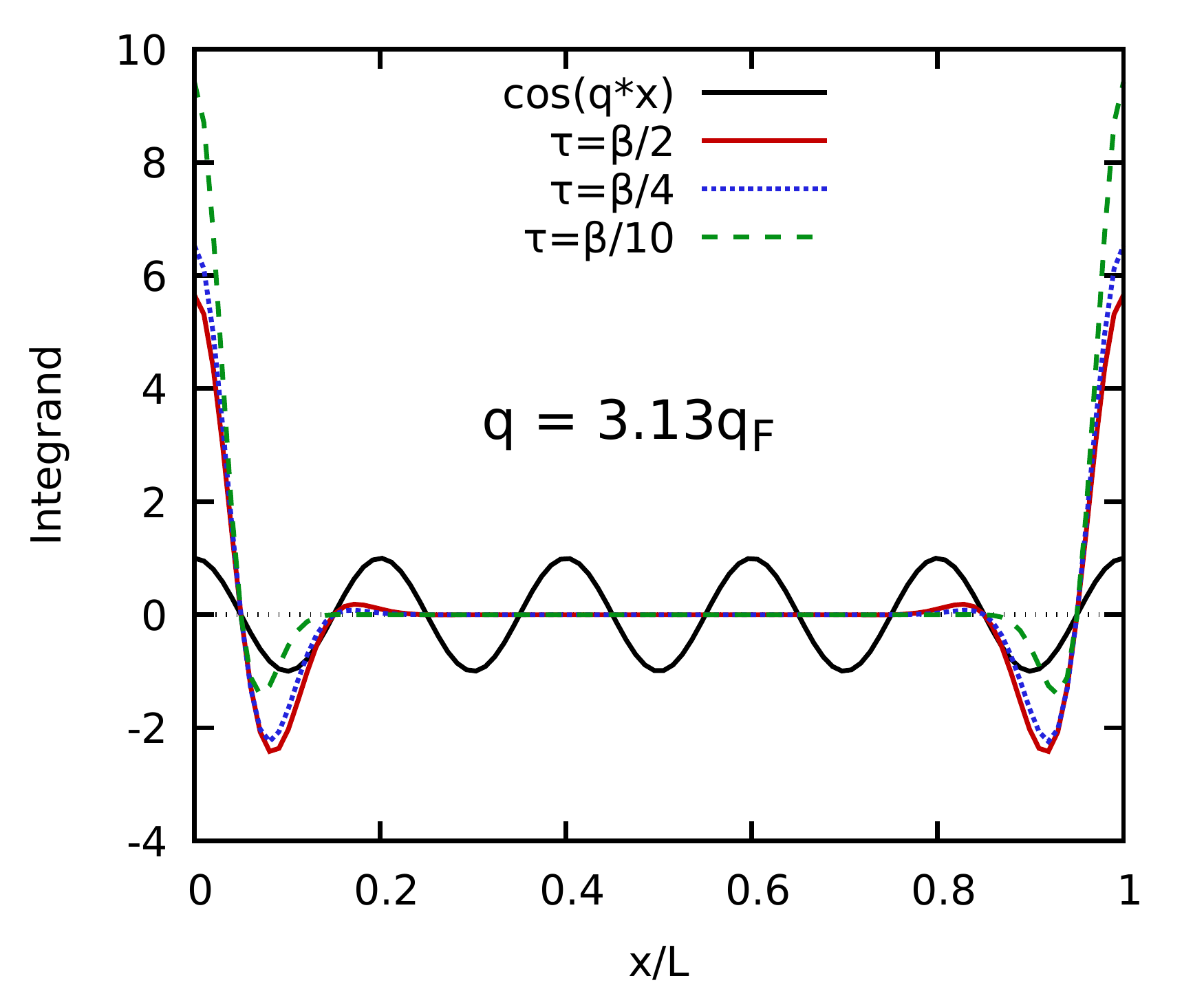}\includegraphics[width=0.475\textwidth]{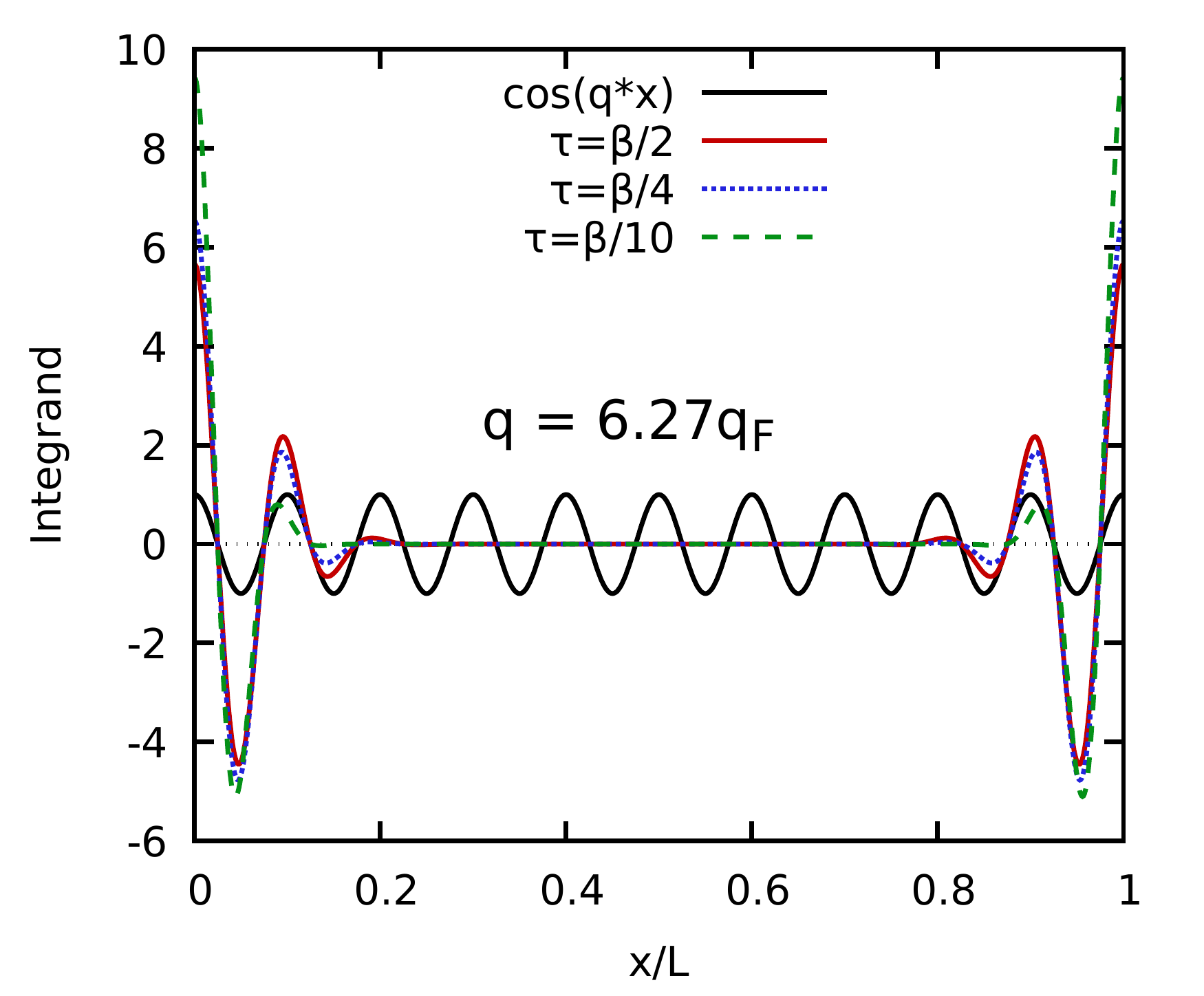}
\caption{\label{fig:Integrand_theta1_rs4}
Integrand of the estimator for the single-particle ITCF $F_\textnormal{SP}(\mathbf{q},\tau)$, Eq.~(\ref{eq:F_SP}), for $N=34$ and $\Theta=1$ for a) $q=0.63q_\textnormal{F}$, b) $q=1.25q_\textnormal{F}$, c) $q=3.13q_\textnormal{F}$, and d) $q=6.27q_\textnormal{F}$. The solid black sinusoidal lines show the respective cosine function, and the solid red, dotted blue, and dashed green curves show the full integrand for $\tau=\beta/2$, $\tau=\beta/4$, and $\tau=\beta/10$, respectively. 
}
\end{figure*}

\subsection{Single-particle delocalization model\label{sec:delocalization}}

To gain further insight into the physical meaning of $F(\mathbf{q},\tau)$, we will present a simple model for its $\tau$-dependence. In particular, it has recently been reported~\cite{Dornheim_PRR_2022} based on extensive, spin-resolved PIMC results for the ITCF that the $\tau$-dependence is almost exclusively due to single-particle imaginary-time diffusion effects; density--density correlations between electrons of different spin-orientation are almost unaffected by the thermal delocalization of the paths. Based on this empirical observation, we decompose the total ITCF into
\begin{eqnarray}\label{eq:decompose}
F(\mathbf{q},\tau) = S(\mathbf{q}) + \Delta F(\mathbf{q},\tau)\ ,
\end{eqnarray}
where the static structure factor (SSF) corresponds to the $\tau\to0$ limit of the ITCF, $S(\mathbf{q})=F(\mathbf{q},0)$. The full $\tau$-dependence is thus, by definition, contained in the function $\Delta F(\mathbf{q},\tau)=F(\mathbf{q},\tau)-F(\mathbf{q},0)$.

Let us next consider the imaginary-time diffusion of a single particle at the inverse temperature $\beta$ in a volume $\Omega=L^3$, with $L\gg \lambda_\beta$. 
In this case, the appropriate imaginary-time propagator is simply given by the diagonal elements of the ideal thermal density matrix $\rho_0(\mathbf{r},\mathbf{r};\beta)$, see Eq.~(\ref{eq:rho_ideal}) above. To estimate the corresponding single-particle ITCF $F_\textnormal{SP}(\mathbf{q},\tau')$, we insert a single intermediate time slice at $\tau'$. This can be expressed as
\begin{eqnarray}\label{eq:intermediate}
\rho_0(\mathbf{r},\mathbf{r};\beta) = \int_\Omega \textnormal{d}\mathbf{r}'\ \rho_0(\mathbf{r},\mathbf{r}';\tau')\rho_0(\mathbf{r}',\mathbf{r};\beta-\tau')\ .
\end{eqnarray}
Inserting the corresponding density operators $\hat{n}(\mathbf{q})$ and $\hat{n}(-\mathbf{q})$ at the appropriate imaginary times and performing a few straightforward transformation then gives a simple expression for the single-particle ITCF,
\begin{eqnarray}\label{eq:F_SP}
F_\textnormal{SP}(\mathbf{q},\tau') =  \int_\Omega \textnormal{d}\mathbf{\Delta r}\ P(\mathbf{\Delta r},\tau')\ \textnormal{cos}\left(\mathbf{q}\cdot\mathbf{\Delta r}\right)\ ,
\end{eqnarray}
with $\mathbf{\Delta r}$ being the displacement between the position of the particle at $\tau=0$ (which is the same as at $\tau=\beta$ due to the diagonal nature of the trace, leading to closed trajectories in the path-integral picture) and $\tau=\tau'$. 
The probability of a particular displacement is given by the three-dimensional Gaussian distribution
\begin{eqnarray}\label{eq:Gaussian_modified}
P(\mathbf{\Delta r},\tau) = \left(\frac{\lambda_\beta}{\lambda_\tau\lambda_{\beta-\tau}}\right)^3 \textnormal{exp}\left(-\pi \frac{\lambda_\beta^2}{\lambda_\tau^2\lambda_{\beta-\tau}^2} \mathbf{\Delta r}^2 \right)\ ,
\end{eqnarray}
which takes into account the distance in imaginary-time to the time-slice of interest both starting from $\tau=0$ and from $\tau=\beta$.

The basic idea behind Eq.~(\ref{eq:F_SP}) is schematically illustrated in Fig.~\ref{fig:ITCF_diffusion} a). Specifically, the blue beads correspond to the start and end points at $\tau=0$ and $\tau=\beta$, which are always equal. The green curves depict the Gaussian probability distributions from Eq.~(\ref{eq:Gaussian_modified}) for different values of $\tau$; they are symmetric with respect to $\tau=\beta/2$, as it is expected. In other words, the symmetry relation (\ref{eq:symmetry}) of the ITCF can not only be seen as a consequence of the detailed balance of the DSF, but also naturally emerges as a consequence of the $\beta$-periodicity within the imaginary-time path-integral formalism.
The yellow star at $\tau=4\epsilon$ depicts a particular realization of the displacement $\mathbf{\Delta r}(\tau)$, which is weighted with the probability $P(\mathbf{\Delta r},\tau)$. The final expectation value for $F_\textnormal{SP}(\mathbf{q},\tau)$ then follows from an integral over all possible values of $\mathbf{\Delta r}$, which we evaluate numerically.

Let us next analyse the integrand of Eq.~(\ref{eq:F_SP}), which is depicted in Fig.~\ref{fig:Integrand_theta1_rs4} along the $x$-component of $\mathbf{\Delta r}$ at $y=z=0$ for $N=34$ and $\Theta=1$. For completeness, we note that such single-particle expressions do, by definition, not depend on the density parameter $r_s$ for a given value of $\Theta$.
The different panels of Fig.~\ref{fig:Integrand_theta1_rs4} correspond to different wave vectors $\mathbf{q}=(q,0,0)^T$, and the sinusoidal black lines depict the cosine from Eq.~(\ref{eq:F_SP}). More specifically, panel a) has been obtained for the minimum possible wave number $q=2\pi/L$, panel b) for $q=2 q_\textnormal{min}$, c) for $q=5q_\textnormal{min}$ and d) for $q=10q_\textnormal{min}$. In addition, the solid red, dotted blue, and dashed green lines show the product of the cosine with the thermal probability function $P(\mathbf{\Delta r},\tau)$ for different values of $\tau$.
For completeness, we note that we employ standard periodic boundary conditions such that $x\to x+L$ for $x<0$.
Evidently, the Gaussian becomes increasingly narrow with decreasing $\tau$, since the corresponding path has only a shorter imaginary-time interval to diffuse away from its position at $\tau=0$ in that case. In the limit of $\tau\to0$, $P(\mathbf{\Delta r},\tau)$ becomes a delta distribution. Consequently, the single-particle ITCF attains unity for $\tau=0$ and $\tau=\beta$ for all $\mathbf{q}$.

\begin{figure}\centering
\includegraphics[width=0.475\textwidth]{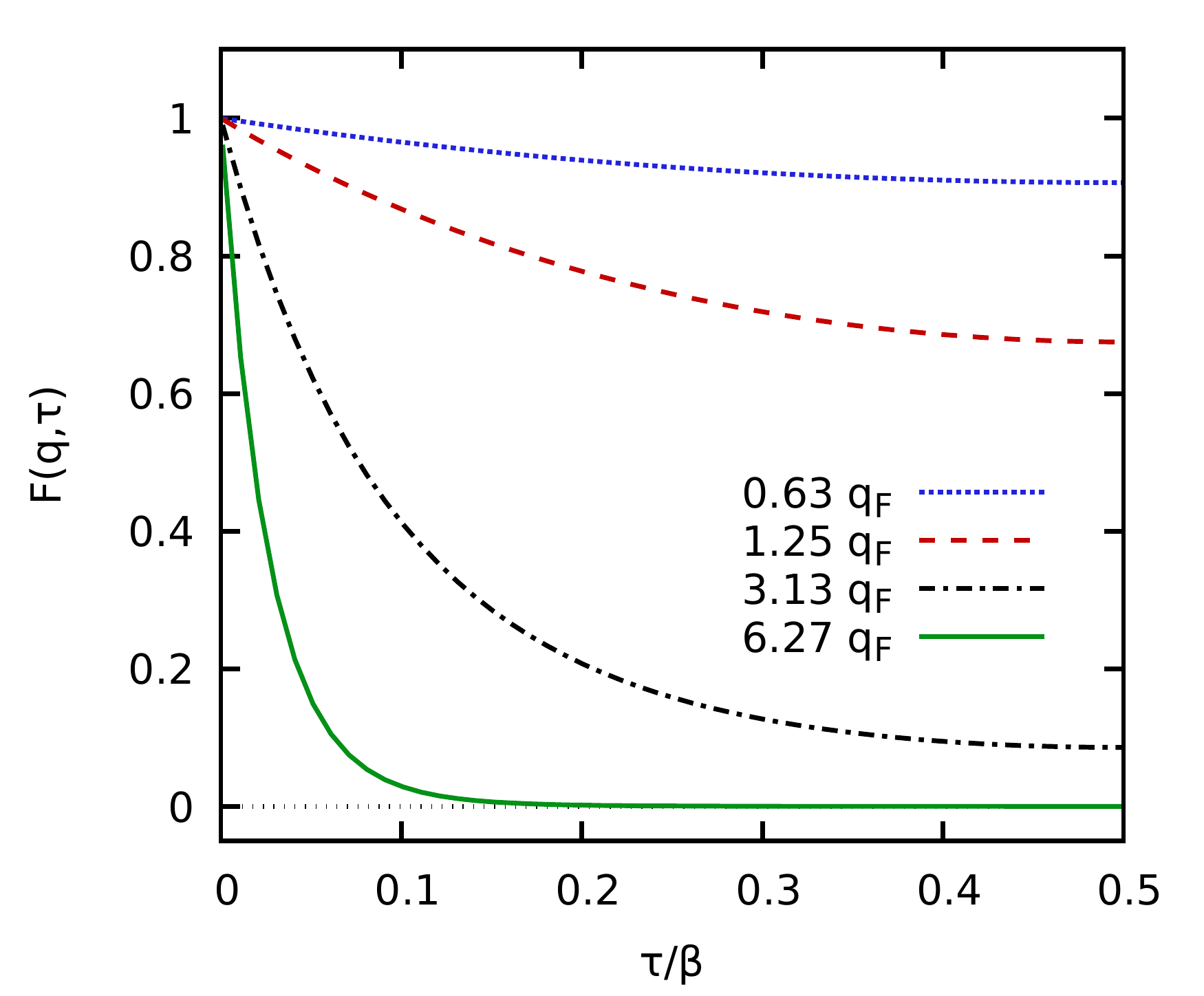}
\caption{\label{fig:SP_ITCF}
Single-particle ITCF [Eq.~(\ref{eq:F_SP})] for different wave vectors $\mathbf{q}$ for $N=34$ and $\Theta=1$.
}
\end{figure}

For $q=q_\textnormal{min}$, the cosine is positive over the entire $x$-range where the thermal distribution has significant values. Consequently, the oscillating nature of the cosine function has a small impact, as it is almost constant in the relevant $x$-interval. Still, the small decay of the cosine function for $x\leq L/2$ becomes more important with increasing $\tau\leq\beta/2$, since the thermal distribution is spread out to larger $x$ in this case. This is the origin of the comparably small $\tau$-decay of the corresponding single-particle ITCF, which is depicted as the dotted blue line in Fig.~\ref{fig:SP_ITCF}.

In Fig.~\ref{fig:Integrand_theta1_rs4} b), we show the integrand of Eq.~(\ref{eq:F_SP}) for a moderate wave vector, $q=1.25q_\textnormal{F}$. In this case, the cosine function oscillates twice as fast as for $q=0.63q_\textnormal{F}$, which has a noticeable impact on the integrand. Indeed, the latter even becomes slightly negative for $x\gtrsim L/8$; the resulting cancellation becomes more pronounced with increasing $\tau$, which explains the comparably larger $\tau$-decay of the corresponding single-particle ITCF (dashed red curve) in Fig.~\ref{fig:SP_ITCF}. Further increasing the wave numbers to $q=3.13q_\textnormal{F}$ and $q=6.27q_\textnormal{F}$ leads to the integrands shown in Figs.~\ref{fig:Integrand_theta1_rs4} c) and d), which exhibit an even more substantial cancellation of positive and negative contributions. Consequently, the $\tau$-decay of $F_\textnormal{SP}(\mathbf{q},\tau)$, which is shown in Fig.~\ref{fig:SP_ITCF} as the dash-dotted black and solid green curves, is even sharper. 
The sharp decay in the single particle limit of $q\gg q_\textnormal{F}$ is thus the result of a competition between the delta-distribution limit of the thermal distribution Eq.~(\ref{eq:Gaussian_modified}) for $\tau\to0$ and the cancellation of the integrand as a consequence of the rapidly oscillating cosine function for large $q$ at any finite value of $\tau$.


\section{Results\label{sec:results}}

All PIMC results have been obtained using the extended ensemble approach introduced in Ref.~\cite{Dornheim_PRB_nk_2021}, which is a canonical adaption of the worm algorithm by Boninsegni \emph{et al.}~\cite{boninsegni1,boninsegni2}. We typically employ $P=200$ imaginary-time slices, which is sufficient with respect to the convergence of the factorization error~\cite{dynamic_folgepaper}, and sufficient to resolve $F(\mathbf{q},\tau)$ on a useful $\tau$-grid.

\subsection{Dependence on the imaginary-time\label{sec:tau}}

\begin{figure*}\centering
\includegraphics[width=0.475\textwidth]{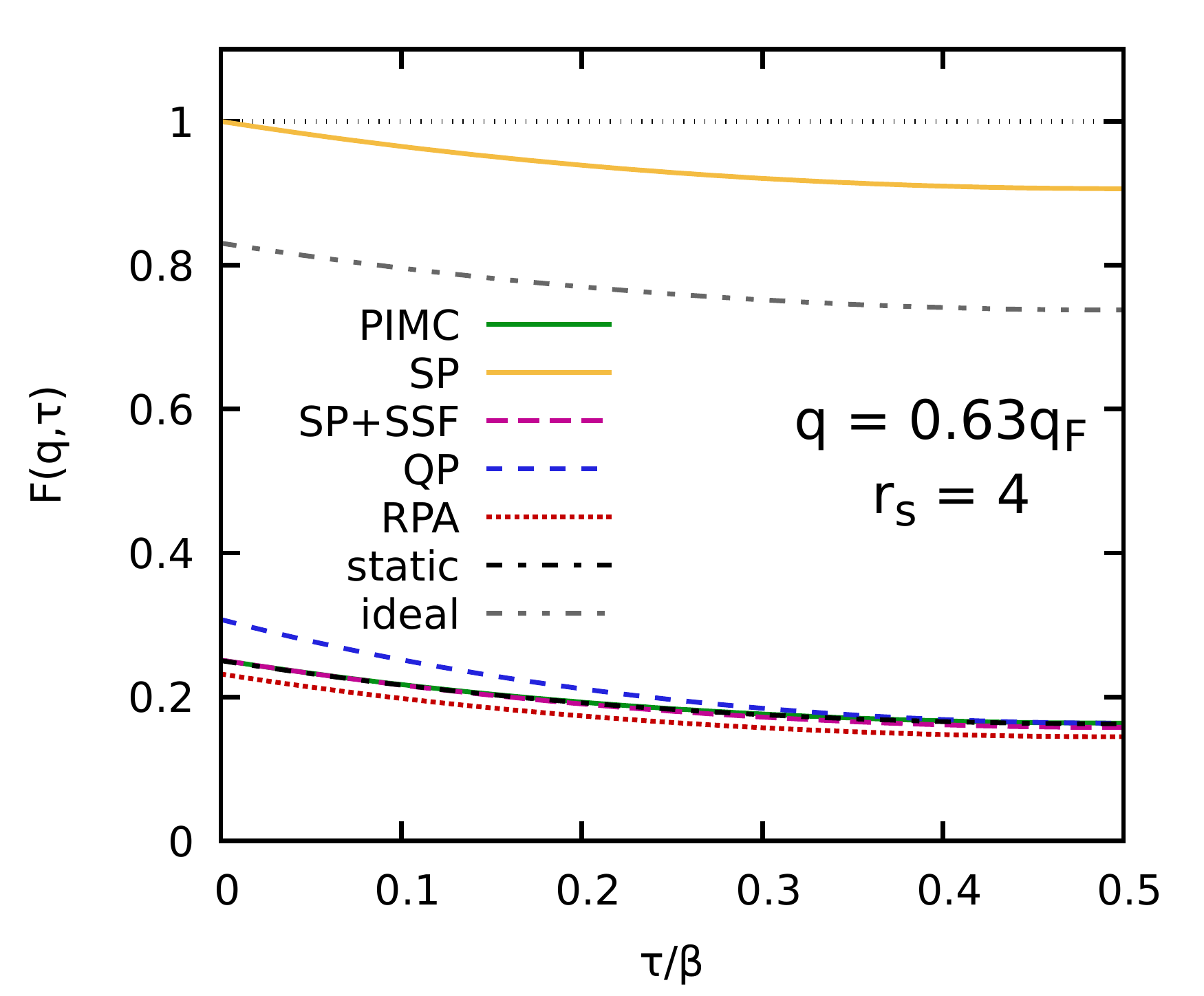}\includegraphics[width=0.475\textwidth]{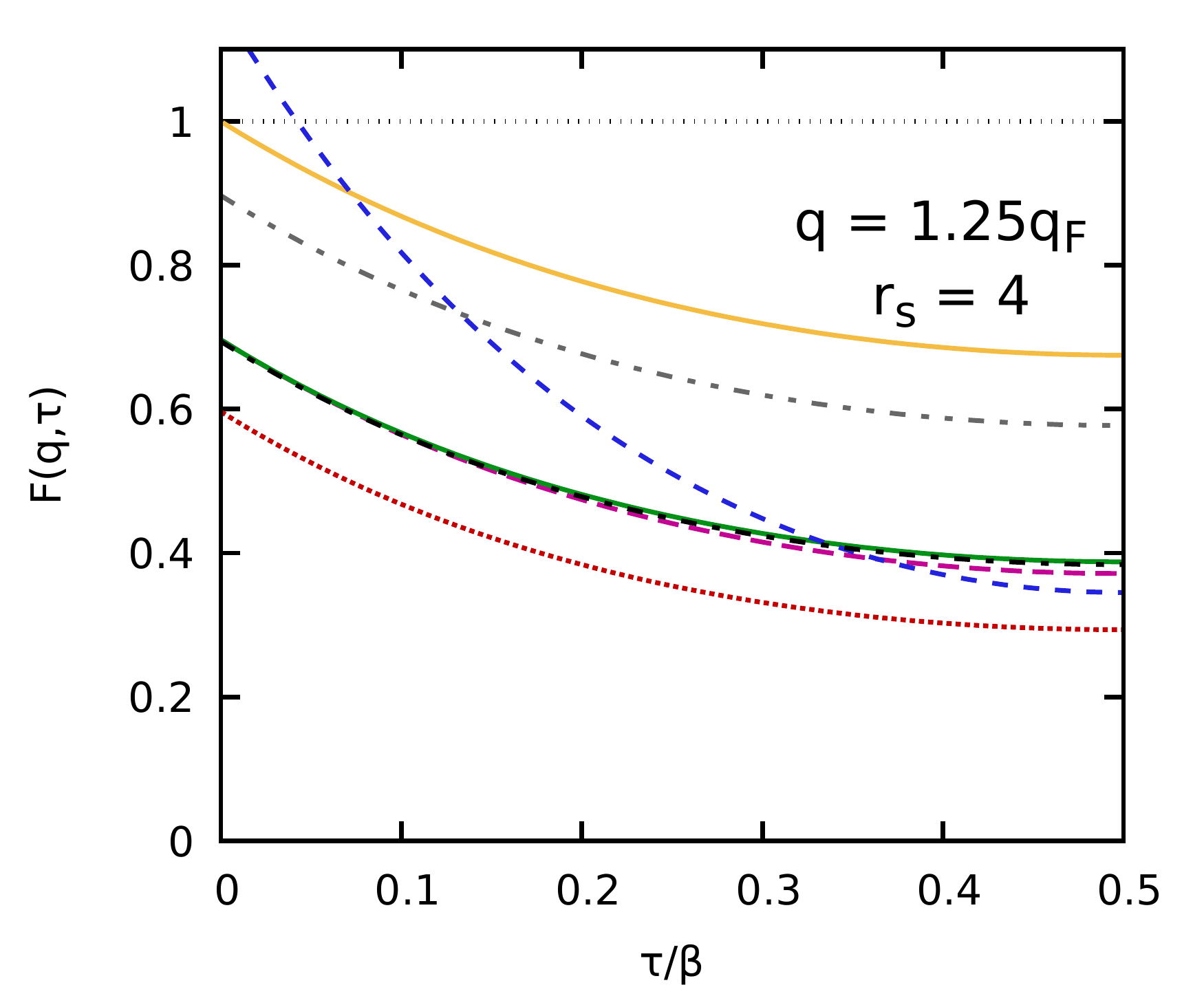}\\\vspace*{-0.5cm}\includegraphics[width=0.475\textwidth]{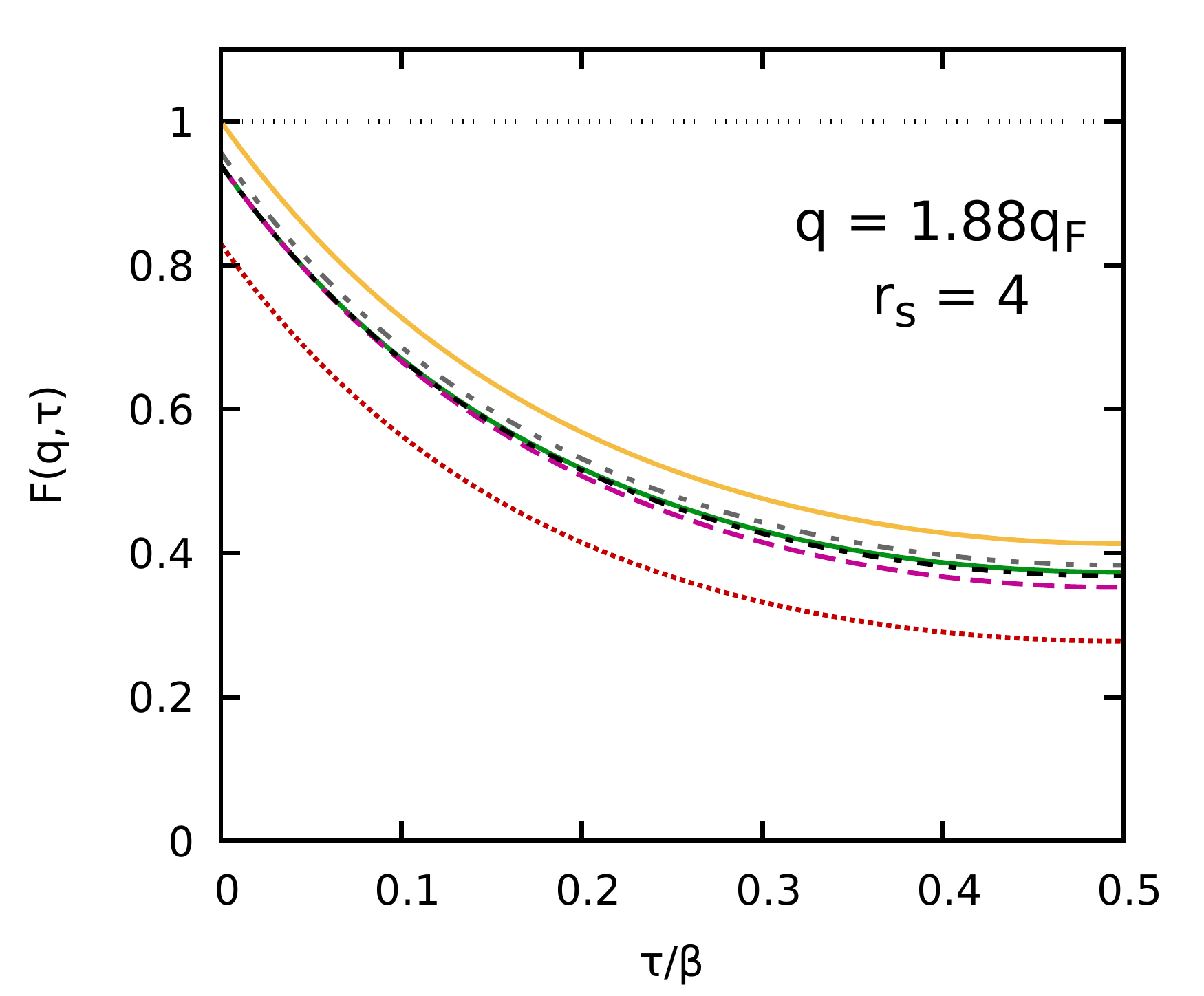}\includegraphics[width=0.475\textwidth]{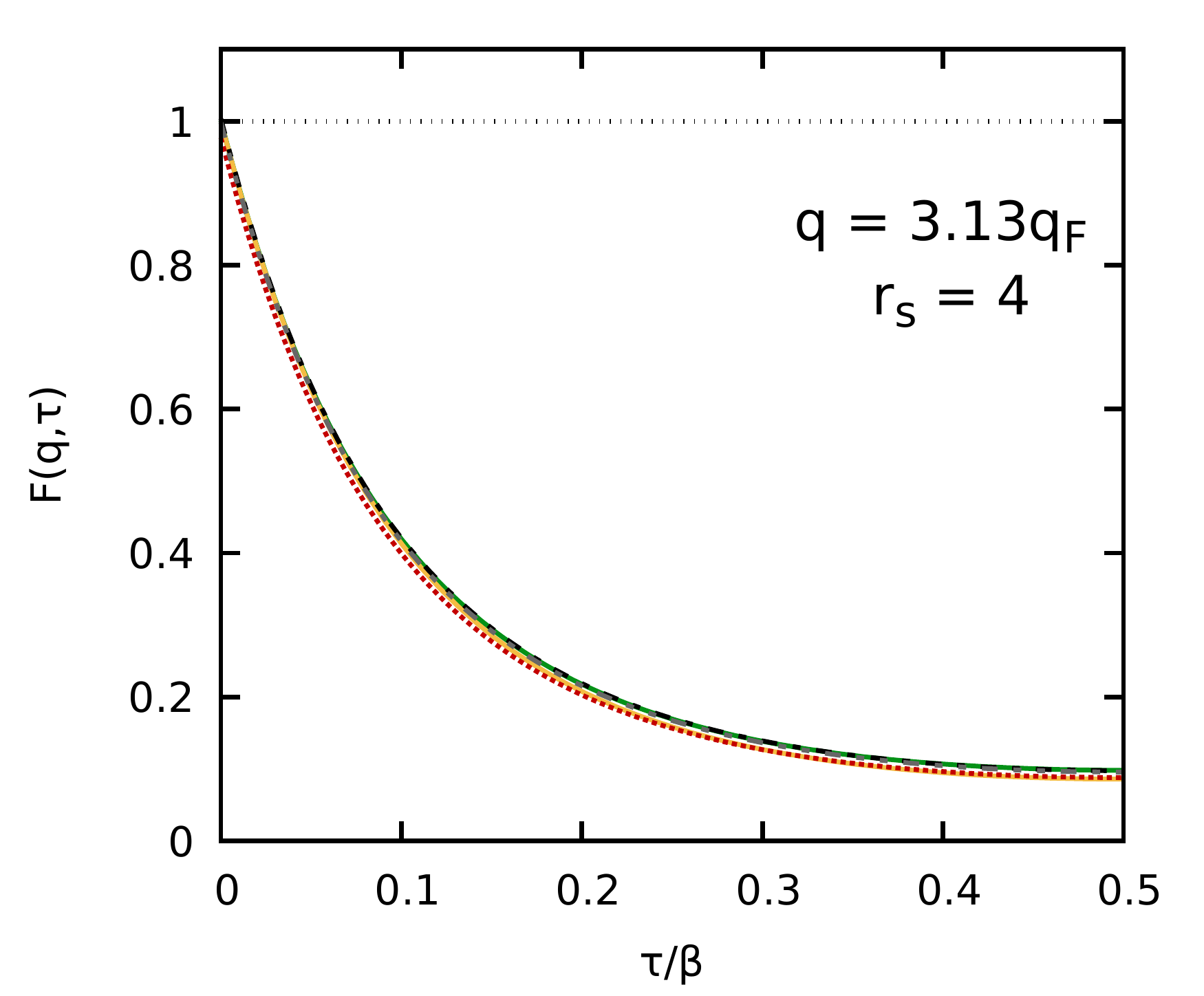}\\\vspace*{-0.5cm}\includegraphics[width=0.475\textwidth]{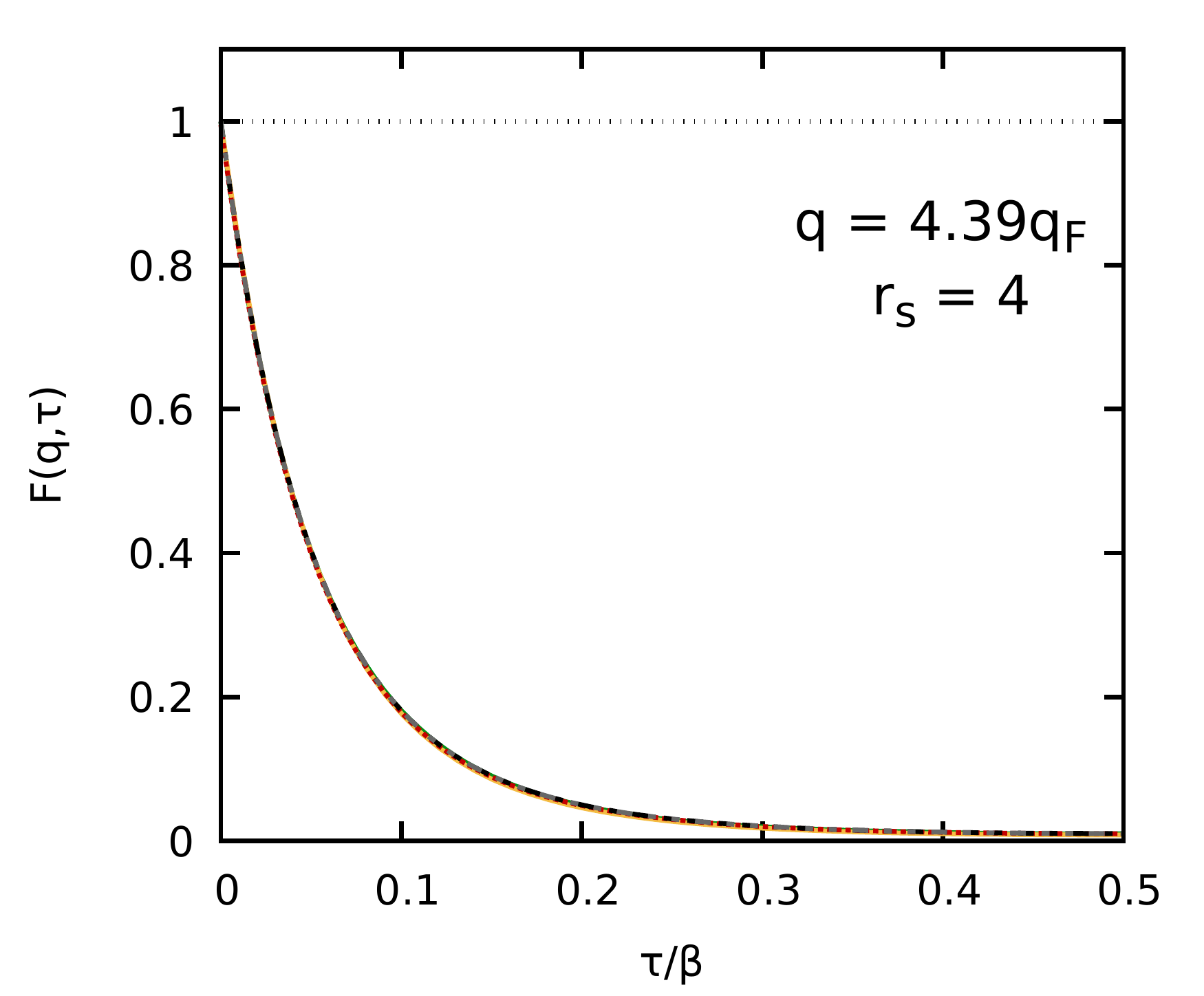}\includegraphics[width=0.475\textwidth]{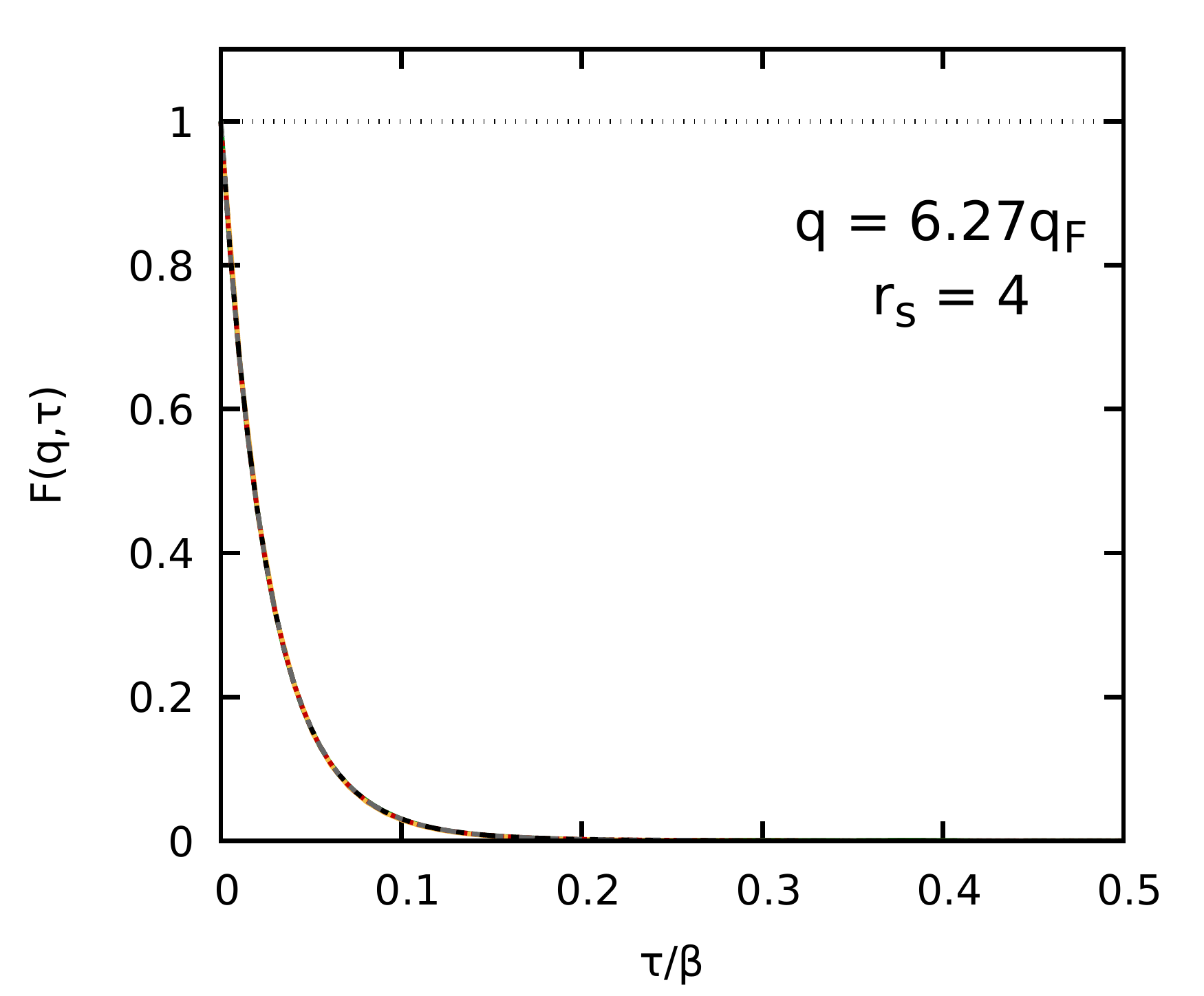}
\caption{\label{fig:ITCF_PIMC_theta1_rs4}
Imaginary-time dependence of the ITCF $F(\mathbf{q},\tau)$ for different wave numbers $q$ for the unpolarized UEG with $N=34$, $r_s=4$, and $\Theta=1$. Solid green: exact PIMC results; solid yellow: single-particle ITCF $F_\textnormal{SP}(\mathbf{q},\tau)$ [Eq.~(\ref{eq:F_SP})]; long-dashed purple: combination of SSF with single-particle ITCF, Eq.~(\ref{eq:decompose}); dashed blue: quasi-particle expansion, Eq.~(\ref{eq:F_QP}); dotted red: RPA; dash-dotted black: static approximation, i.e., $G_\textnormal{static}(\mathbf{q},\omega) \equiv G(\mathbf{q},0)$; dash-double-dotted grey: ideal Fermi gas.
}
\end{figure*}

Let us begin our in-depth analysis of the $\tau$-dependence of the imaginary-time density--density correlation function of the warm dense UEG at the electronic Fermi temperature (i.e., $\Theta=1$) for the metallic density of $r_s=4$, which is shown in Fig.~\ref{fig:ITCF_PIMC_theta1_rs4}. We note that it is sufficient to show $F(\mathbf{q},\tau)$ in the half-range of $0\leq\tau\beta/2$, since all curves are symmetric around $\tau=\beta/2$, cf.~Eq.~(\ref{eq:symmetry}) above.
The solid green line shows our exact PIMC results; we note that the statistical uncertainty of these data are smaller than the width of the curve, and can be neglected here. Let us next consider the dash-dotted black curve that depicts the \emph{static approximation}~\cite{dornheim_dynamic}, i.e., to setting $G_\textnormal{static}(\mathbf{q},\omega)\equiv G(\mathbf{q},0)$ in Eq.~(\ref{eq:LFC}). For completeness, we note that we have used the neural-net parametrization of $G(\mathbf{q},\omega=0;r_s,\theta)$ from Ref.~\cite{dornheim_ML}, but the analytic parametrization of the local field factor~\cite{Dornheim_PRB_ESA_2021} within the recently introduced effective static approximation (ESA)~\cite{Dornheim_PRL_2020_ESA} would have worked equally well for this purpose.
Evidently, the \emph{static approximation} provides a very accurate description of $F(\mathbf{q},\tau)$ for all depicted wave vectors, and over the entire $\tau$-range. This is consistent to previous investigations of the dynamic structure factor $S(\mathbf{q},\omega)$, which have reported the same trend~\cite{dornheim_dynamic}. The only deviations that are visible on the depicted scale appear for the intermediate wave numbers of $q=1.25q_\textnormal{F}$ and $q=1.88q_\textnormal{F}$ that are shown in panels b) and c), but the systematic errors are very small.

\begin{figure}\centering
\includegraphics[width=0.475\textwidth]{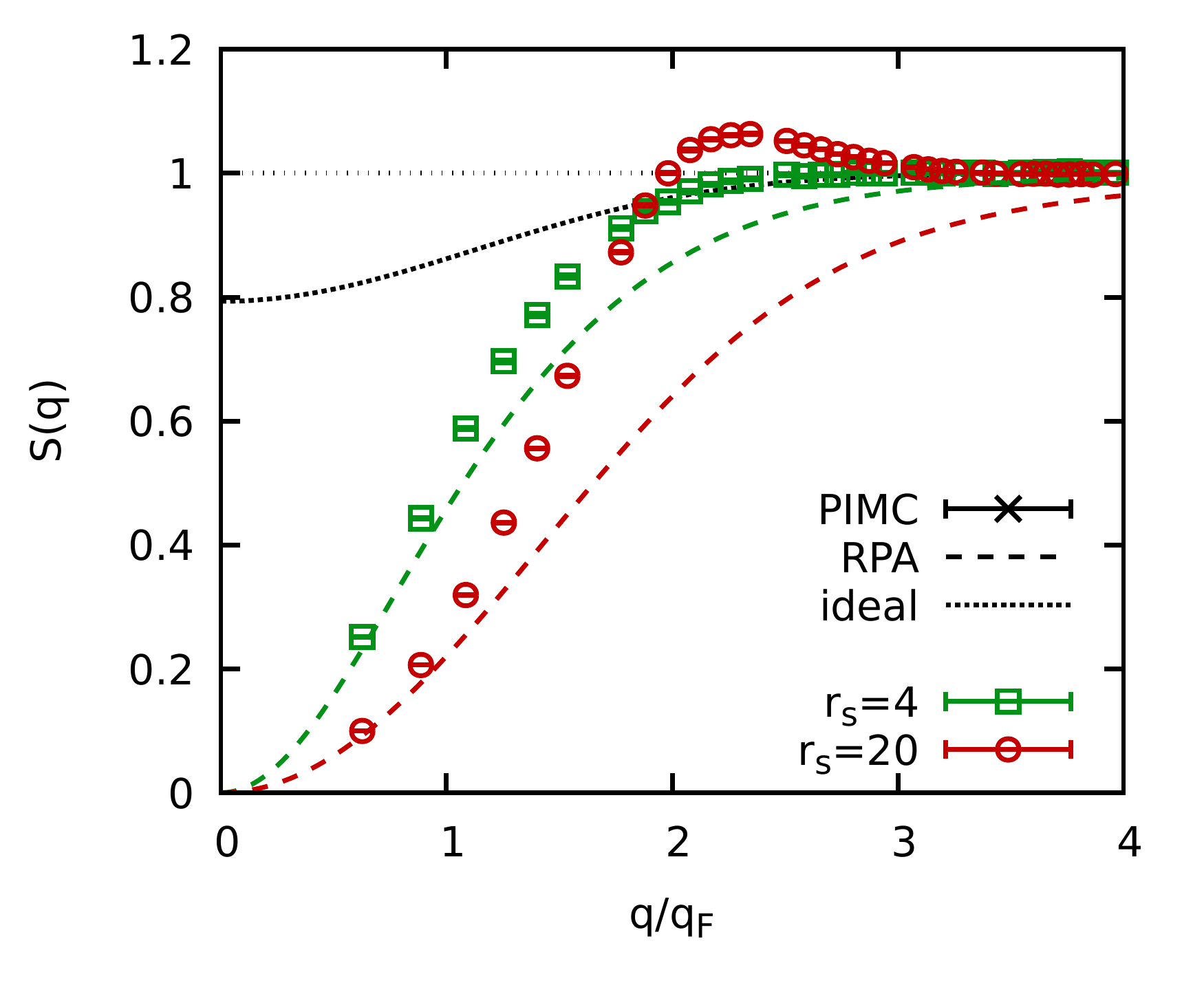}
\caption{\label{fig:SSF_theta1}
PIMC results for the static structure factor of the UEG with $N=34$ at $\Theta=1$ for $r_s=4$ (green squares) and $r_s=20$ (red circles). The dashed curves show corresponding results within the RPA, and the dotted line depicts the SSF of the ideal Fermi gas, which is independent of $r_s$.
}
\end{figure}

Let us next consider the dotted red curve, which corresponds to the random phase approximation. It is well known that the RPA becomes exact in the limit of $q\to0$, where the UEG is dominated by the collective plasmon feature, cf.~Eq.~(\ref{eq:S_QP}) below. For $q=0.63 q_\textnormal{F}$, which is the smallest wave number that is accessible for simulations with $N=34$ particles, one is already within the pair continuum~\cite{Dornheim_Nature_2022} and the DSF substantially differs from the plasmon shape both with respect to its peak position and the peak shape. In this regime, the RPA becomes increasingly inaccurate, and the dotted red curve noticeably deviates from the solid green PIMC results. With increasing $q$, the systematic error of the RPA becomes more pronounced, and attains a maximum around intermediate wave numbers, $q\sim1.5q_\textnormal{F}$. In the single-particle limit of $q\gg q_\textnormal{F}$, the RPA becomes exact again, and no systematic deviations are visible for $q=4.39q_\textnormal{F}$ and $q=6.27q_\textnormal{F}$ (bottom row of Fig.~\ref{fig:ITCF_PIMC_theta1_rs4}).
Interestingly, the main source of error of the RPA at intermediate wave numbers seems to be a constant shift in $F(\mathbf{q},\tau)$ that does not depend on the imaginary time. 
In fact, it is easy to see that the first derivative of $F(\mathbf{q},\tau)$ around the origin is exactly reproduced by the RPA, since it fulfills the f-sum rule, cf.~Eqs.~(\ref{eq:moments_derivative}) and (\ref{eq:f_sum_rule}) above.
From Eq.~(\ref{eq:decompose}), it is then clear that the bulk of the RPA error should already be present in the SSF $S(\mathbf{q})$, which is shown in Fig.~\ref{fig:SSF_theta1}.
More specifically, the green squares show our exact PIMC results for $r_s=4$, and the green dashed line the corresponding RPA. 
Evidently, the RPA underestimates the true magnitude of $S(\mathbf{q})$ for intermediate $q$, which explains the observed trends in Fig.~\ref{fig:ITCF_PIMC_theta1_rs4}.

Let us next consider the dashed blue curves in Fig.~\ref{fig:ITCF_PIMC_theta1_rs4}, which have been computed from a simple-quasi particle ansatz. In this model, it is assumed that the DSF consists of a single delta-like peak around the quasi-particle excitation energy $\omega_\textnormal{QP}$,
\begin{eqnarray}\label{eq:S_QP}
S_\textnormal{SP}(\mathbf{q},\omega) = A\left\{
\delta(\omega - \omega_\textnormal{QP}) + e^{-\beta\omega_\textnormal{QP}}\delta(\omega+\omega_\textnormal{QP})
\right\}\ ;\quad
\end{eqnarray}
the second term on the RHS corresponds to the respective contribution at negative frequencies and obeys the the detailed balance relation Eq.~(\ref{eq:detailed_balance}).
Inserting Eq.~(\ref{eq:S_QP}) into the two-sided Laplace transform Eq.~(\ref{eq:Laplace}) then gives the corresponding ITCF,
\begin{eqnarray}\label{eq:F_QP}
F_\textnormal{QP}(\mathbf{q},\tau) = A \left\{
e^{-\tau\omega_\textnormal{QP}} + e^{-(\beta-\tau)\omega_\textnormal{QP}}
\right\}\ .
\end{eqnarray}
For the case of a plasmon excitation in the UEG, we have $\omega_\textnormal{QP}=\omega_\textnormal{p}$ (with $\omega_\textnormal{p}=\sqrt{3/r_s^3}$ being the usual plasma frequency~\cite{quantum_theory}) for $q\to0$, and the normalization $A$ of Eqs.~(\ref{eq:S_QP}) and (\ref{eq:F_QP}) is determined by the static structure factor $S(\mathbf{q})=F(\mathbf{q},0)$,
\begin{eqnarray}\label{eq:A_normalization}
A = \frac{S(\mathbf{q})}{1+e^{-\beta\omega_\textnormal{QP}}}\ .
\end{eqnarray}
In the limit of small wave vectors, the latter can be computed analytically from the parabolic expansion~\cite{kugler_bounds}
\begin{eqnarray}\label{eq:S_RPA}
S(\mathbf{q}) = \frac{q^2}{2\omega_\textnormal{p}} \textnormal{coth}\left(
\frac{\beta\omega_\textnormal{p}}{2}
\right)\ .
\end{eqnarray}
As a final ingredient to compute the quasi-particle ITCF $F_\textnormal{QP}(\mathbf{q},\tau)$, we require some information about the wave-number dependence of the excitation energy, which can be expanded as~\cite{Hamann_CPP_2020}
\begin{eqnarray}\label{eq:omega_dispersion}
\frac{\omega^2(q)}{\omega_\textnormal{p}^2} = 1 + B_2(r_s, \Theta) \left(
\frac{q}{q_\textnormal{F}}
\right)^2\ 
\end{eqnarray}
within the RPA in the limit of small $q$.
For completeness, we note that the coefficient $B_2(r_s,\Theta)$ has been parametrized by 
Hamann \emph{et al.}~\cite{Hamann_CPP_2020}.
For $q=0.63q_\textnormal{F}$, Eq.~(\ref{eq:F_QP}) gives the correct qualitative description of $F(\mathbf{q},\tau)$, but it substantially overestimates the static limit of $\tau\to0$; this is a direct consequence of the overestimation of the true SSF by the $q\to0$ expansion given in Eq.~(\ref{eq:S_RPA}). For $q=1.25q_\textnormal{F}$, the validity of this simple plasmon approximation breaks down, and neither the $\tau$-decay nor the static limit are described accurately. Therefore, we omit the corresponding results from the other panels for even larger values of $q$.

Let us next consider the results for the semi-analytical single-particle model $F_\textnormal{SP}(\mathbf{q},\tau)$ that we have introduced in Sec.~\ref{sec:delocalization} above, and which is depicted by the solid yellow curves in Fig.~\ref{fig:ITCF_PIMC_theta1_rs4}. First and foremost, we find that $F_\textnormal{SP}(\mathbf{q},\tau)$ exhibits the correct qualitative $\tau$-decay for all $q$. Yet, it always exhibits a static limit of $F_\textnormal{SQ}(\mathbf{q},0)\equiv1$, which leads to large systematic errors for small $q$. From a physical perspective, this deficiency can be attributed to the lack of screening effects in the underlying single-particle imaginary-time diffusion process. 
Combining these results for the thermal delocalization with the exact description of static correlation effects (i.e., by using our PIMC results for $S(\mathbf{q})=F(\mathbf{q},\tau)$) via Eq.~(\ref{eq:decompose}) leads to the long-dashed purple curves. Clearly, Eq.~(\ref{eq:decompose}) leads to a spectacular improvement over the bare $F_\textnormal{SP}(\mathbf{q},\tau)$
for $q\lesssim2q_\textnormal{F}$, i.e., in the regime where the static structure factor $S(\mathbf{q})$ has not yet reached the single-particle limit of $S(q\to\infty)=1$.
This finding constitutes one of the central observations of the present work, as it implies that the complicated dynamic behaviour of the quantum UEG can be accurately approximated by a phenomenologically simple combination of static correlations, which can easily be estimated from equilibrium simulations, with the simple Gaussian imaginary-time diffusion of a single 
particle.

\begin{figure}\centering
\includegraphics[width=0.475\textwidth]{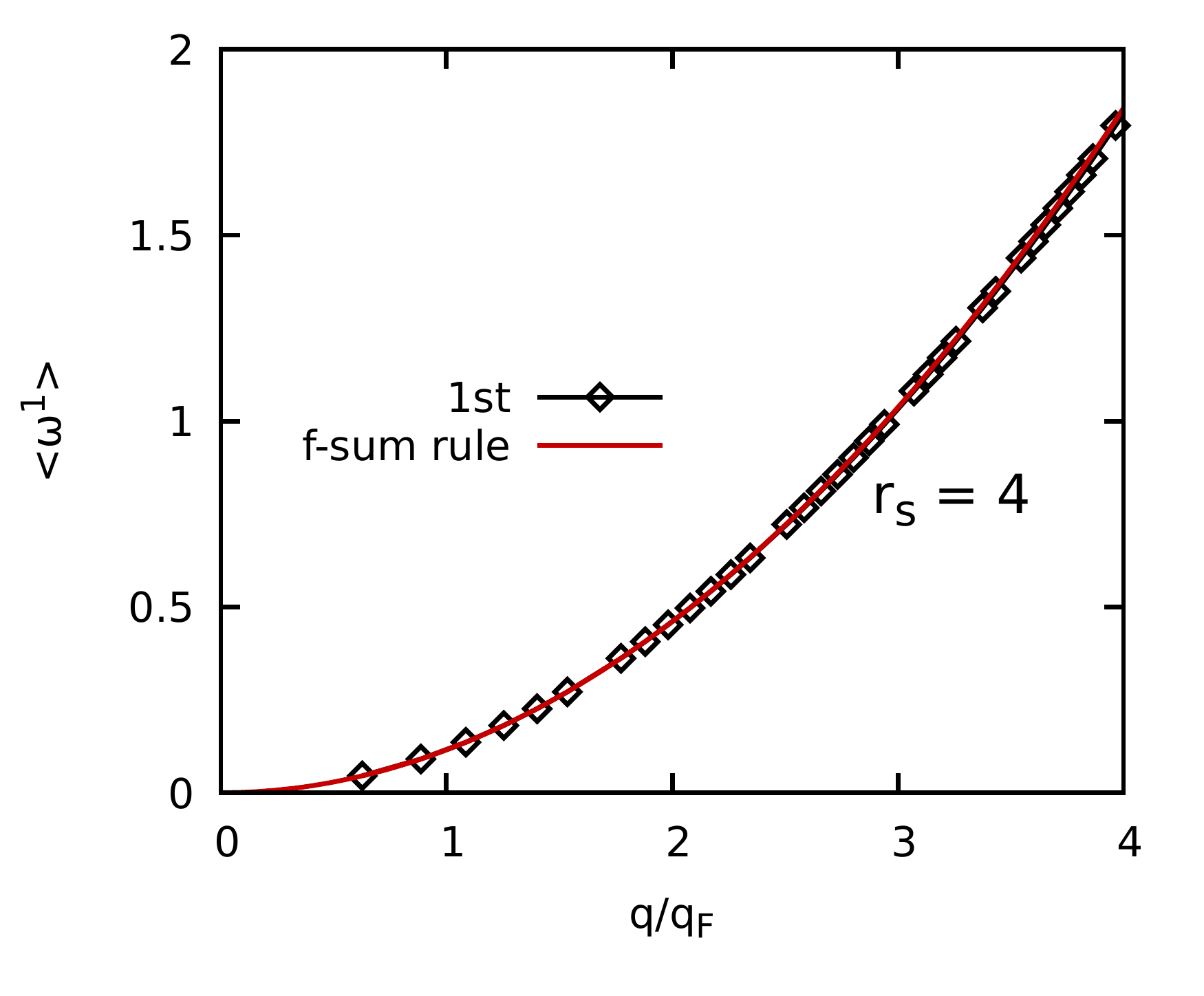}
\caption{\label{fig:Derivative}
First frequency-moment of the UEG at $r_s=4$ and $\Theta=1$. Solid red: exact f-sum rule, Eq.~(\ref{eq:f_sum_rule}); black diamonds: numerical derivative of the single-particle ITCF $F_\textnormal{SP}(\mathbf{q},\tau)$, cf.~Eq.~(\ref{eq:moments_derivative}).
}
\end{figure} 
In fact, we can show that the $\tau$-dependence of the single-particle ITCF, $\Delta F_\textnormal{SP}(\mathbf{q},\tau)$,
does become exact in the limit of $\tau\to0$ by numerically evaluating the derivative of $F_\textnormal{SP}(\mathbf{q},\tau)$ with respect to $\tau$ around the origin. The results are shown as the black squares in Fig.~\ref{fig:Derivative} and compared to the exact f-sum rule [Eq.~(\ref{eq:f_sum_rule})] that is depicted by the solid red line via Eq.~(\ref{eq:moments_derivative}).
We find perfect agreement between the numerical results obtained from the single-particle ITCF and the exact sum rule for all $q$.
Returning once more to Fig.~\ref{fig:ITCF_PIMC_theta1_rs4}, we find that $F_\textnormal{SP}(\mathbf{q},\tau)$ even becomes exact for large wave numbers where $S(\mathbf{q})\to1$, and we find perfect agreement with the PIMC reference data for $q=6.27q_\textnormal{F}$.

Finally, the grey dash-double-dotted curves in Fig.~\ref{fig:ITCF_PIMC_theta1_rs4} show results for the ITCF of the ideal Fermi gas~\cite{Baerwinkel1971}, $F_0(\mathbf{q},\tau)$ and we observe similar, though not equal trends as in $F_\textnormal{SP}(\mathbf{q},\tau)$.
In particular, fermionic exchange effects lead to density correlations between electrons of the same spin-orientation, and the static limit $F_0(\mathbf{q},0)$ that is depicted as the dotted black curve in Fig.~\ref{fig:SSF_theta1} deviates from unity for $q\lesssim2q_\textnormal{F}$. Therefore, $F_0(\mathbf{q},\tau)$ is somewhat closer to the PIMC results for $q\lesssim2q_\textnormal{F}$ compared to $F_\textnormal{SP}(\mathbf{q},\tau)$, even though they exhibit a nearly identical dependence on the imaginary time.

\begin{figure*}\centering
\includegraphics[width=0.475\textwidth]{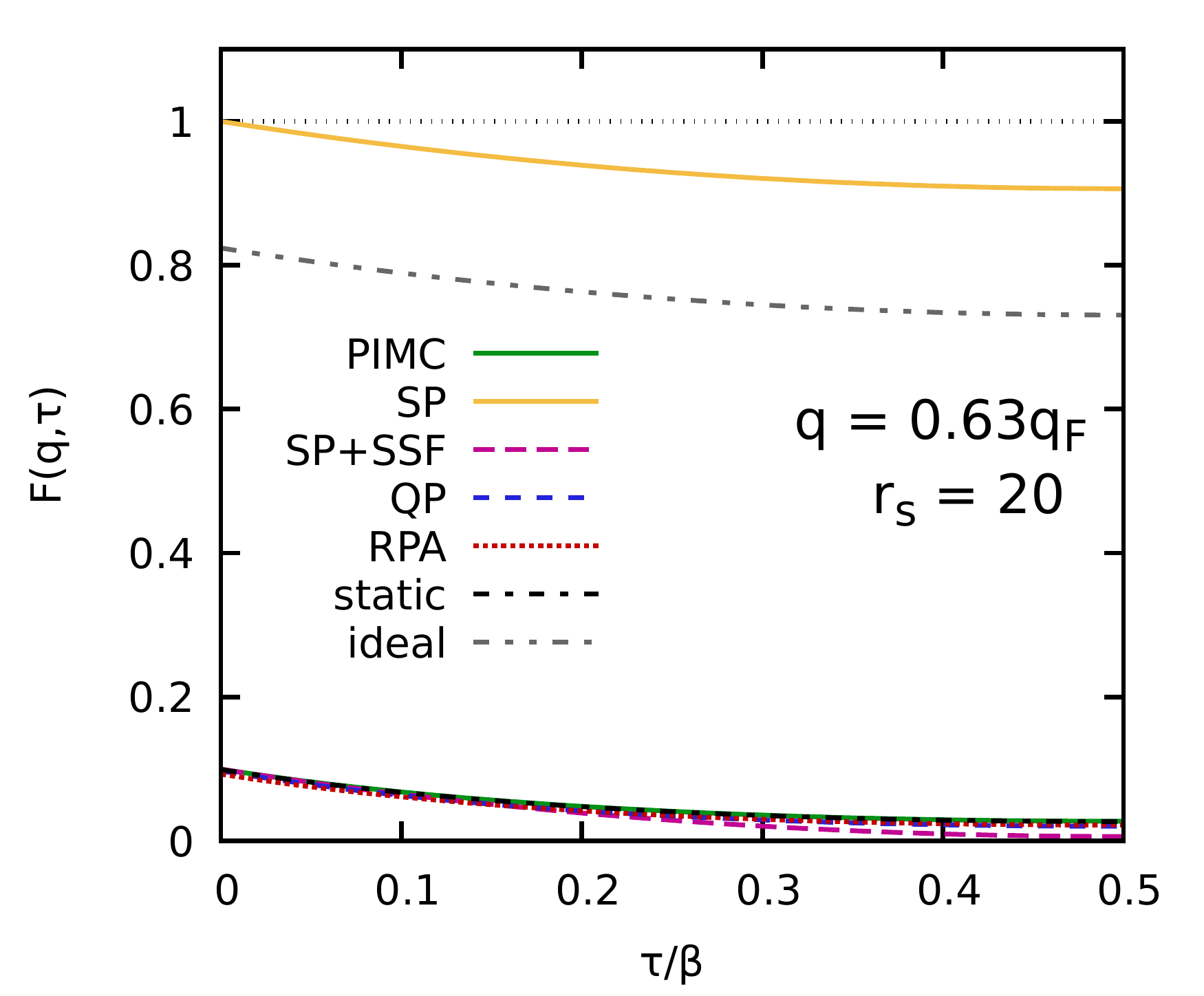}\includegraphics[width=0.475\textwidth]{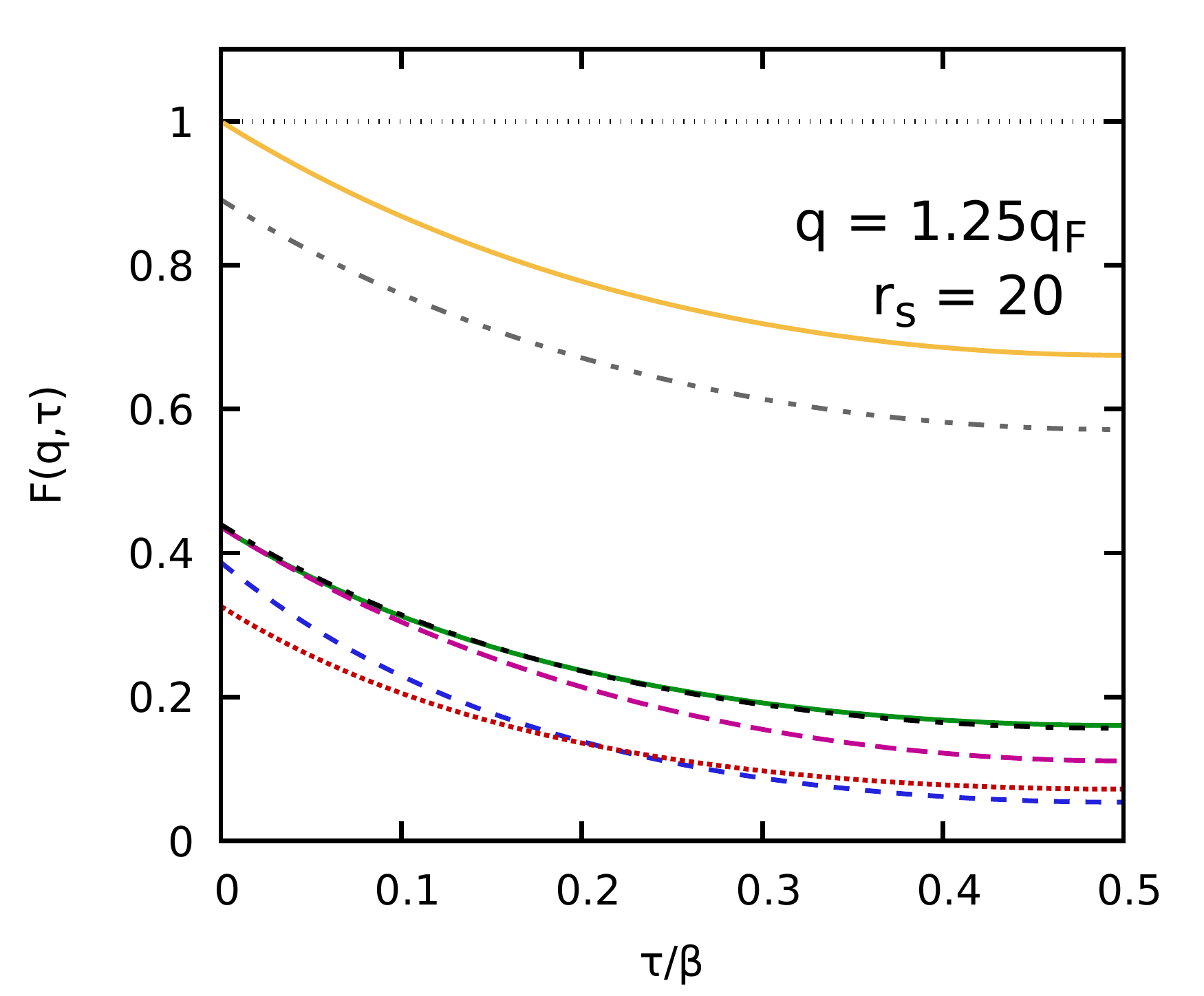}\\\vspace*{-0.5cm}\includegraphics[width=0.475\textwidth]{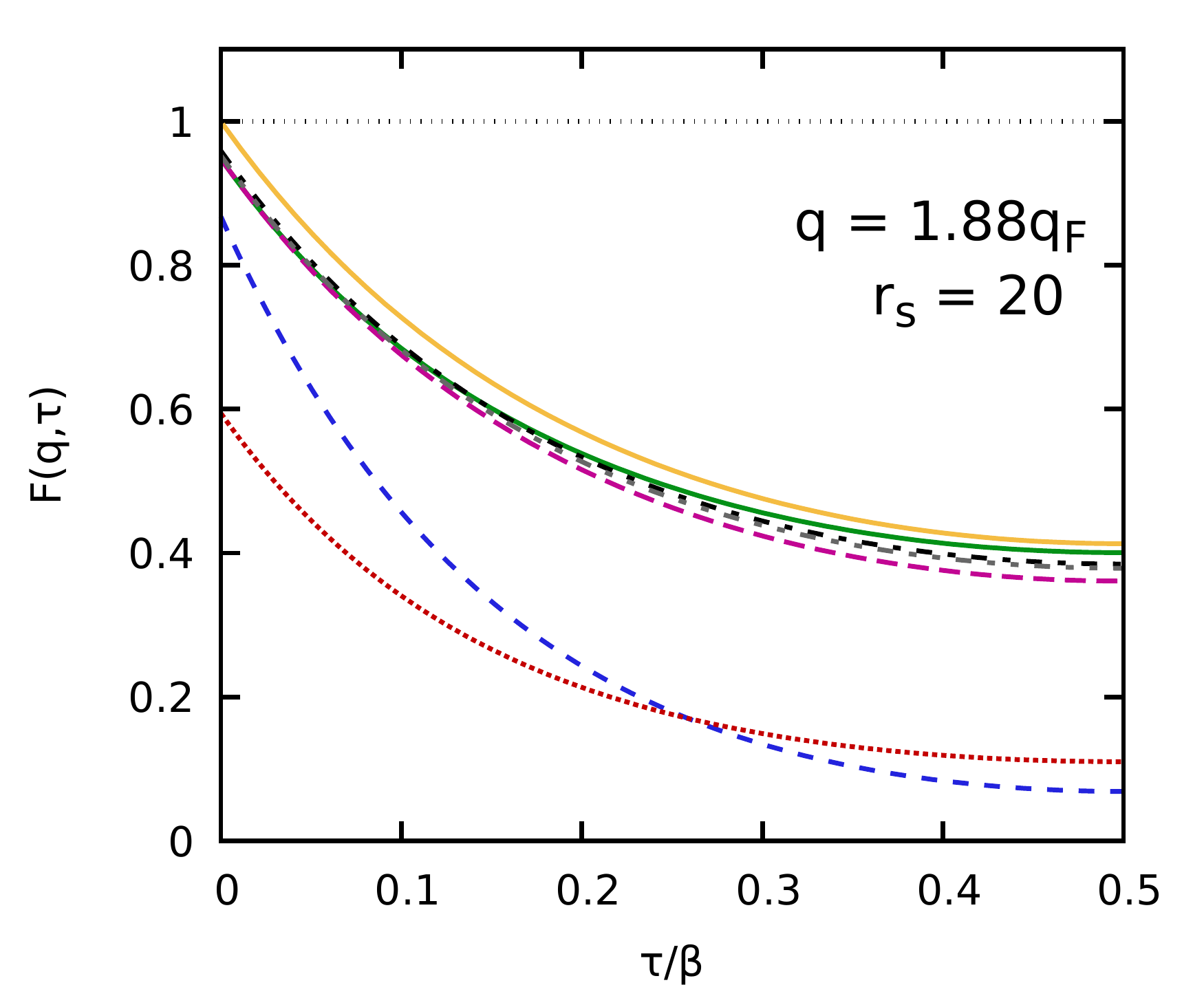}\includegraphics[width=0.475\textwidth]{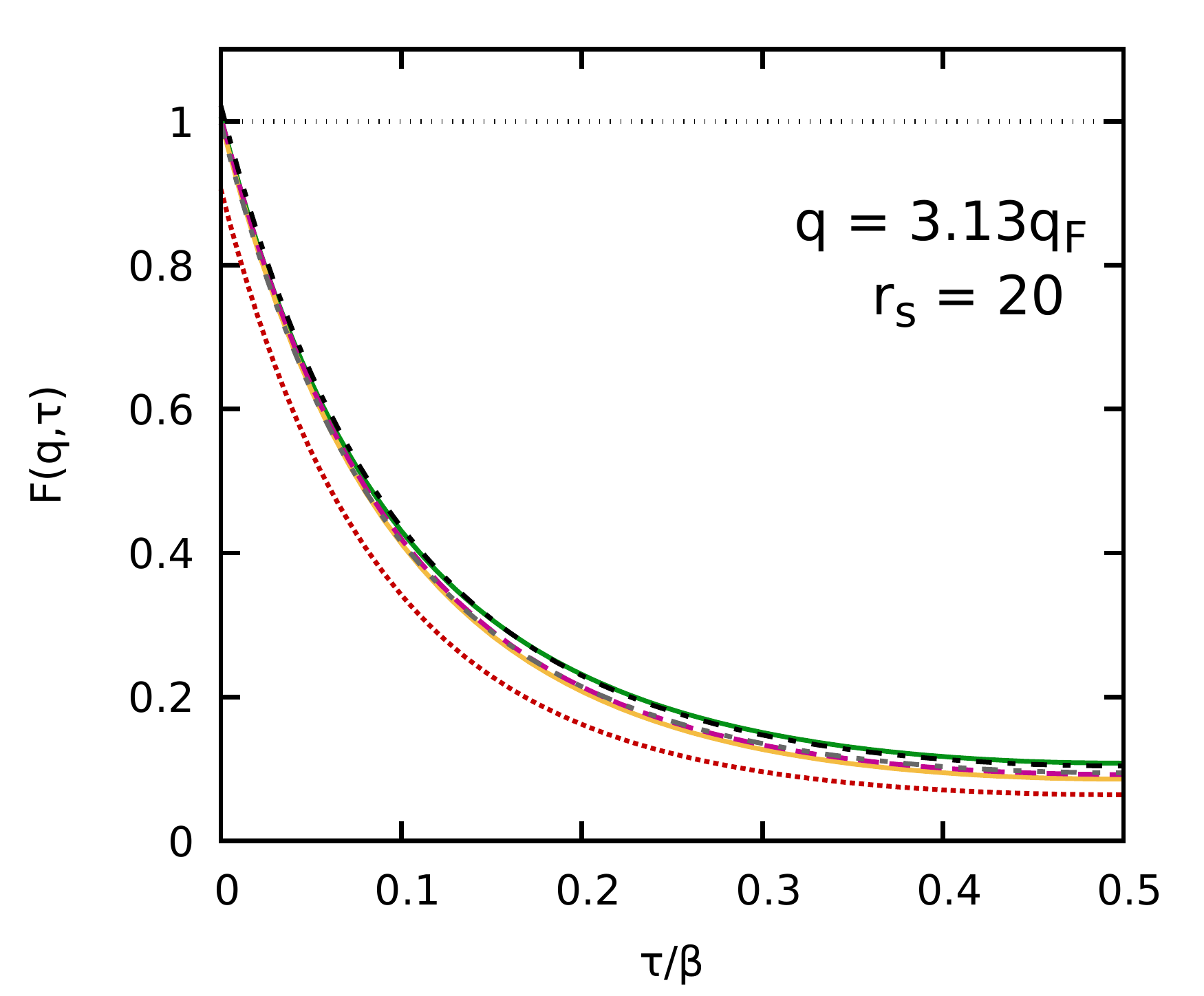}\\\vspace*{-0.5cm}\includegraphics[width=0.475\textwidth]{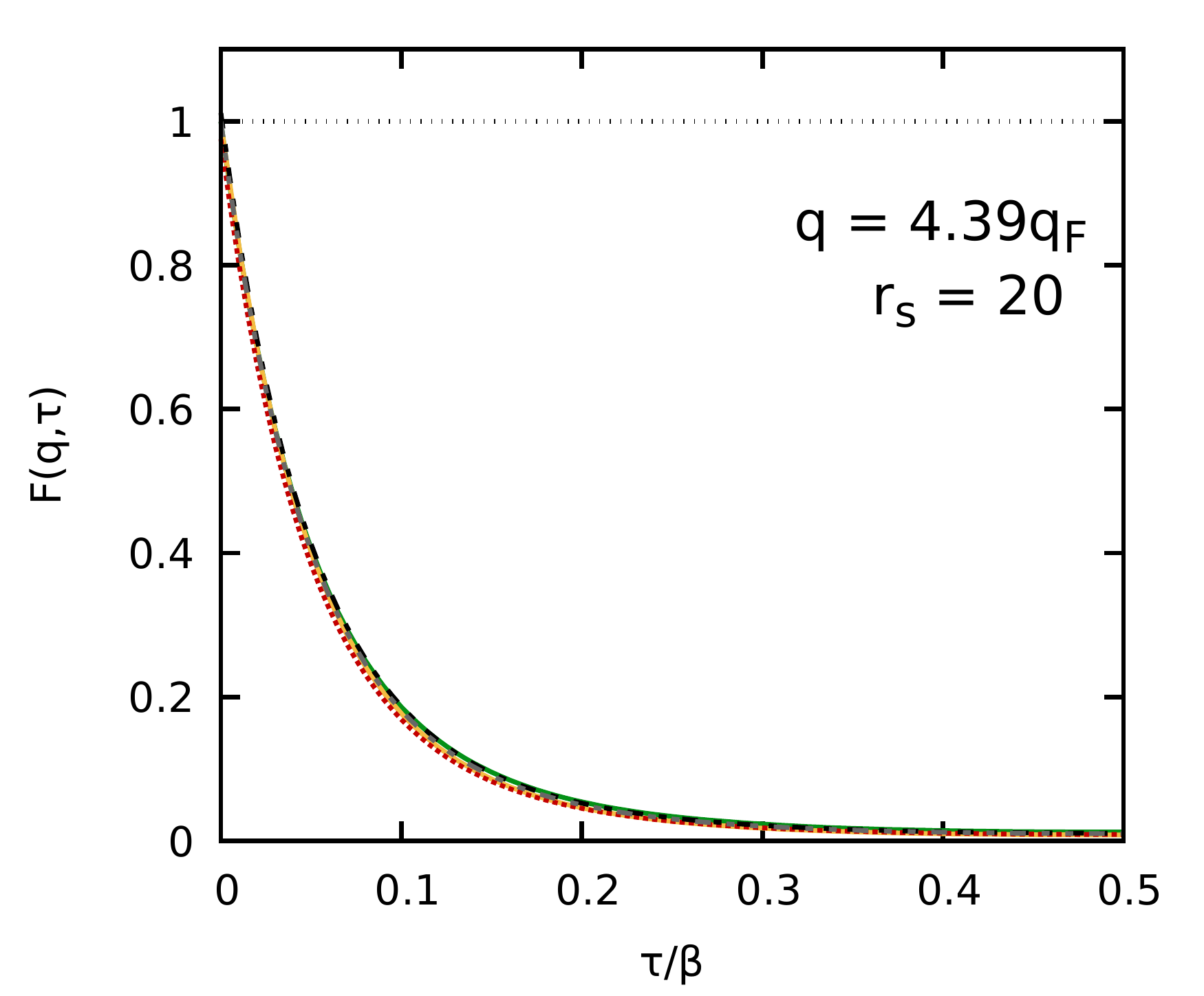}\includegraphics[width=0.475\textwidth]{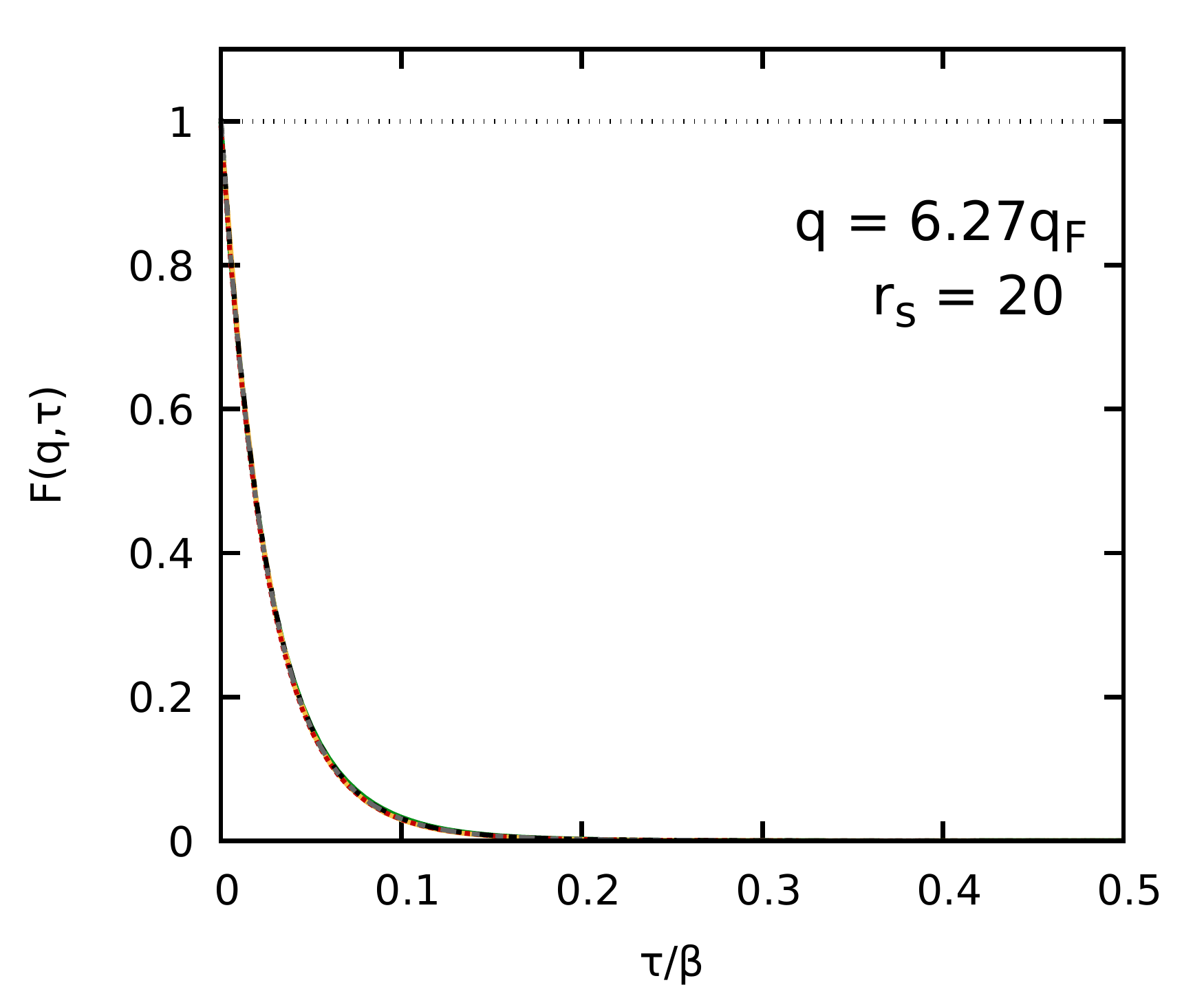}
\caption{\label{fig:ITCF_PIMC_theta1_rs20}
Imaginary-time dependence of the ITCF $F(\mathbf{q},\tau)$ for different wave numbers $q$ for the unpolarized UEG with $N=34$, $r_s=20$, and $\Theta=1$. Solid green: exact PIMC results; solid yellow: single-particle ITCF $F_\textnormal{SP}(\mathbf{q},\tau)$ [Eq.~(\ref{eq:F_SP})]; long-dashed purple: combination of SSF with single-particle ITCF, Eq.~(\ref{eq:decompose}); dashed blue: quasi-particle expansion, Eq.~(\ref{eq:F_QP}); dotted red: RPA; dash-dotted black: static approximation, i.e., $G_\textnormal{static}(\mathbf{q},\omega) \equiv G(\mathbf{q},0)$; dash-double-dotted grey: ideal Fermi gas.
}
\end{figure*}

Let us next repeat this analysis at a lower density, $r_s=20$, which is shown in Fig.~\ref{fig:ITCF_PIMC_theta1_rs20}. From a physical perspective, these conditions are characterized by the onset of strong electronic correlations and constitute the boundary to the strongly coupled electron liquid regime~\cite{dornheim_electron_liquid}. As a consequence, the dispersion relation of the DSF exhibits a pronounced roton feature~\cite{dornheim_dynamic,Takada_PRB_2016} that has very recently been explained by Dornheim \emph{et al.}~\cite{Dornheim_Nature_2022} in terms of a new pair alignment model. 
Since we overall find similar trends of $F(\mathbf{q},\tau)$ as compared to $r_s=4$, we here restrict ourselves to a brief discussion of the main differences between the two regimes.
In fact, the most pronounced difference comes from the static limit $F(\mathbf{q},0)=S(\mathbf{q})$, which is shown in Fig.~\ref{fig:SSF_theta1} as the red circles and red dashed line for PIMC and RPA. In particular, we find that $S(\mathbf{q})$ attains smaller values compared to $r_s=4$ for $ q \lesssim 2q_\textnormal{F}$, which is directly reflected by our results for $F(\mathbf{q},\tau)$ in Fig.~\ref{fig:ITCF_PIMC_theta1_rs20}. In addition, the RPA is less accurate as electronic exchange--correlation effects are substantially more important in the electron liquid regime.
Surprisingly, we find that the RPA is even less accurate than the bare single-particle ITCF $F_\textnormal{SP}(\mathbf{q},\tau)$, or the corresponding function of the ideal Fermi gas $F_0(\mathbf{q},\tau)$ for intermediate wave numbers; this is a direct consequence of the overestimation of the Coulomb repulsion in the RPA, which is discussed in more detail, e.g., in Ref.~\cite{Dornheim_Nature_2022}.

In addition, we note that the combination of our new single-particle imaginary-time diffusion model for $F_\textnormal{SP}(\mathbf{q},\tau)$ with the static structure factor via Eq.~(\ref{eq:decompose})
overall exhibits a comparable accuracy as for $r_s=4$. An exception is only given for the smallest depicted wave number, $q=0.63 q_\textnormal{F}$, where the corresponding long-dashed purple curve exhibits a too pronounced decay towards $\tau=\beta/2$.
Finally, we find exactly the same functional form of all models for $F(\mathbf{q},\tau)$ as for $r_s=4$ at $q=6.27q_\textnormal{F}$, as electron--electron correlations do not influence density correlations in the single-particle regime.

\subsection{Dependence on the wave number\label{sec:q}}

\begin{figure}\centering
\hspace*{0.0\textwidth}\includegraphics[width=0.475\textwidth]{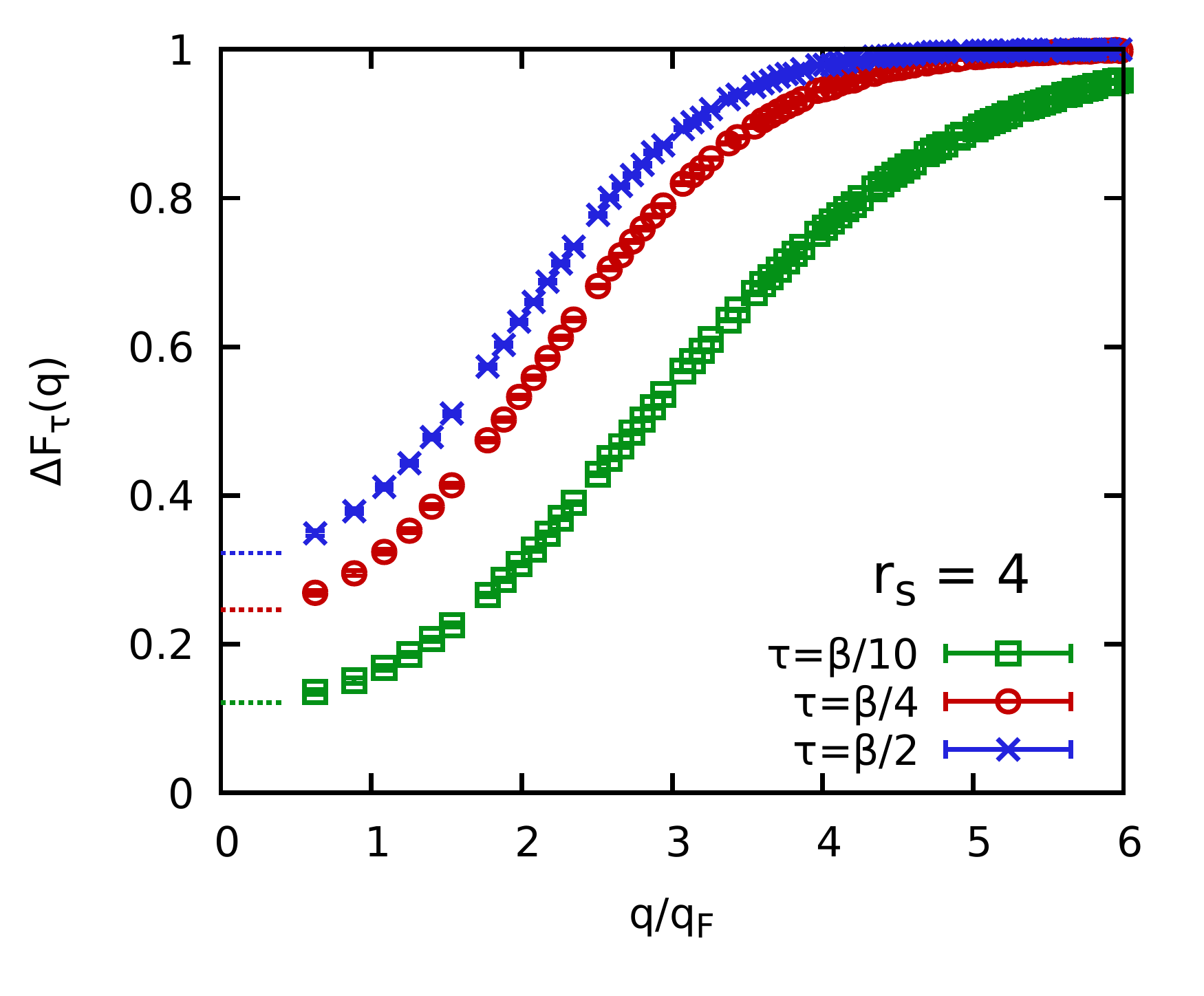}\\\vspace*{-0.75cm}\includegraphics[width=0.475\textwidth]{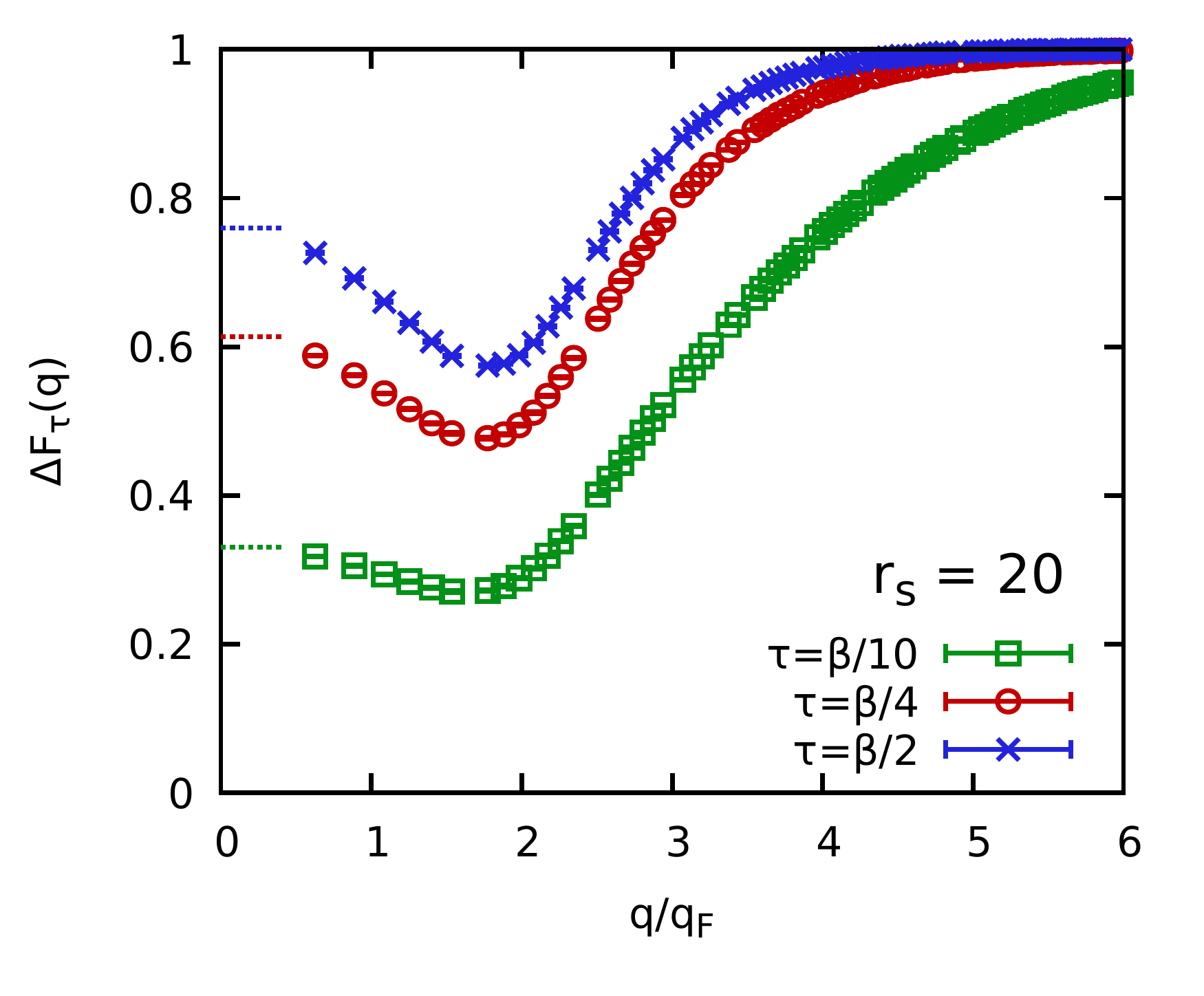}
\caption{\label{fig:decay_PIMC_theta1_rs}
\emph{Ab initio} PIMC results for the relative decay measure $\Delta F_\tau(\mathbf{q})$ for $\tau=\beta/2$ (blue crosses), $\tau=\beta/4$ (red circles) and $\tau=\beta/10$ (green squares) of the UEG at $\Theta=1$ for $r_s=4$ (top) and $r_s=20$ (bottom).
}
\end{figure}

To get a better qualitative insight into the dependence of imaginary-time density--density correlations on the wave number $q$, Dornheim \emph{et al.}~\cite{Dornheim_insight_2022} have very recently suggested to analyse the relative decay measure $\Delta F_\tau(\mathbf{q})$ defined in Eq.~(\ref{eq:decay_measure}) above. The results are shown in Fig.~\ref{fig:decay_PIMC_theta1_rs} for $r_s=4$ (top) and $r_s=20$ (bottom) based on exact PIMC simulation data for $F(\mathbf{q},\tau)$.
More specifically, the green squares, red circles, and blue crosses show $\Delta F_\tau(\mathbf{q})$ for $\tau=\beta/10$, $\tau=\beta/4$, and $\tau=\beta/2$, respectively. 
For $r_s=4$, the three depicted data sets exhibit a qualitatively similar behaviour. In the limit of small $q$, they approach the exact plasmon quasi-particle limit that directly follows from $F_\textnormal{QP}(\mathbf{q}\to0,\tau)$, cf.~Eq.~(\ref{eq:F_QP}) above. Upon increasing $q$, all three curves monotonically increase and eventually attain the single-particle limit of $\Delta F_\tau(\mathbf{q}\to\infty)=1$ for large $q$; this is a direct consequence of the steeper decay of $F(\mathbf{q},\tau)$ with $\tau$ for $0\leq\tau\leq\beta/2$ that is depicted in Figs.~\ref{fig:ITCF_PIMC_theta1_rs4} f) and \ref{fig:ITCF_PIMC_theta1_rs20} f). Furthermore, it is easy to see that the onset of the single-particle limit 
happens for increasingly large $q$ for decreasing imaginary-time distances $\tau$ in $\Delta F_\tau(\mathbf{q})$.

For $r_s=20$, the relative deviation measure exhibits an even more interesting dependence on $q$. While again all three data sets attain their respective plasmon limit for $q\to0$, $\Delta F_\tau(\mathbf{q})$ has a minimum for intermediate wave numbers of $q\sim2q_\textnormal{F}$ for all depicted values of $\tau$; for larger $q$, we then recover the same single-particle limit as for $r_s=4$.
From a physical perspective, the minimum in the relative decay measure signals a reduced decay of correlations along the imaginary-time; in other words, electronic correlations remain more stable throughout the imaginary-time diffusion process described in Sec.~\ref{sec:delocalization} above in the strongly coupled case of $r_s=20$. Indeed, Dornheim \emph{et al.}~\cite{Dornheim_insight_2022} have suggested that such a reduced $\tau$-decay of $F(\mathbf{q},\tau)$ can directly be translated to a shift of spectral weight in the DSF towards lower frequencies; this effect manifests as a \emph{roton-type} minimum in the dispersion relation $\omega(q)$ of $S(\mathbf{q},\omega)$, cf.~Fig.~\ref{fig:rescale_PIMC_theta1_rs} below. For completeness, we note that this effect appears when the wave length $\lambda_q=2\pi/q$ is comparable to the average interparticle distance $d$; it has been explained in Ref.~\cite{Dornheim_Nature_2022} in terms of the spontaneous alignment of pairs of electrons, which was further substantiated in the subsequent Ref.~\cite{Dornheim_JCP_2022}.

\begin{figure*}\centering
\hspace*{0.01\textwidth}\includegraphics[width=0.46\textwidth]{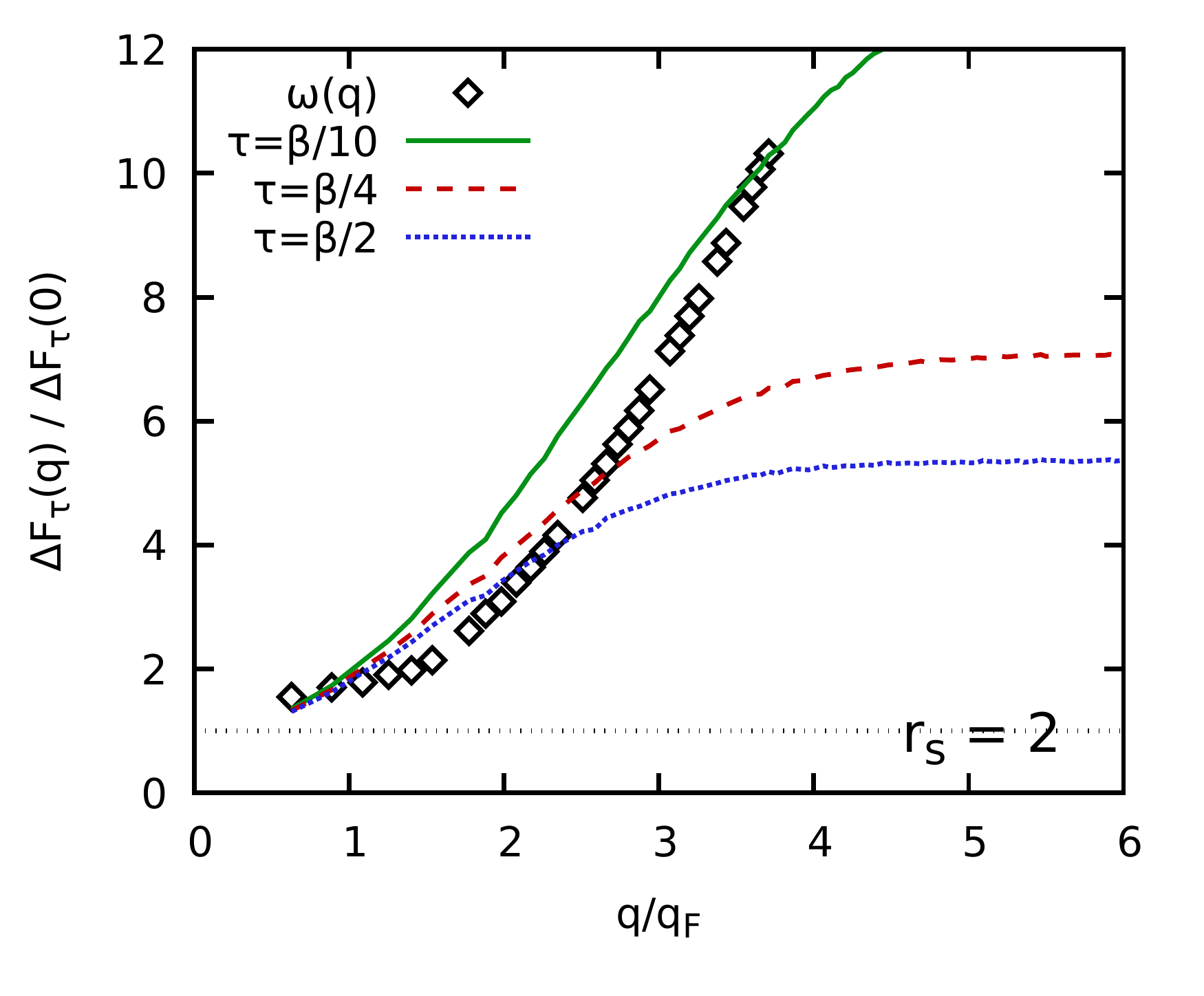}\hspace*{0.02\textwidth}\includegraphics[width=0.448\textwidth]{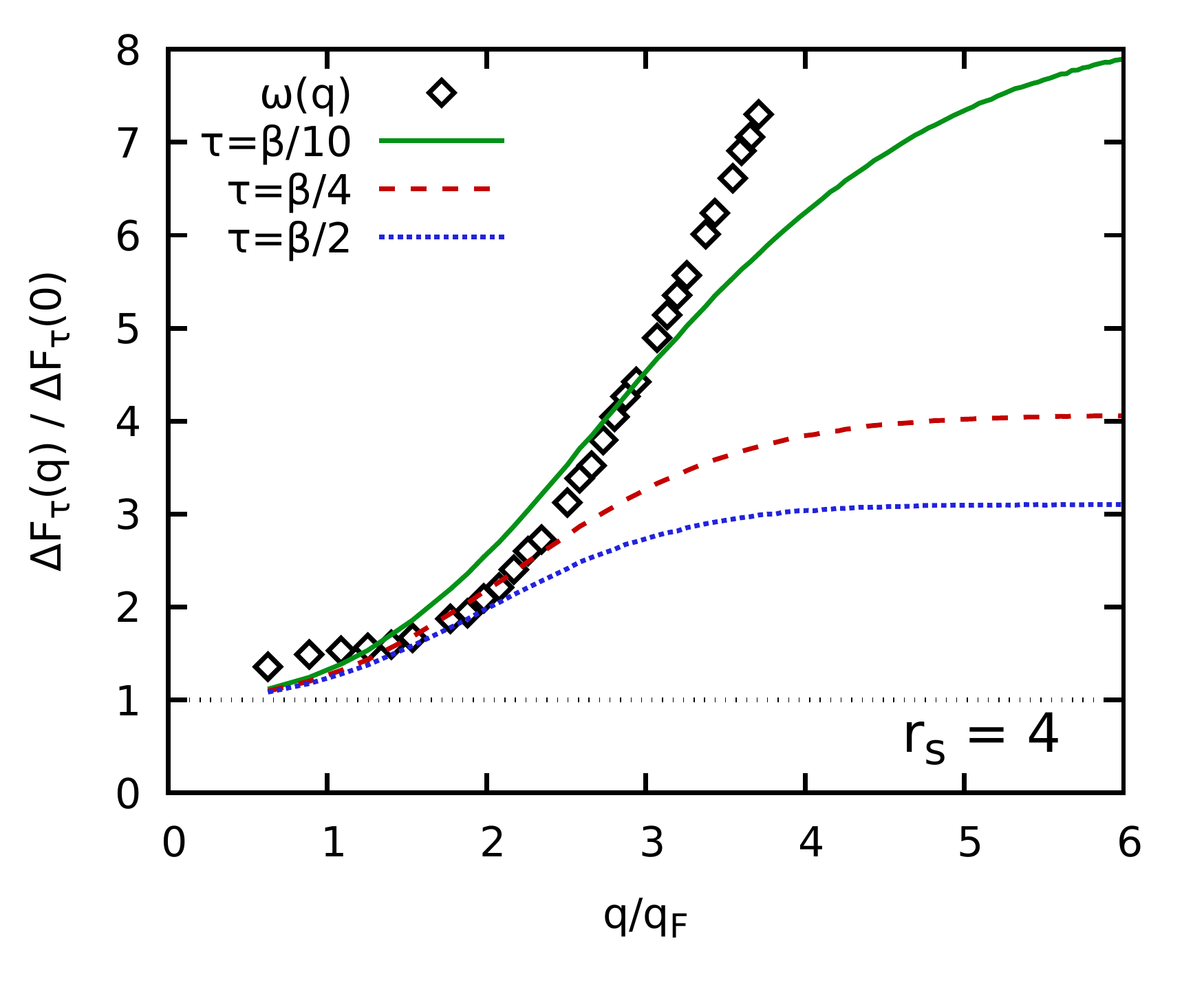}\\\vspace*{-0.75cm}\includegraphics[width=0.475\textwidth]{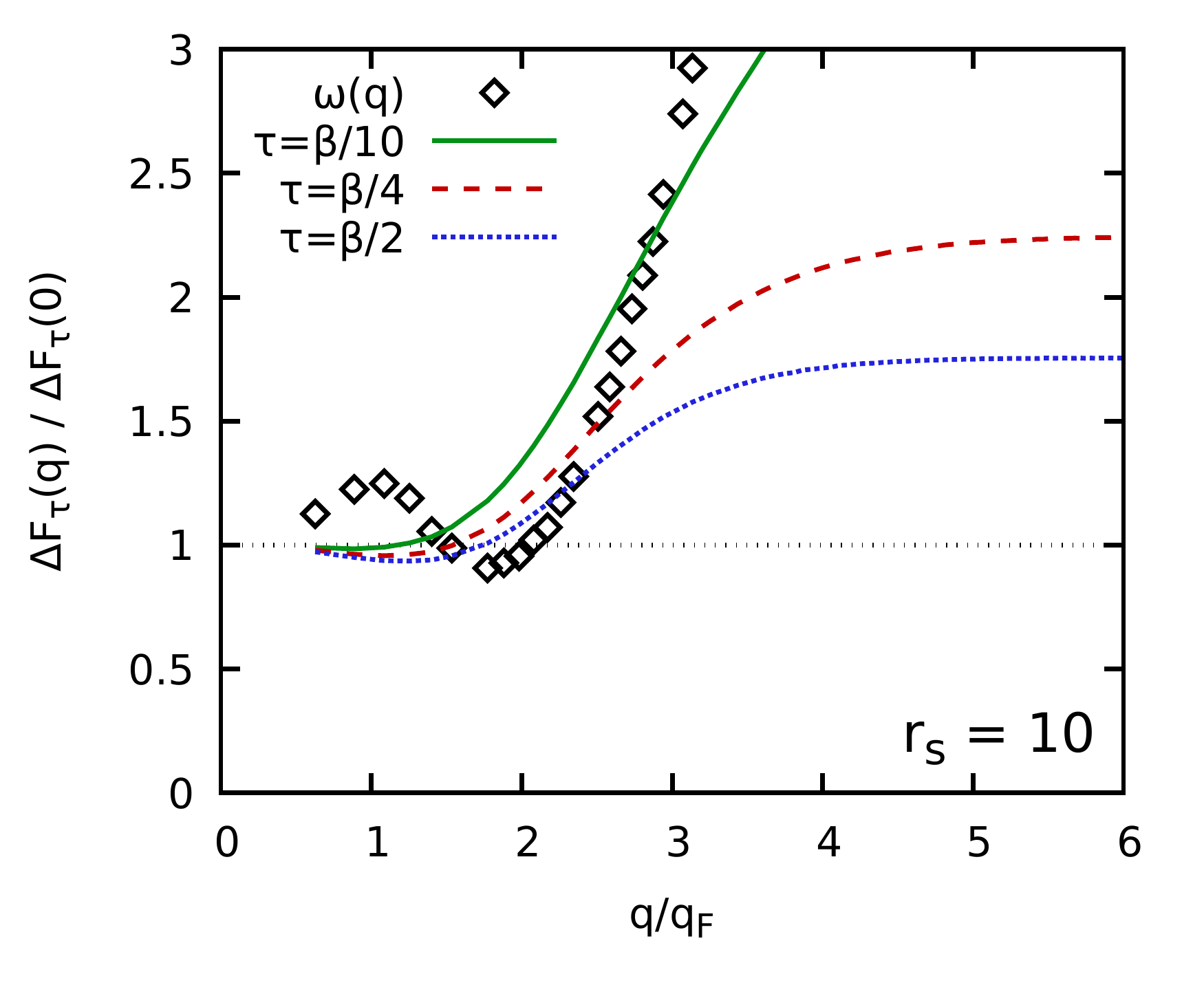}\includegraphics[width=0.475\textwidth]{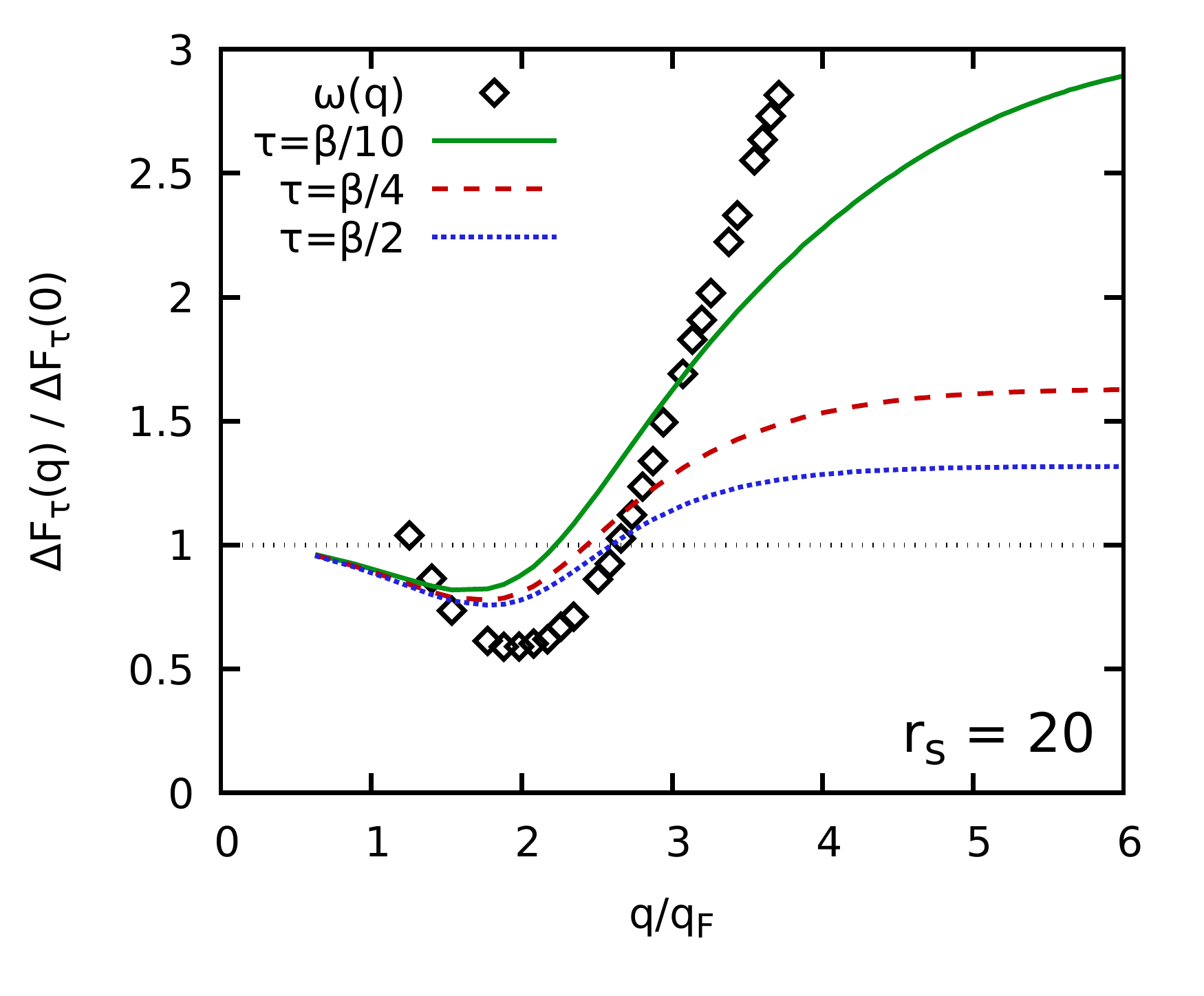}
\caption{\label{fig:rescale_PIMC_theta1_rs}
PIMC results for the dispersion relation $\omega(\mathbf{q})$ of the dynamic structure factor (black diamonds, taken from Ref.~\cite{dornheim_dynamic}) and relative decay measure of the ITCF $\Delta F_\tau(\mathbf{q})$
for $\tau=\beta/2$ (dotted blue), $\tau=\beta/4$ (dashed red), and $\tau=\beta/10$ (solid green) for the unpolarized UEG with $N=34$ at $\Theta=1$ for different values of the density parameter $r_s$. All curves have been re-normalised to their respective $q\to0$ limit, which is given by the plasma frequency $\omega_\textnormal{p}$ in the case of $\omega(\mathbf{q})$.
}
\end{figure*}

The connection between the relative measure of $\tau$-decay $\Delta F_\tau(\mathbf{q})$ and the dispersion $\omega(\mathbf{q})$ of $S(\mathbf{q},\omega)$ is investigated in more detail in Fig.~\ref{fig:rescale_PIMC_theta1_rs}, where we overlay both quantities at $\Theta=1$ for different values of the density parameter $r_s$; we note that all curves have been rescaled by their respective $q\to0$ limit, which is given by the plasma frequency $\omega_\textnormal{p}$ in the case of the dispersion relation. For the metallic density of $r_s=2$, the PIMC-based results for $\omega(\mathbf{q})$ taken from Ref.~\cite{dornheim_dynamic} 
exhibit a smooth, monotonous crossover between the collective regime where the DSF corresponds to a sharp plasmon feature [cf.~Eq.~(\ref{eq:S_QP})], and the single-particle regime with $\lambda_q\ll d$, where it holds $\omega(\mathbf{q})\sim q^2$.
The corresponding results for the decay measure $\Delta F_\tau(\mathbf{q})$ exhibit a very similar behaviour; the main difference to $\omega(\mathbf{q})$ is given by the fact that the single-particle limit is given by a constant instead of a parabolic divergence.

For $r_s=4$, we find a very similar picture compared to $r_s=2$ for both $\omega(\mathbf{q})$ and the three data sets for $\Delta F_\tau(\mathbf{q})$.
Upon further decreasing the density to $r_s=10$, the dispersion starts to exhibit the incipient \emph{roton minimum} around intermediate wave numbers. This effect is qualitatively reflected by the relative $\tau$-decay measure $\Delta F_\tau(\mathbf{q})$ by a minimum around $q\sim 2q_\textnormal{F}$, although its overall behaviour is less complicated compared to $\omega(\mathbf{q})$, as it does not exhibit the local maximum of the latter around $q=q_\textnormal{F}$.

Finally, we show results for the strongly coupled case of $r_s=20$ in Fig.~\ref{fig:rescale_PIMC_theta1_rs} d). In this case, both the \emph{roton minimum} in $\omega(\mathbf{q})$ and the reduced decay in $\Delta F_\tau(\mathbf{q})$ are more pronounced, as it is expected. Interestingly, the physical mechanism of the roton---the spontaneous alignment of pairs of electrons~\cite{Dornheim_Nature_2022}---manifests in $\Delta F_\tau(\mathbf{q})$ with a similar magnitude for all three considered values of $\tau$. This finding is of practical importance for the discussion of different approximate theories for $F(\mathbf{q},\tau)$, which are discussed next.

\begin{figure*}\centering
\hspace*{0.0\textwidth}\includegraphics[width=0.475\textwidth]{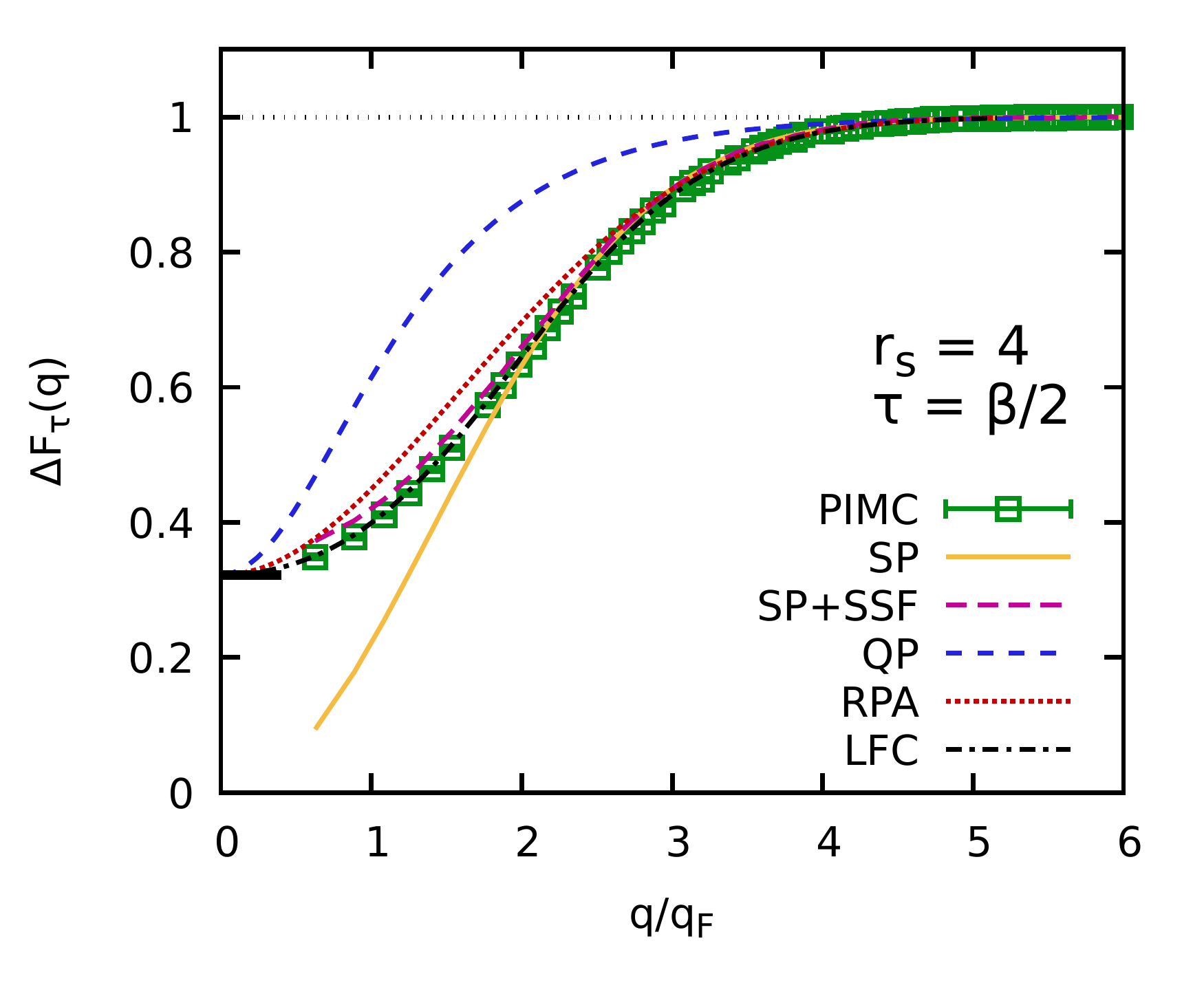}\includegraphics[width=0.475\textwidth]{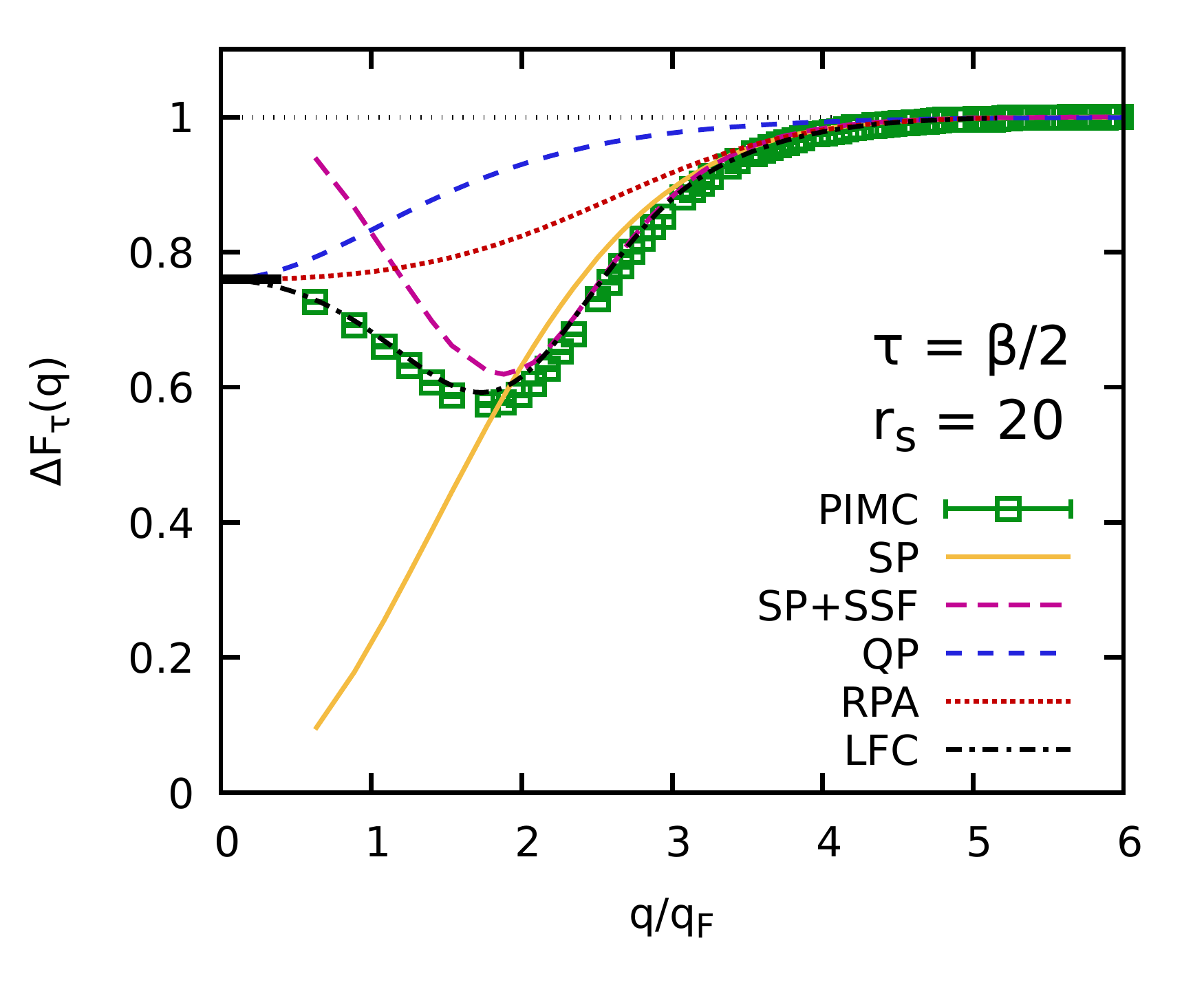}\\\vspace*{-0.75cm}\includegraphics[width=0.475\textwidth]{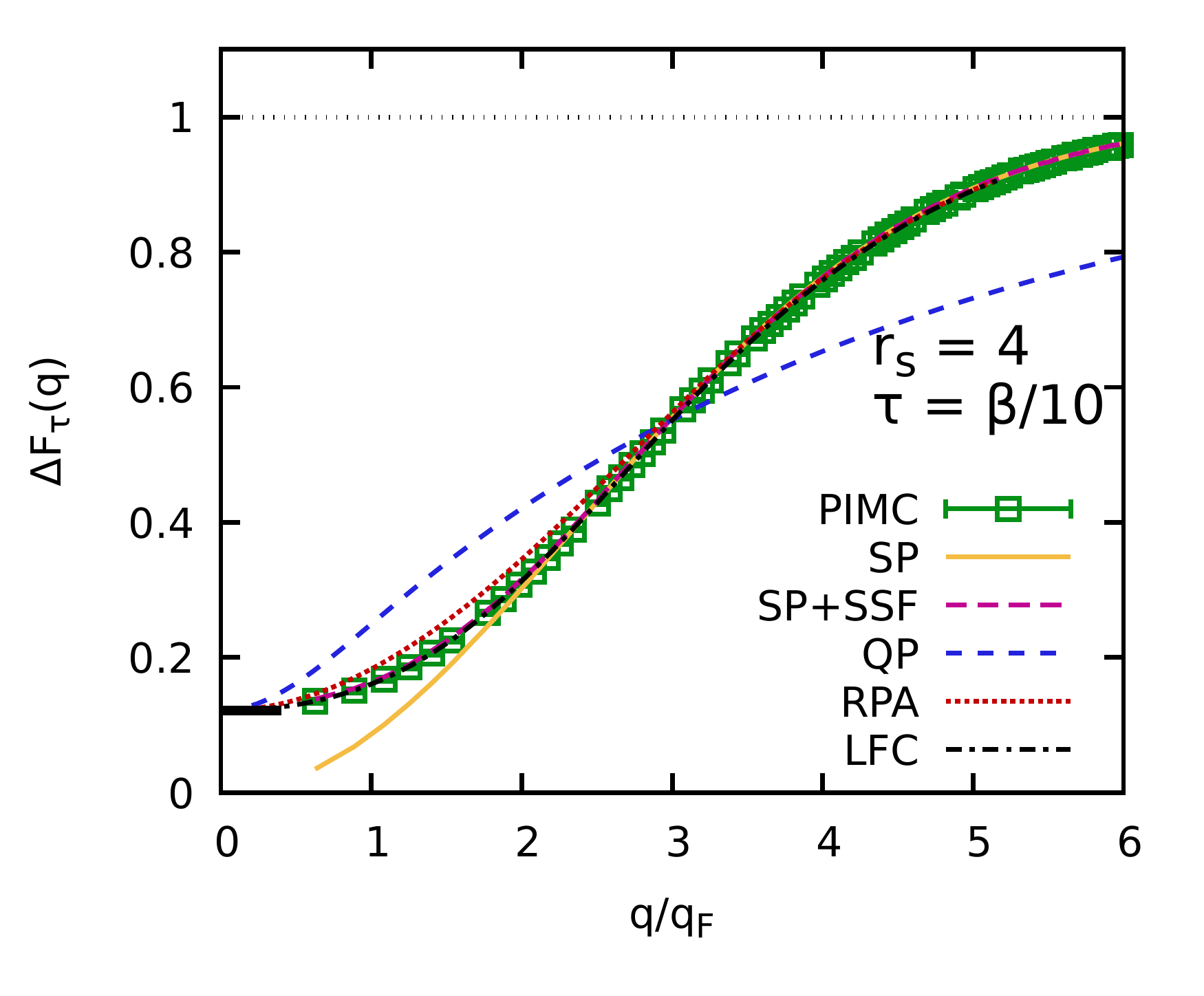}\includegraphics[width=0.475\textwidth]{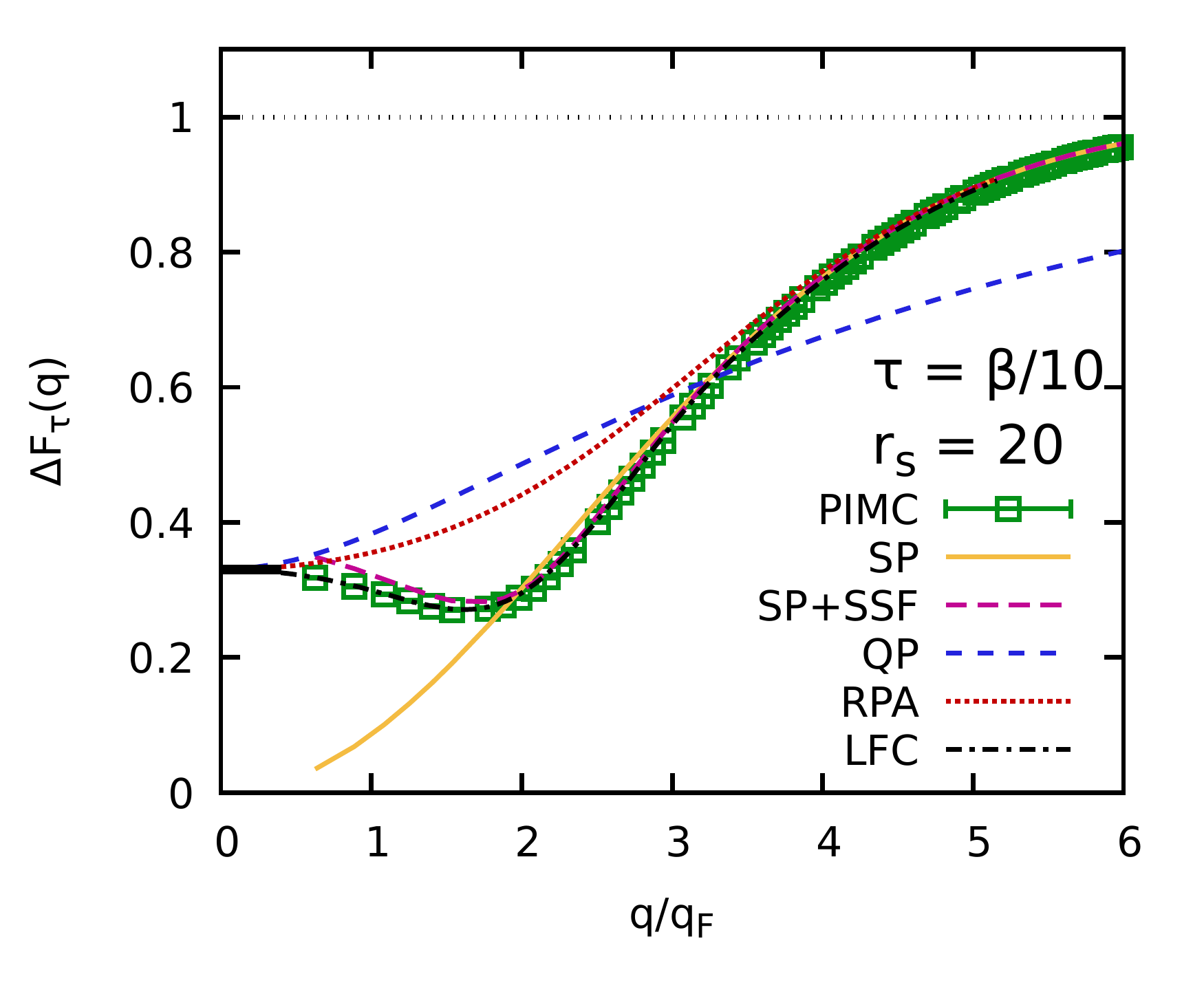}
\caption{\label{fig:compare_decay_PIMC_theta1_rs}
Relative imaginary-time decay measure $\Delta F_\tau(\mathbf{q})$ [cf.~Eq.~(\ref{eq:decay_measure})] of the unpolarized UEG at $r_s=4$ (left) and $r_s=20$ (right) for $\Theta=1$. The top and bottom rows have been obtained for $\tau=\beta/2$ and $\tau=\beta/10$. Green squares: \emph{ab initio} PIMC results; solid yellow: single-particle ITCF model $F_\textnormal{SP}(\mathbf{q},\tau)$ [Eq.~(\ref{eq:F_SP})]; long-dashed purple: single-particle $\tau$-decay combined with PIMC results for the SSF $S(\mathbf{q})$ via Eq.~(\ref{eq:decompose}); dashed blue: quasi-particle ITCF $F_\textnormal{QP}(\mathbf{q},\tau)$ [Eq.~(\ref{eq:F_QP})]; dotted red: RPA; dash-dotted black: static approximation using as input the LFC from Ref.~\cite{dornheim_ML}.
}
\end{figure*}

In Fig.~\ref{fig:compare_decay_PIMC_theta1_rs}, we show the wave-number dependence of $\Delta F_\tau(\mathbf{q})$ for both $r_s=4$ (left column) and $r_s=20$ (right column) for $\tau=\beta/2$ (top row) and $\tau=\beta/10$. The green squares have been obtained from our \emph{ab initio} PIMC simulations and provide an unassailable benchmark for the other depicted theoretical approaches.

Let us start our comparative analysis by considering the dashed blue curves, which have been computed from the quasi-particle approximation for the ITCF, $F_\textnormal{QP}(\mathbf{q},\tau)$, defined in Eqs.~(\ref{eq:F_QP}) and (\ref{eq:omega_dispersion}) above. For $r_s=4$, the corresponding decay measure qualitatively follows the PIMC data; both the limits of large- and small wave numbers are correctly reproduced, although this happens only in the limit for very large $q$ in the case of $\tau=\beta/10$.
In addition, $\Delta F_\tau(\mathbf{q})$ exhibits a smooth and monotonous transition between these collective and single-particle limits, which is phenomenologically correct, but quantitatively strongly deviates from the green squares. In particular, we find that while $F_\textnormal{QP}(\mathbf{q},\tau)$ becomes, by definition, exact in the plasmon limit for $q\to0$, the change in the position of the delta peak described by Eq.~(\ref{eq:omega_dispersion}) does not constitute a good first-order correction to this limit. In other words, the finite width of the true DSF for $q>0$ appears to be more important than the plasmon shift away from the plasma frequency $\omega_\textnormal{p}$.
This trend can be seen particularly well for the more strongly coupled case of $r_s=20$, where the true $\Delta F_\tau(\mathbf{q})$ has a negative slope for small $q$; in contrast, the quasi-particle approximation $F_\textnormal{QP}(\mathbf{q},\tau)$ exhibits the opposite behaviour for both depicted values of $\tau$.

Let us next consider the full RPA, which has been included as the dotted red curves in Fig.~\ref{fig:compare_decay_PIMC_theta1_rs}. Clearly, the RPA is exact both in the limits of $q\to0$ and $q\gg q_\textnormal{F}$, and constitutes a substantial improvement over the simple QP ansatz in between. For $r_s=4$, the RPA performs comparably well over the entire $q$-range, with a maximum relative error around intermediate wave numbers $q\sim2q_\textnormal{F}$. In addition, we note that the $\tau$-decay measure within the RPA is more accurate for $\tau=\beta/10$ compared to $\tau=\beta/2$. This can be explained by the fact that, while the SSF $S(\mathbf{q})=F(\mathbf{q},0)$ in the RPA exhibits systematic errors for intermediate q, cf.~Fig.~\ref{fig:SSF_theta1} above, it does attain the correct first derivative given by the well-known f-sum rule [Eq.~(\ref{eq:f_sum_rule})].
For $r_s=20$, the qualitative accuracy of the RPA noticeably deteriorates as electron--electron exchange--correlation effects become more important. In particular, the RPA, too, does not reproduce the negative slope of the exact PIMC data for $\Delta F_\tau(\mathbf{q})$ for small $q$. This is consistent to the inability of the RPA to capture the \emph{roton minimum} in the dispersion  relation of the DSF, which has been investigated in detail in Refs.~\cite{dornheim_dynamic,Dornheim_Nature_2022}.

This systematic shortcoming of the mean-field based RPA is remedied by including exact PIMC results for the static local field correction via Eq.~(\ref{eq:LFC}), and the results for $\Delta F_\tau(\mathbf{q})$ are shown as the dash-dotted black curve in Fig.~\ref{fig:compare_decay_PIMC_theta1_rs}.
For the metallic density of $r_s=4$, no systematic deviations between the \emph{static approximation} and the PIMC reference results can be resolved on the depicted scale. 
For the electron liquid at $r_s=20$, too, we find that including the static local field correction leads to a spectacular improvement, and the dash-dotted black curve nicely captures the \emph{roton minimum} of $\Delta F_\tau(\mathbf{q})$ around $q=2q_\textnormal{F}$. At the same time, small deviations to the green squares can be resolved in particular for $\tau=\beta/2$ in this regime; yet, the effect is significantly smaller than the corresponding underestimation of the true depth of the \emph{roton minimum} of $S(\mathbf{q},\omega)$ due to the \emph{static approximation}, which is discussed in more detail in Ref.~\cite{Dornheim_Nature_2022}.

\begin{figure}\centering
\includegraphics[width=0.45\textwidth]{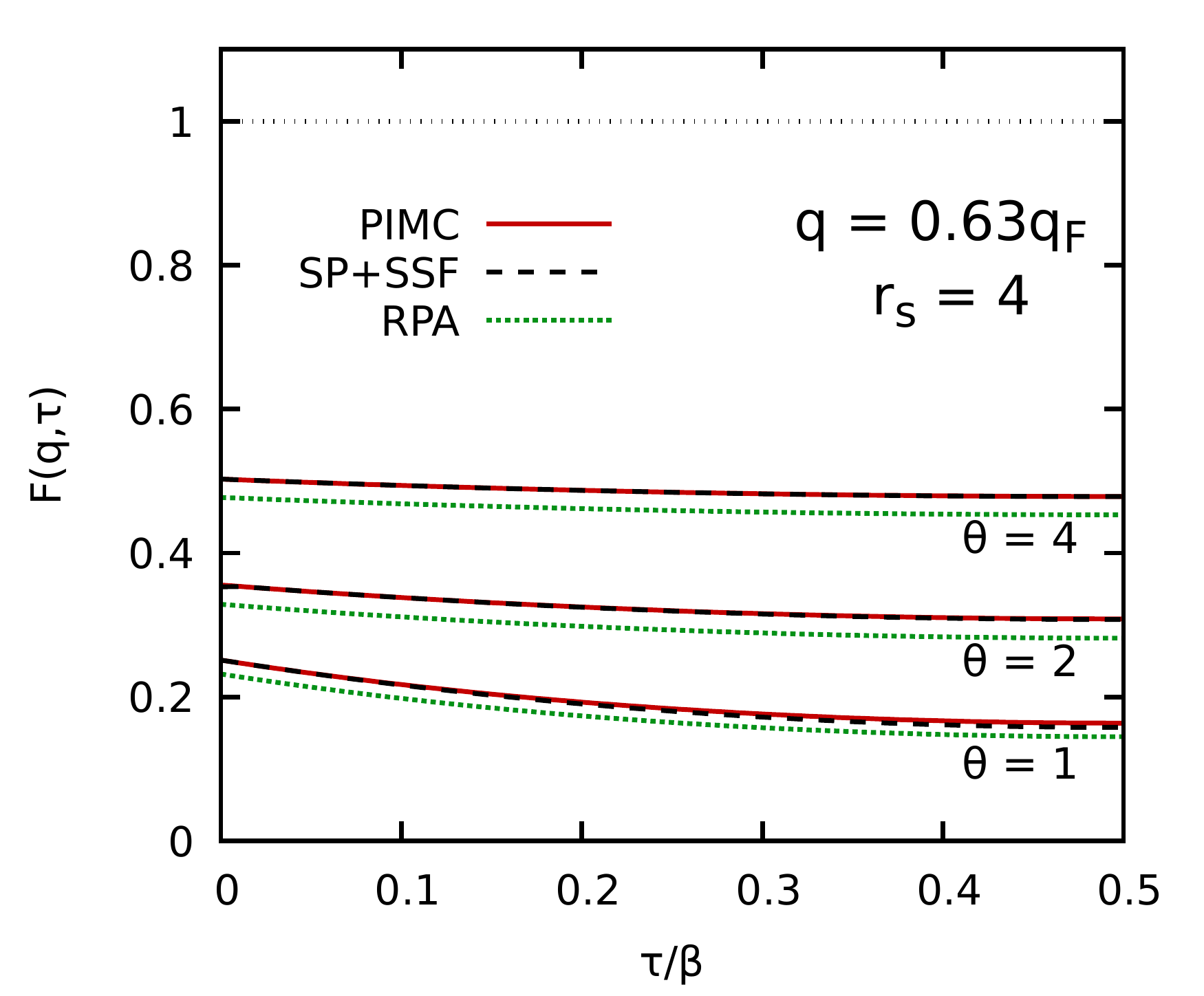}\\\vspace*{-0.965cm}\includegraphics[width=0.45\textwidth]{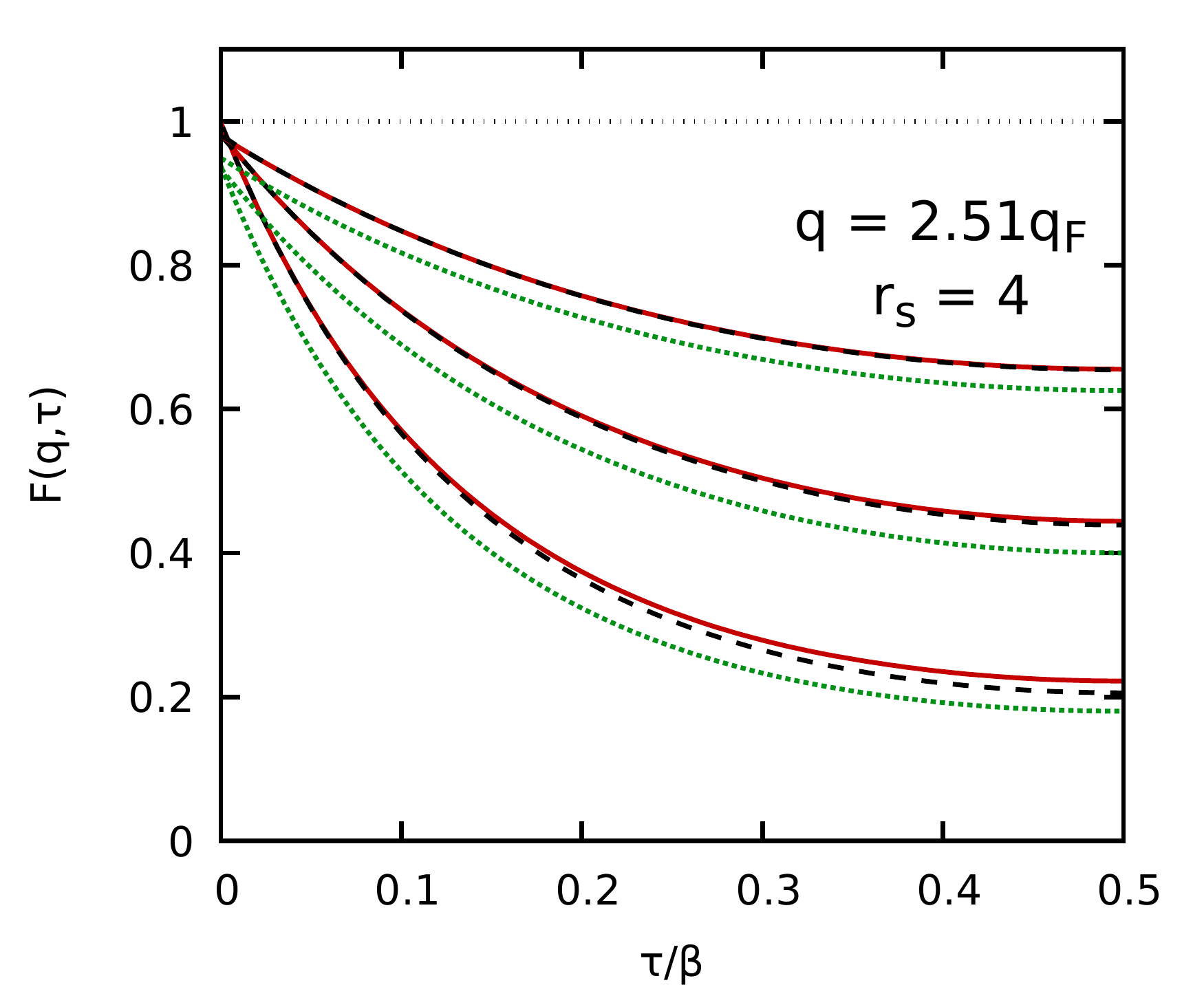}\\\vspace*{-0.96cm}\includegraphics[width=0.45\textwidth]{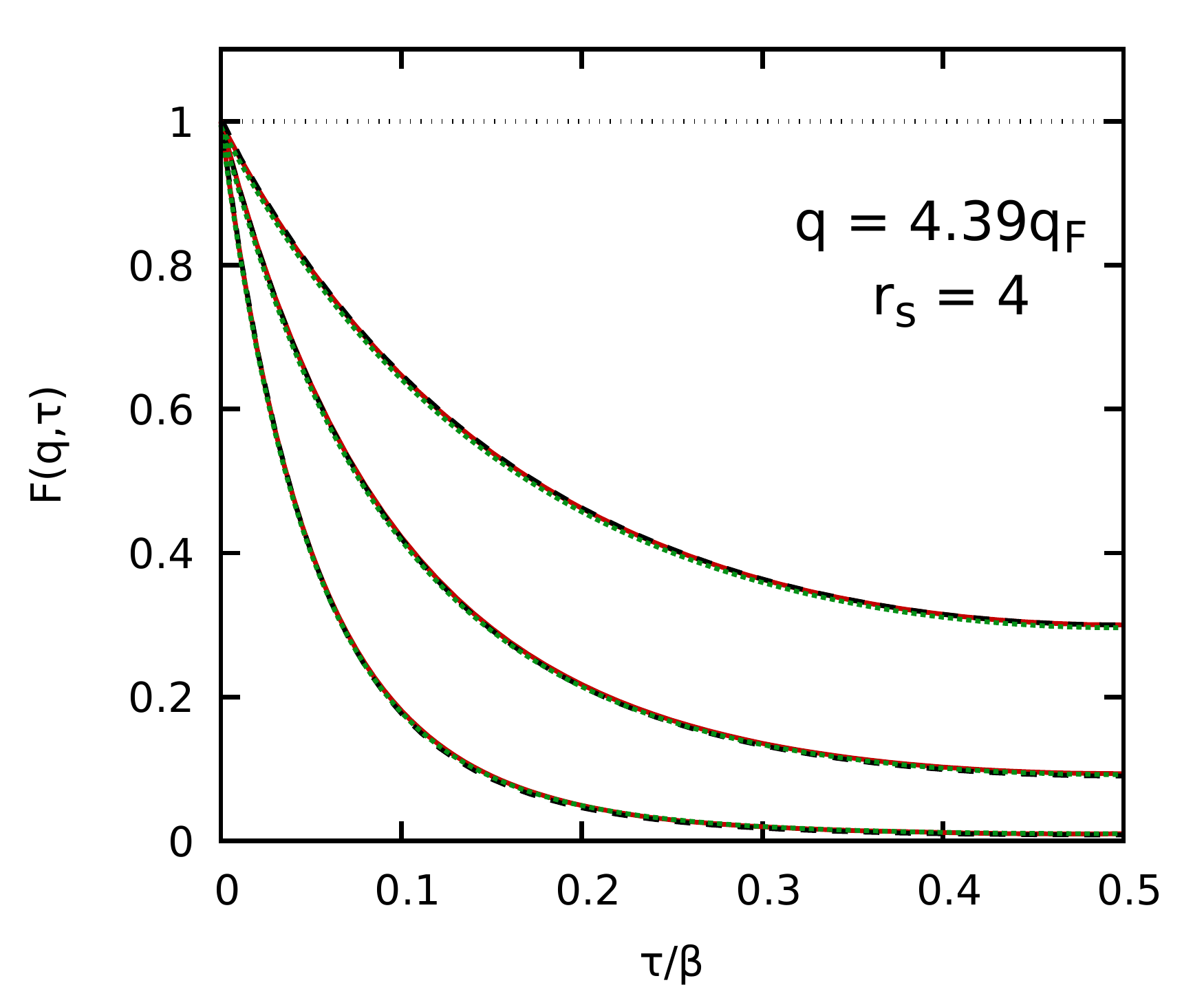}
\caption{\label{fig:theta}
ITCF for different $\Theta$. Solid red: PIMC; dotted green: RPA; dashed black: combination of $F_\textnormal{SP}(\mathbf{q},\tau)$ with $S(\mathbf{q})$ via Eq.~(\ref{eq:decompose}).
}
\end{figure}

Let us conclude this analysis by considering our new single-particle imaginary-time diffusion model $F_\textnormal{SP}(\mathbf{q},\tau)$ [see Eq.~(\ref{eq:F_SP}) in Sec.~\ref{sec:delocalization}], which is included in Fig.~\ref{fig:compare_decay_PIMC_theta1_rs} a the solid yellow curve.  Evidently, this simple, semi-analytical model becomes exact in the single-particle regime of $q\geq 3q_\textnormal{F}$, but does not reproduce either the collective plasmon limit, nor the \emph{roton minimum} for intermediate $q$ at $r_s=20$.
The combination of $F_\textnormal{SP}(\mathbf{q},\tau)$ with the correct static structure $F(\mathbf{q},0)=S(\mathbf{q})$ via Eq.~(\ref{eq:decompose}), see the long-dashed purple curves in Fig.~\ref{fig:compare_decay_PIMC_theta1_rs}, substantially mitigates this deficiency. More specifically, our model is very accurate at $r_s=4$ over the entire $q$-range. Indeed, no systematic errors can be resolved on the depicted scale for the smaller value of the imaginary time $\tau=\beta/10$, since Eq.~(\ref{eq:decompose}) becomes exact for $\tau=0$ both with respect to $F(\mathbf{q},0)$, and with respect to the first slope, which is given by the f-sum rule that has been investigated in detail in Fig.~\ref{fig:Derivative} above. 
Even in the more complicated case of $r_s=20$, the combination of the correct static structure with the simple single-particle model for the imaginary-time diffusion discussed in Sec.~\ref{sec:delocalization} is capable to reproduce the \emph{roton minimum} of $\Delta F_\tau(\mathbf{q})$ with remarkable precision; it only breaks down for $q\lesssim q_\textnormal{F}$, where it fails to attain the true plasmon limit in the collective regime. 
For $\tau=\beta/10$, even this flaw is basically removed, and the long-dashed purple curve gives a very good description of the entire wave-number range.

\subsection{Dependence on the temperature\label{sec:temperature}}

Let us conclude this investigation of the imaginary-time density--density correlation function of the warm dense UEG by briefly touching upon its dependence on the temperature. In Fig.~\ref{fig:theta}, we show $F(\mathbf{q},\tau)$ at $r_s=4$ for $q=0.63q_\textnormal{F}$ (top), $q=2.51q_\textnormal{F}$ (center), and $q=4.39q_\textnormal{F}$ (bottom). The solid red lines correspond to exact PIMC simulation data for $\Theta=4$, $\Theta=2$, and $\Theta=1$, and the curves are ordered with respect to temperature as it is indicated in panel a). Overall, we find a similar trend for all depicted wave numbers: the RPA (dotted green) underestimates the static limit of $F(\mathbf{q},0)=S(\mathbf{q})$, but exhibits the correct qualitative decay with respect to $\tau$. The combination of our new single-particle model for the imaginary-time diffusion and the exact static limit via Eq.~(\ref{eq:decompose}) overall fits even better to the PIMC reference data and captures the correct trends for all depicted temperatures and wave numbers. In fact, it becomes more accurate with increasing temperature, which can be understood intuitively in the following way. As we have previously shown in this work, Eq.~(\ref{eq:decompose}) becomes exact in the limit of $\tau\to0$ both with respect to $F(\mathbf{q},0)$ and also the slope at the origin. For higher temperatures, the imaginary-time propagation described in Sec.~\ref{sec:delocalization} above is carried out over a shorter distance in the $\tau$-domain. Therefore, systematic errors have a shorter distance to accumulate, which explains the increased relative accuracy for larger values of $\Theta$.

\section{Discussion\label{sec:summary}}

In this work, we have introduced a simple, semi-analytic model that gives new insights into the dependence of the ITCF $F(\mathbf{q},\tau)$ on the imaginary time $\tau$. In practice, we find that the combination of the static structure $S(\mathbf{q})=F(\mathbf{q},0)$ with the simple single-particle imaginary-time diffusion function $F_\textnormal{SP}(\mathbf{q},\tau)$ [cf.~Eq.~(\ref{eq:decompose}) above] gives a highly accurate description of the ITCF over a broad range of densities, temperatures and wave numbers. Remarkably, our model is even capable to capture the \emph{roton feature} of the strongly coupled electron liquid~\cite{dornheim_dynamic,Dornheim_Nature_2022,Dornheim_insight_2022}, which emerges due to the exchange--correlation induced alignment of pairs of electrons. Even in this regime, the effect of electron--electron correlations on the imaginary-time diffusion is comparably small. 

In addition, our model also very accurately captures the behaviour of the ITCF with respect to the temperature parameter $\Theta$, and generally agrees well with the PIMC results for the decay measure $\Delta F_\tau(\mathbf{q})$.

We are convinced that our new results constitute an important first step towards a more thorough theoretical understanding of dynamic quantum many-body effects formulated in the imaginary time. Such an improved understanding will be of direct use for the interpretation of \emph{ab initio} PIMC simulations of WDM~\cite{https://doi.org/10.48550/arxiv.2207.14716,Bohme_PRL_2022,review,dornheim_ML}, which give direct access to the ITCF. While previous direct PIMC simulations---without the fixed-node approximation, which prevents access to $F(\mathbf{q},\tau)$---have mostly been restricted to the UEG, future efforts will give results for the ITCF of real materials such as hydrogen~\cite{https://doi.org/10.48550/arxiv.2207.14716,Bohme_PRL_2022}. In addition, we note that previous models for $S(\mathbf{q},\omega)$ such as the Chihara decomposition~\cite{Chihara_1987,Gregori_PRE_2003}, but also time-dependent density functional theory calculations~\cite{Ramakrishna_PRB_2021,dynamic2}, can easily be translated to the imaginary-time domain, and benchmarked against upcoming PIMC results.
In addition, having an improved understanding of the physical properties of $F(\mathbf{q},\tau)$ will be helpful for the interpretation of XRTS experiments with WDM. As mentioned above, the latter give straightforward access to the ITCF~\cite{Dornheim_T_2022}, and contain the same physical information as the DSF. Performing the corresponding analysis of the observed XRTS signal in the $\tau$-domain, therefore, has the potential to give physical insights beyond temperature diagnostics without any models or approximations~\cite{Dornheim_insight_2022}.


\section*{Acknowledgments}
This work was partly funded by the Center of Advanced Systems Understanding (CASUS) which is financed by Germany's Federal Ministry of Education and Research (BMBF) and by the Saxon Ministry for Science, Culture and Tourism (SMWK) with tax funds on the basis of the budget approved by the Saxon State Parliament.
The PIMC calculations were carried out at the Norddeutscher Verbund f\"ur Hoch- und H\"ochstleistungsrechnen (HLRN) under grant shp00026, and on a Bull Cluster at the Center for Information Services and High Performance Computing (ZIH) at Technische Universit\"at Dresden.

\bibliography{bibliography}
\end{document}